\documentclass{article}

\usepackage[english]{babel}

\usepackage[letterpaper,margin=1in]{geometry}

\usepackage{amsmath}
\usepackage{amsthm}
\usepackage{amsfonts}
\usepackage{graphicx}
\usepackage{algorithm}
\usepackage{algorithmic}
\usepackage{subcaption}
\usepackage{enumitem}

\usepackage[colorlinks=true, allcolors=blue]{hyperref}
\usepackage{booktabs}
\usepackage{multirow}
\usepackage{xcolor}

\usepackage{tikz}

\def\supp{{\rm supp}}
\def\<{{\langle}}
\def\>{{\rangle}}

\def\cA{{\mathcal A}}
\def\cX{{\mathcal X}}
\def\cY{{\mathcal Y}}
\def\cZ{{\mathcal Z}}
\def\cL{{\mathcal L}}
\def\E{{\mathbb{E}}}
\def\P{{\mathbb{P}}}
\def\R{{\mathbb{R}}}
\def\cF{{\mathcal F}}
\def\pinball{{\sf pinball}}

\def\N{{\mathbb N}}
\def\E{{\mathbb E}}
\def\sT{{\sf{T}}}
\def\Var{{\rm Var}}
\def\Cov{{\rm Cov}}
\def\lo{{\sf lo}}
\def\hi{{\sf hi}}
\def\Ktrun{{K_{\sf trun}}}

\def\nsim{{n_{\sf sim}}}
\def\cP{{\mathcal P}}

\def\Z{{\mathbb Z}}
\def\SE{{\rm SE}}
\def\nobs{{n}}
\def\ntest{{n_{\sf te}}}
\def\ntrain{{n_{\sf tr}}}

\def\nval{{n_{\sf val}}}
\def\ERRTar{\textup{Err}_{\rm sto}}
\def\ERR{\textup{Err}}

\def\FCV{{\rm FCV}}

\def\val{{\rm val}}
\def\test{{\rm test}}
\def\QFCV{{\rm QFCV}}
\def\AQFCV{{\rm AQFCV}}

\def\hatSE{{\widehat{\rm SE}}}
\def\hatERR{{\widehat{\rm Err}}}

\def\hatPI{{\rm PI}}
\def\hatCI{{\rm CI}}
\def\hatRI{{\rm RI}}

\def\fea{{\rm val}}
\def\cN{{\mathcal N}}
\def\bphi{{\boldsymbol \phi}}
\def\btheta{{\boldsymbol \theta}}

\newcommand\blfootnote[1]{%
  \begingroup
  \renewcommand\thefootnote{}\footnote{#1}%
  \addtocounter{footnote}{-1}%
  \endgroup
}

\newtheorem{theorem}{Theorem}[section]

\newtheorem{lemma}[theorem]{Lemma}
\newtheorem{proposition}[theorem]{Proposition}
\newtheorem{assumption}{Assumption}

\begin{document}

\title{Uncertainty Intervals for Prediction Errors in Time Series Forecasting} 

\author{Hui Xu\thanks{Department of Statistics, Stanford University. E-mail: \url{huixu18@stanford.edu}} \and Song Mei\thanks{Department of Statistics and EECS, University of California, Berkeley. E-mail: \url{songmei@berkeley.edu}} \and Stephen Bates\thanks{Department of Statistics and EECS, University of California, Berkeley. E-mail: \url{stephenbates@cs.berkeley.edu}} \and Jonathan Taylor\thanks{Department of Statistics, Stanford University. E-mail: \url{jtaylo@stanford.edu}} \and Robert Tibshirani\thanks{Departments of Biomedical Data Science and  Statistics, Stanford University. E-mail: \url{tibs@stanford.edu}}}
\date{}

\maketitle

\begin{abstract}
% This is an awesome abstract.

Inference for prediction errors is critical in time series forecasting pipelines. However, providing statistically meaningful uncertainty intervals for prediction errors remains relatively under-explored. 
Practitioners often resort to forward cross-validation (FCV) for obtaining point estimators and constructing confidence intervals based on the Central Limit Theorem (CLT). The naive version assumes independence, a condition that is usually invalid due to time correlation. These approaches lack statistical interpretations and theoretical justifications even under stationarity. 

This paper systematically investigates uncertainty intervals for prediction errors in time series forecasting. We first distinguish two key inferential targets: the stochastic test error over near future data points, and the expected test error as the expectation of the former. The stochastic test error is often more relevant in applications needing to quantify uncertainty over individual time series instances. To construct prediction intervals for the stochastic test error, we propose the quantile-based forward cross-validation (QFCV) method. Under an ergodicity assumption, QFCV intervals have asymptotically valid coverage and are shorter than marginal empirical quantiles. In addition, we also illustrate why naive CLT-based FCV intervals
%,as commonly used by practitioners, 
fail to provide valid uncertainty intervals,
even with 
certain corrections. For non-stationary time series, we further provide rolling intervals by combining QFCV with adaptive conformal prediction to give time-average coverage guarantees. Overall, we advocate the use of QFCV procedures and demonstrate their coverage and efficiency through simulations and real data examples.

\blfootnote{Code for our experiments is available at~\url{https://github.com/huixu18/QFCV}.}

% Inference for prediction error is of critical importance in machine learning pipelines. However, it remains a relatively less explored area especially for time series forecasting. While point estimators in the form of forward cross-validation are more extensively studied, no prior work provides statistically meaningful interpretations for inference methods of prediction error. We start by formulating the problem with two different potential inference targets and argue that prediction interval for stochastic test error is a more useful quantity in time series. We examine two straightforward improvements of naive CLT (Central Limit Theorem) based confidence intervals by autocovariance correction and scaling transformation, and give analysis to understand why they are suboptimal. Our main contribution is that we propose a new framework of quantile-based forward cross-validation (QFCV) method with provable asymptotic coverage guarantee under ergodicity. QFCV methods give valid and well-calibrated coverages, and at the same time achieve efficiency gains in terms of shorter average interval widths than fixed oracle intervals. We also provide rolling prediction intervals based on QFCV by incorporating adaptive conformal inference (ACI) with delayed feedback, resulting in asymptotic time-average coverage guarantee under arbitrary non-stationarity. We compare the performance of our proposed methods with different baselines in both simulation and real data examples.
\end{abstract}

\tableofcontents

\section{Introduction}

% \sm{$\nobs$: current observed sample size. $\ntest$: size of the test dataset. $\ntrain$: size of training dataset. $\nval$: size of validation dataset. }

Inference for prediction errors is crucial in time series forecasting \cite{hamilton2020time}, with applications in a variety of domains including economics and inance \cite{franses2000non, cochrane1997time}, healthcare \cite{liu2015efficient}, supply chain \cite{aviv2003time, gilbert2005arima}, and energy forecasting \cite{deb2017review}. While many point estimators for prediction error have been proposed \cite{racine2000consistent, bergmeir2011forecaster,cerqueira2020evaluating, petropoulos2022forecasting, hewamalage2023forecast}, uncertainty quantification for prediction error in time series remains relatively under-explored. Indeed, existing naive methods for uncertainty quantifications of prediction errors tend to lack statistically meaningful interpretations. 

 % since training data and test data are both random draws from some super-population (We will defer the formal definition of $\ERRTar$ and relevant notations to Section \ref{sec:setting}). 
 %Uncertainty quantification for the target of $\ERRTar$ is a prediction interval and is a more natural goal, especially in time series applications \sm{this}. 
 %Uncertainty quantification for the expected prediction error $\ERR$ is a confidence interval. 

In particular, a widely used approach for inference for prediction error is cross-validation (CV) \cite{allen1974relationship, geisser1975predictive, stone1974cross}. Cross-validation is a resampling method that uses the prediction error of sub-sampled folds to infer the true prediction error. Although widely studied for independent, identically distributed (IID) data, cross-validation is often adapted to time series settings. To account for the temporal correlation in time series, each fold usually comprises continuous time windows instead of IID-resampled time indices. After dividing into multiple folds, practitioners usually use the mean and standard deviation of the validation errors to give an uncertainty interval for the prediction error, similar to the IID case. We call these CLT-based uncertainty intervals the ``naive forward cross-validation'' (FCV) intervals. However, it is not clear whether naive FCV intervals provide statistically meaningful coverage for prediction errors. 

To provide statistically meaningful uncertainty quantification, we should first clarify our inferential target and define prediction error precisely in the time series setting. Suppose we are given a forecaster trained on historical data. A natural notion of the inferential target is the forecaster's average loss over near future test points, a random variable called the \emph{stochastic test error} ($\ERRTar$). Another natural target is $\ERR = \E[\ERRTar]$, the expectation of $\ERRTar$ over all randomness including training and test data, called the \emph{expected test error}. The mathematical formulations of these targets are in Section \ref{sec:setting}. Note that these two prediction errors are often very close in the IID setting per the law of large numbers (LLN), since we are often interested in predicting a large number of IID test points. However, in times series settings, we are not interested in the long-term performance of forecasters, and LLN often does not apply. To distinguish, we call uncertainty intervals for $\ERRTar$ \emph{prediction intervals}, and for $\ERR$ \emph{confidence intervals}. 

%In general, it is more difficult to provide a prediction interval than a confidence interval due to higher variability in $\ERRTar$ \sm{This}. %\sm{Here maybe we don't want to talk about relative challenge, but which quantity makes more sense? }

% In particular, we propose adaptations of naive cross-validation and nested cross-validation confidence intervals in i.i.d setting \cite{bates2021cross} to time series context, called forward cross-validation (FCV) method. While FCV is designed to provide confidence intervals, naive applications may still undercover for $\ERR$ due to time correlation. Autocovariance adjustments are needed in order to restore valid coverage guarantee. 

In time series applications, providing coverage of the stochastic test error $\ERRTar$ is often more intuitive and useful than coverage of  the expected test error $\ERR$. For example, energy companies are typically interested in the fluctuation of the actual prediction error of power demand for the next few days, rather than the range of the hypothetical expected error. So this paper will focus more on providing prediction intervals to $\ERRTar$, but will also discuss confidence intervals for $\ERR$. 

% While confidence intervals may be comparable to prediction intervals when the test size is large enough, the gap can be non-negligible especially when test size is small. Existing CLT based confidence intervals cannot be adapted to produce prediction intervals due to potential violation of central limit theorem with small test size. 

The second important ingredient in providing statistically meaningful uncertainty quantification is to make appropriate statistical assumptions on the time series. While strong parametric assumptions like ARMA models are undesirable, certain distributional assumptions are needed for statistical inference. In this paper, we start with the assumption that the time series is \emph{stationary}. This assumption seems very strong, but to our knowledge, there are no existing statistically meaningful uncertainty intervals for prediction errors in stationary time series. Starting with stationarity allows an initial meaningful step towards uncertainty quantification for prediction errors in time series.

After clarifying the inferential target and statistical assumptions, we are ready to discuss inferential procedures for prediction errors. In this paper, we introduce and advocate quantile-based forward cross-validation (QFCV) prediction intervals, which provide statistically meaningful intervals with asymptotically valid coverage for prediction errors (Section \ref{sec:QFCV}). We will also illustrate why naive FCV intervals fail to provide valid uncertainty intervals, even with certain corrections (Section \ref{sec:FCV}). Finally, we will consider non-stationary time series, and provide rolling intervals by combining QFCV with adaptive conformal prediction methods to give certain coverage guarantees under non-stationarity (Section \ref{sec:rolling}).

% Notice that, any meaningful inference methods in time series should at least be valid/meaningful under stationarity. Therefore we will first assume stationarity in motivating our method and proving theoretical results based on ergodicity. Modifications by adapting to nonstationarity with online learning techniques will be discussed in later sections. 

\subsection{An illustrating example}\label{sec:illustrating_example}
\begin{figure}[t]
\centering
  \includegraphics[width = 0.46\linewidth]{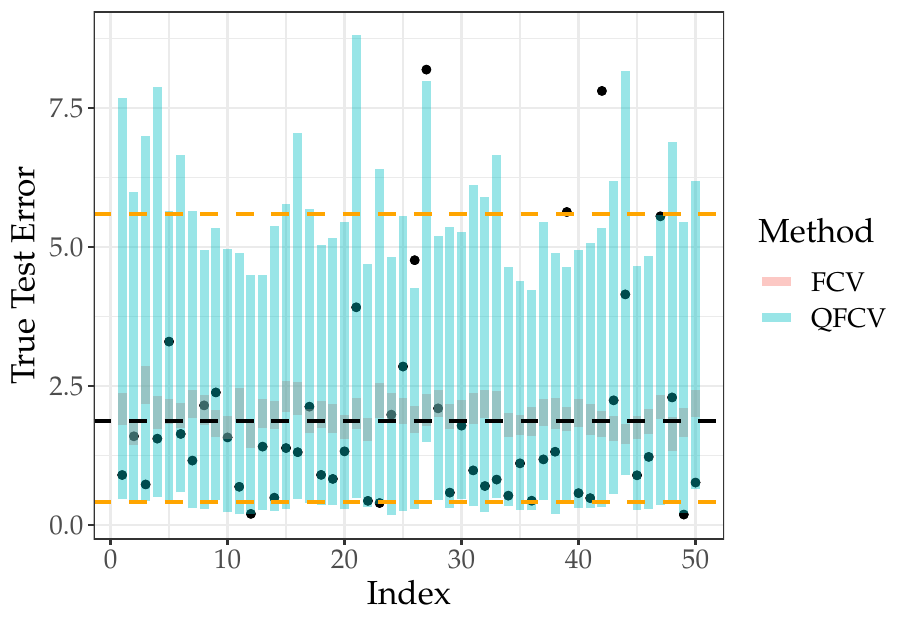}
\caption{Stochastic and expected test errors, along with naive FCV and QFCV intervals for 50 simulation instances. Black points show the stochastic test error per instance. The black dashed line is the expected test error. Red and blue bars represent naive FCV and QFCV intervals, respectively, for each run. Orange dashed lines indicate the $5\%$ and $95\%$ empirical quantiles of stochastic test errors over $500$ instances. %The black and orange lines are computed over $500$ simulation instances as populations. 
}\label{fig:motivation}
\end{figure}

We provide a simulation example to illustrate the difference between stochastic and expected test errors in time series forecasting, and compare naive FCV and QFCV methods, where the former is a naive CLT-based uncertainty interval (Eq \eqref{eqn:naive_fcv_interval} in Section \ref{sec:FCV_procedure}) as a baseline and the latter is our proposed uncertainty interval that will be elaborated in Algorithm \ref{alg:fcv_qc} in Section \ref{sec:QFCV}. At each time step $t \in \{1, \ldots, 1000\}$, we observe feature vector $x_t \in \R^p$ where $p=20$, with IID standard normal entries, and outcomes $y_t = x_t^\sT \beta +\varepsilon_t$ following a linear model. Here $\beta = (1,1,1,1,0,\ldots)^\sT \in \mathbb{R}^p$, and noise $\{\varepsilon_t\}_{t=1}$ is an AR(1) process with parameter $\phi = 0.5$. The forecaster is a Lasso regression trained on the last $40$ time indices ($961$ to $1000$). Our target is the prediction error over the unseen future $20$ data points ($1001$ to $1020$).

We calculate the coverage of naive FCV and QFCV intervals over $500$ simulation instances. Figure \ref{fig:motivation} shows these intervals for the first 50 instances. Numerical results demonstrate $90\%$ nominal coverage by QFCV for the stochastic test error, which we theoretically justify in Section \ref{sec:QFCV}. In contrast, FCV severely undercovers the stochastic test error at only $12.8\%$ coverage, and slightly undercovers the expected test error at $78.0\% coverage$. In Section \ref{sec:FCV}, we explain the issues with naive FCV and investigate certain correction approaches. Overall, this example illustrates the difference between stochastic and expected test errors in time series, and shows that QFCV provides valid coverage while naive FCV does not.

\subsection{Our contribution}

This paper provides a systematic investigation of inference for prediction errors in time series forecasting. We identify two inferential targets, the stochastic test error and the expected test error, and discuss different methods for providing uncertainty intervals. Our contribution is three-fold. 
\begin{itemize}[leftmargin=*]
\item We propose quantile-based forward cross-validation (QFCV) methods to provide prediction intervals for stochastic test error (Section \ref{sec:QFCV}). We prove that QFCV methods have asymptotically valid coverage under the assumption of ergodicity. Through numerical simulations, we show that QFCV methods give well-calibrated coverages. In particular, when choosing an appropriate auxiliary function, QFCV provides shorter intervals than the naive empirical quantile method, especially for smooth time series. 

\item We examine forward cross-validation (FCV) methods, which are CLT-based procedures adapted from cross-validation in the IID setting (Section \ref{sec:FCV}). These methods seem like a natural choice for practitioners. We identify issues with the naive FCV method that underlie its failure to cover the stochastic and expected test errors. We also investigate two modifications to naive FCV that partially address these problems. However, these modifications do not perfectly restore valid coverage without strong assumptions. 
\item For non-stationary time series, we provide rolling intervals by combining the QFCV method with a variant of the adaptive conformal inference, which we name the AQFCV method (Section \ref{sec:rolling}). AQFCV method produces rolling intervals with asymptotically valid time-average coverage under arbitrary non-stationarity. We articulate the difference between time-average coverage and the frequentist's notion of instance-average coverage. Finally, we demonstrate AQFCV's performance on both simulated and real-world time series data. 
\end{itemize}

%due to wrong target of inference and biased standarad error estimates with time correlation. This includes methods based on cross-validation, such as forward cross-validation (FCV) with lag size as a tunable hyperparameter. We propose two modifications to naive FCV to address the problems of time correlation and target mismatch respectively. However, modifications cannot perfectly restore valid coverage due to inherent limitation based on the central limit theorem. 

\subsection{Related work}

\paragraph{Cross validation with independent data}

Cross-validation (CV) \cite{allen1974relationship, geisser1975predictive, stone1974cross} is used ubiquitously to estimate the prediction error of a model. A comprehensive review of CV methods is presented in \cite{arlot2010survey}. Despite the simplicity of CV, the seemingly basic question ``what is the inferential target of cross-validation?” has engendered considerable debate \cite{zhang1995assessing,hastie2009elements,yousefa2020leisurely, rosset2019fixed, wager2020cross, bates2021cross}. 

Turning to the question of inference for prediction error, extensive studies provided various intervals of prediction error based on CV \cite{dietterich1998approximate, nadeau1999inference, markatou2005analysis, efron1983estimating, efron1997improvements, yousef2021estimating, ledell2015computationally, benkeser2020improved, austern2020asymptotics, bayle2020cross, bates2021cross}. In particular, \cite{dietterich1998approximate, nadeau1999inference, markatou2005analysis} provides estimators of the CV standard error based on either the sample splitting method or the moment method. Works such as \cite{efron1983estimating, efron1997improvements, yousef2021estimating} provide estimators of the standard error of bootstrap estimate of the prediction error, based on the influence function method. Asymptotically consistent estimators of CV standard error were provided in \cite{ledell2015computationally, benkeser2020improved, austern2020asymptotics, bayle2020cross}. Furthermore, \cite{bates2021cross} introduced the nested CV method, which produces an unbiased estimator of the mean squared error for the CV point estimator. However, all these estimators are designed for datasets with independent data points and do not extend to time series datasets, the setting considered in this work. 

Finally, we remark that CV is often used for comparing predictive models, such as in model selection or determining optimal values for hyperparameters of learning algorithms \cite{stoica1986model, shao1993linear, zhang1993model, yang2007consistency, fong2020marginal, varma2006bias, tibshirani2009bias, yang2007consistency, wager2020cross, lei2020cross}.

\paragraph{Inference for test error in a time series dataset }
In the time series setting, many variants of cross-validation methods have been proposed to provide point estimators for prediction errors \cite{burman1994cross, racine2000consistent, bergmeir2011forecaster, bergmeir2012use, bergmeir2014usefulness, bergmeir2018note, cerqueira2020evaluating, petropoulos2022forecasting, hewamalage2023forecast}. The basic idea  is to only use past data as training data and future data as test data. Time series prediction error estimation has several different setups (see, for example, \cite{hewamalage2023forecast}). The setup we adopt is the ``rolling window" approach, wherein a fixed size of the training dataset is used to fit the forecaster. This setup is quite common in scenarios where the available dataset size is large \cite{mulinka2018remember}. Furthermore, in this setup, inference methods could have meaningful statistical interpretation under only the stationarity assumption. Another setup, the ``expanding window" setting, uses an incrementally expanding training dataset to fit the forecaster over time. In this context, it is challenging to confer meaningful statistical interpretation to any prediction error estimator, even under stationary conditions. 

In this work, instead of providing point estimators, our primary focus is to provide prediction intervals and confidence intervals for prediction errors within the context of time series analysis. To the best of our knowledge, there seems to be no existing literature addressing this particular topic. 

% \sm{Name: FCV? }

% \begin{itemize}
% \item \cite{petropoulos2022forecasting}: review of forecasting methods, theory and practice. 
% \item \cite{hewamalage2023forecast}: Common practice for evaluating the performance of forecasters are flawed. Provide guidance of performance evaluation in practice. No discussion on confidence interval, though mentioned hypothesis testing for model selection. 
%     \item Evaluating time series forecasting models: An empirical study on performance estimation methods \cite{cerqueira2020evaluating}: empirical experiments suggest that blocked cross-validation can be applied to stationary time series
%  \item \cite{bergmeir2011forecaster}: Empirical study, conducted experiments for different performance evaluation methods. Only point estimation.   
% \item \cite{bergmeir2012use}: Empirical study, compare CV and sample splitting. Finding: CV in practice is OK though theoretically invalid. Suggest block CV. 
%     \item \cite{bergmeir2014usefulness}:  Empirical study for blocked CV. 
%     \item A note on the validity of cross-validation for evaluating autoregressive time series prediction \cite{bergmeir2018note}: Theoretically showed that CV on AR process give consistent estimator of the test error (PE), better than sample splitting (OOS). 
% \item \cite{burman1994cross}: Proposed h-"block" cross-validation in time series data. (\cite{burman1992data} propose bias correction when dependent data)
%     \item \cite{racine2000consistent}: showed in consistency of hblock CV. Proposed hv-block CV and showed consistency. 
    
% \end{itemize}

\paragraph{Conformal inference in time series}

A recent line of work focuses on predictive inference (i.e., providing intervals for prediction) in the time series setting \cite{gibbs2021adaptive, gibbs2022conformal, feldman2022conformalized, bhatnagar2023improved, zaffran2022adaptive, xu2022sequential, sun2022copula, lin2022conformal}. These works are mostly based on adaptive conformal inference (ACI) \cite{gibbs2021adaptive, gibbs2022conformal}, a term derived from ``conformal prediction" methods \cite{vovk1999machine, shafer2008tutorial, angelopoulos2021gentle}. Utilizing online learning techniques, the ACI method produces a sequence of prediction intervals with an asymptotic guarantee of valid time-average coverage. 

Our work differs in its focus, with the inferential target being the prediction error, as opposed to the prediction value, the target of the ACI method \cite{gibbs2021adaptive}. However, our {\AQFCV} method borrows the techniques of ACI to provide prediction intervals for prediction error, also providing an asymptotic guarantee of valid time-average coverage. 

% \begin{itemize}
% \item Our method is related to the conformal prediction literature \cite{vovk1999machine, shafer2008tutorial, angelopoulos2021gentle}, since the QCV interval is a prediction interval.  
% \item Adaptive conformal prediction \cite{gibbs2021adaptive, gibbs2022conformal, feldman2022conformalized, bhatnagar2023improved, zaffran2022adaptive, xu2022sequential, sun2022copula, lin2022conformal}
% \end{itemize}

\section{Settings and notations}
\label{sec:setting}

\begin{figure}[t]
    \centering
    \includegraphics[width = 0.65\linewidth]{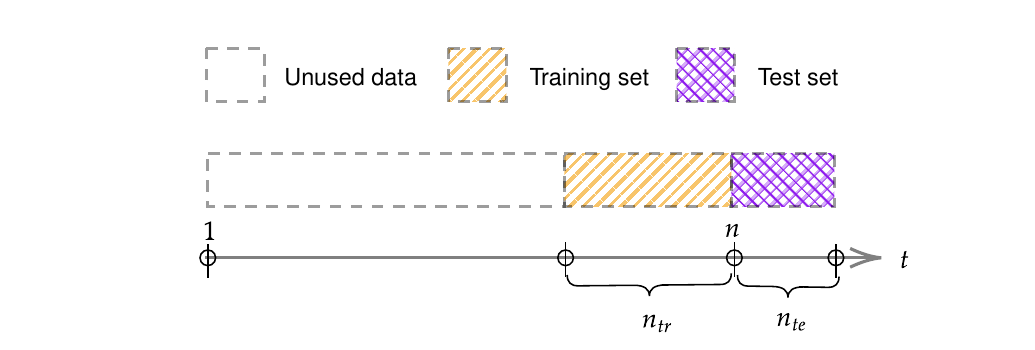}
    \caption{Diagram illustration of prediction error evaluation in time series forecasting.}
    \label{fig:data}
\end{figure}

In the context of time series forecasting, we are provided with a stream of data points $\{ z_t \}_{t \ge 1} \subseteq \cZ \equiv \cX \times \cY$, with $z_t = (x_t, y_t)$. In time series without additional context $x_t$, we can consider the context $x_t$ as a constant value of $1$. By the time index $\nobs$, we have observed the first $\nobs$ pairs $z_{1:\nobs} \in \cZ^\nobs$, and we aim to sequentially predict the forthcoming $\ntest$ outcomes $y_{\nobs + 1: \nobs + \ntest}$ using the contexts $x_{\nobs + 1: \nobs + \ntest}$. 

\paragraph{Stochastic prediction error.} We begin with a function $\hat f: \cX \times \cZ^{\ntrain}$ that forecasts future responses, along with a loss function $\ell: \cY \times \cY \to \R$. An interesting inferential target is the \emph{stochastic prediction error} $\ERRTar$, representing the average losses associated with the next $\ntest$ data points, where the forecaster is trained on the most recent $\ntrain$ data points: 
\begin{equation}\label{eqn:ERRTar_simplified}
\ERRTar = \frac{1}{\ntest} \sum_{t = \nobs +1}^{\nobs + \ntest} \ell(\hat f ( x_t; z_{\nobs - \ntrain + 1 : \nobs}), y_t). 
\end{equation}
See Figure \ref{fig:data} for a diagram illustration of prediction error evaluation in time series forecasting. 

In a more general scenario, we permit the forecaster to provide varying forecasts for different future time steps. In other words, we have a vector function $\hat f: \cX^{\ntest} \times \cZ^\ntrain  \to \cY^{\ntest}$ that is trained on the last $\ntrain$ data points and is designed to predict the next $\ntest$ data points. The stochastic prediction error, in this case, can be similarly expressed as:
\begin{equation}\label{eqn:ERRTar}
\begin{aligned}
\ERRTar =&~ \frac{1}{\ntest} \sum_{t = \nobs +1}^{\nobs + \ntest} \ell(\hat f_{t-\nobs}(x_{\nobs + 1 : \nobs + \ntest}; z_{\nobs - \ntrain + 1 : \nobs}), y_{n+1:n+\ntest}).% = \frac{1}{\ntest} \sum_{t = 1}^{\ntest} \ell(\hat f_{t}(x_{\nobs + 1 : \nobs + \ntest}; z_{\nobs - \ntrain + 1 : \nobs}), y_{t + \nobs}). 
\end{aligned}
\end{equation}
Note that (\ref{eqn:ERRTar}) is slightly more general than (\ref{eqn:ERRTar_simplified}), albeit with a more complicated notation. Throughout the paper, we will assume that our forecasters are identical at each future prediction step and primarily focus on the (\ref{eqn:ERRTar_simplified}) formulation of the stochastic prediction error. However, it is important to keep in mind that our methods can accommodate the general case of a non-identical prediction function.

\paragraph{Expected prediction error.} Another potential inferential target is the \emph{expected prediction error}, denoted as $\ERR$. Assuming a distribution over $\{ z_t \}_{t \ge 1}$, the expected prediction error is defined as the expectation of the stochastic prediction error, $\ERRTar$, 
\begin{equation}\label{eqn:ERR}
\ERR = \E[\ERRTar].
\end{equation}
The expectation is taken with respect to all the randomness variables, namely, $z_{1:\nobs + \ntest}$ and possible randomness present in the forecaster $\hat f$. Notably, due to time correlation, it is probable that
\begin{align*}
\ERR \neq \E\Big[\ell(\hat f ( x_{n+1}; z_{\nobs - \ntrain + 1 : \nobs}), y_{n+1})\Big], ~~~~
\ERR \neq \lim_{T \to \infty} \frac{1}{T-t+1}\sum_{t = n+1}^T \E\Big[\ell(\hat f ( x_{t}; z_{\nobs - \ntrain + 1: \nobs}), y_{t})\Big]. 
\end{align*}
Indeed, the expected prediction error $\ERR$ also depends on the number of test points $\ntest$, different from the IID setting. Additionally, we should not expect $\ntest$ to be a significantly large number, since we are typically concerned with the forecaster's performance in the near future. In such a scenario, we should not expect $\ERRTar$ to concentrate around $\ERR$: as is illustrated in Figure \ref{fig:motivation}, the black dashed line represents the expected prediction error $\ERR$, while the black dots represent the stochastic prediction error $\ERRTar$. 

\paragraph{Prediction intervals and confidence intervals.} In this paper, our primary focus is quantifying the uncertainty for both stochastic and expected prediction errors. When the stochastic prediction error $\ERRTar$ is our inferential target, we aim to offer a \emph{prediction interval} (PI), denoted as $\hatPI^{\alpha} \subseteq \R$, based on the observed data $z_{1:\ntrain}$. Conversely, when the expected prediction error $\ERR$ is our target, our goal is to provide a \emph{confidence interval} (CI), denoted as $\hatCI^\alpha \subseteq \R$, also based on the observed data $z_{1:\ntrain}$: 
\begin{equation}\label{eqn:PI_CI}
\P\Big(\ERRTar \in \hatPI^{\alpha}\Big) \stackrel{\cdot}{=} 1 - \alpha,~~~~~~~~~~~~~\P\Big(\ERR \in \hatCI^{\alpha}\Big) \stackrel{\cdot}{=} 1 - \alpha. 
\end{equation}
In many practical scenarios, our interest leans more towards providing a prediction interval for the stochastic prediction error $\ERRTar$. This preference stems from the fact that the prediction interval of $\ERRTar$ incorporates the fluctuation of the realized loss that we will incur on future unseen data. Capturing this fluctuation is crucial for risk control and uncertainty quantification. In contrast, a confidence interval for the expected prediction error $\ERR$ marginalizes over the randomness in $\ERRTar$, providing less information about the variability of the test error in a single time series instance.

% \sm{Discussion: why $\E[\ell] \neq \ERR$ for a single $\ell$; why $\ERRTar$ is a better inferential target (refer to figure 1). }

% \begin{assumption}[Stationarity and Ergodicity]
% For all $k, t \in \mathbb{N}_+$, we have 
% \[
% z_{1 + k : t+k} \deq z_{1:t}, 
% \]
% and the process $\{ z_t\}_{t \ge 1}$ is ergodic.  
% \end{assumption}

\section{Quantile-based procedures for prediction intervals}\label{sec:QFCV}
In this section, we first present the Quantile-based Forward Cross-Validation (QFCV) method used for generating prediction intervals for the stochastic prediction error $\ERRTar$ (Section \ref{sec:QFCV_algorithm}). We then demonstrate in Section \ref{sec:validity_QFCV},  under the stationary and ergodicity assumptions, that the QFCV prediction interval guarantees an asymptotically valid coverage of $\ERRTar$. Following this, Section \ref{sec:simulation_QFCV} provides a series of numerical simulations, while Section \ref{sec:discussion_QFCV} offers a detailed discussion. 

% In this section, we introduce Quantile-based forward cross-validation (QFCV) methods for providing prediction intervals for the inference target $\ERRTar$, as defined in Eq. (\ref{eqn:ERRTar_simplified}). Notice that our target $\ERRTar$ for QFCV is a random variable that depends on future observations. This differs from its expectation $\ERR = \E[\ERRTar]$, which is the target of FCV and NFCV methods. We will show that any QFCV method will provide a prediction interval $\hatPI$ for $\ERRTar$ with asymptotically valid coverage guarantee, i.e., $\P( \ERRTar \in  \hatPI ) \approx 1 - \alpha$. 

\subsection{The QFCV prediction interval}\label{sec:QFCV_algorithm}

\begin{algorithm}[t]
  \caption{Quantile-based forward cross-validation (QFCV)}
  \label{alg:fcv_qc}
  \small
  \begin{algorithmic}[1]
    \REQUIRE Dataset $\{ z_t\}_{t \in [n]} \subseteq \cZ$. Auxilliary function $\cA: \cZ^{\ntrain} \times \cZ^{\nval} \to \R^m$. Coverage tolerance $\alpha$. Function class $\cF \subseteq \{ \R \to \R\}$. Let $K = \lfloor ( \nobs - \ntrain - \nval - \ntest )/\Delta \rfloor + 1$. 
    \STATE Compute $\{ (\ERR_{i}^{\fea}, \ERR_{i}^{\test}) \}_{i \in [K]}$ and $\ERR_\star^{\fea}$ using Eq. (\ref{eqn:Err_fea_test_definition}). 
    \STATE For $\beta = \alpha/2$ and $\beta = 1 - \alpha/2$, compute 
    \[
    \hat f^{\beta} = \arg\min_{f \in \cF} \frac{1}{K} \sum_{i = 1}^{K} \pinball_\beta\Big(\ERR_{i}^{\test} - f(\ERR_{i}^{\fea}) \Big), 
    \]
    where $\pinball_\beta(t) = - t \cdot (1\{ t \le 0 \} - \beta)$ is the pinball loss with quantile parameter $\beta$. 
    %$\hat f_n^{\alpha/2}(e)$ (and $\hat f_n^{1-\alpha/2}(e)$) be the output of quantile regression on the  $\{ (\textup{Err}_{i}^{val}, \textup{Err}_{i}^{test}) \}_{i \in [n - D - V - T + 1]}$ with $\alpha/2$ (and $1 - \alpha/2$) quantile. 
    \STATE Output $\hatPI^\alpha_\QFCV = [\hat f^{\alpha/2}(\ERR^{\fea}_\star), \hat f^{1 - \alpha/2}(\ERR^{\fea}_\star)]$. 
  \end{algorithmic}
\end{algorithm}

In stationary time series, a straightforward approach for providing a prediction interval involves estimating the quantile of the stochastic prediction error distribution using the observed dataset. We introduce a more refined method named Quantile-based Forward Cross-Validation (QFCV). In a nutshell, QFCV estimates the quantile function of the stochastic prediction error, $\ERRTar$, based on certain user-specified features. These features are expected to have high predictive power to $\ERRTar$, an example being the validation error. 

\begin{figure}[t]
    \centering
    \includegraphics[width = 0.6\linewidth]{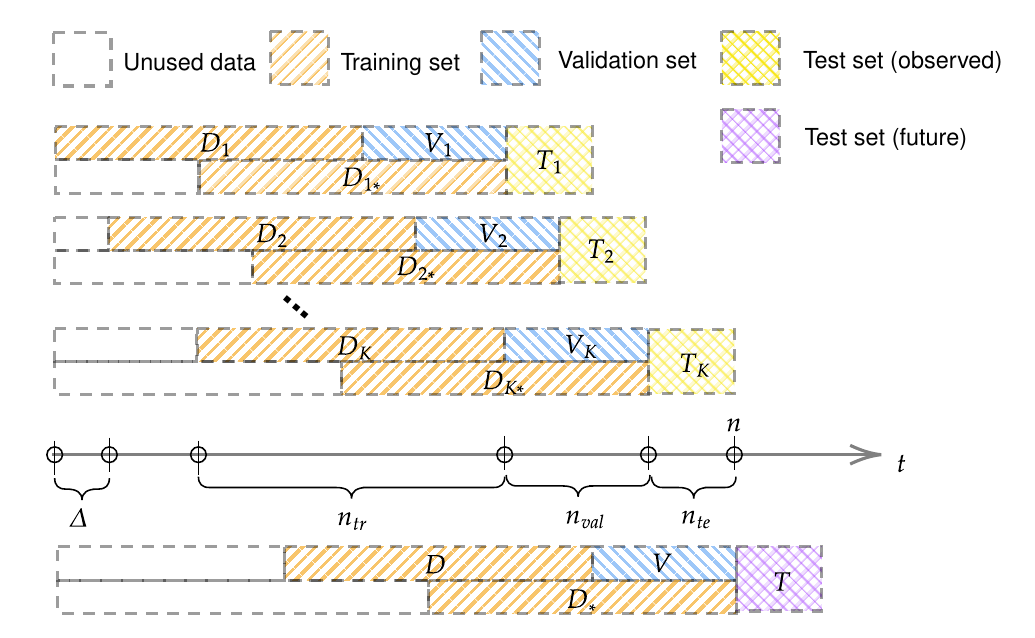}
    \caption{Diagram illustration for an example of QFCV method.}
    \label{fig:diagram_QFCV}
\end{figure}

To introduce QFCV, we organize the time indices into multiple possibly overlapping time windows, which we denote as $\{ D_i, V_i, D_{i \star}, T_i\}_{i \in [K]}$ and $\{ D, V, D_\star, T\}$, as depicted in Figure \ref{fig:diagram_QFCV}. The time windows $\{ D_i, D_{i \star}, D, D_\star\}$ share the same size, denoted as $\ntrain$; $\{ V_i, V\}$ share the same size, denoted as $\nval$; $\{ T_i, T\}$ share the same size, denoted as $\ntest$. In addition, for each $i \in [K]$, each time window in $\{ D_{i+1}, V_{i+1}, D_{(i+1) \star}, T_{i+1}\}$ shifts from $\{ D_i, V_i, D_{i \star}, T_i\}$ by $\Delta$ indices, where $\Delta (K-1) + \ntrain + \nval + \ntest \le \nobs$. Spelling out the elements of each time window, we have 
\begin{equation}
\begin{aligned}
D_i =&~ \{(i-1)\Delta + 1, \ldots, {(i-1)\Delta + \ntrain}\}, ~~~~~~~~~~~~~~~~~~~~~ D = \{\nobs - \ntrain - \nval + 1, \ldots,  \nobs - \nval \}, &~\\
V_i =&~ \{ (i-1)\Delta + \ntrain + 1, \ldots,    (i-1)\Delta + \ntrain + \nval\},~~~~~~~~~~~~~~~~~~~~~ V = \{\nobs - \nval + 1, \ldots,  \nobs \},&~ \\
D_{i\star} =&~ \{(i-1)\Delta + \nval + 1, \ldots, {(i-1)\Delta + \ntrain + \nval}\}, ~~~~~~~~~~~~~~~~~~~~ D_\star = \{\nobs - \ntrain + 1, \ldots,  \nobs\},&~ \\
T_i =&~ \{ (i-1)\Delta + \ntrain + \nval, \ldots, (i-1)\Delta + \ntrain + \nval + \ntest -1 \}, ~~~~~~~ T = \{ \nobs + 1, \ldots, \nobs + \ntest \}. &~
\end{aligned}
\end{equation}
Furthermore, given a user-specified auxiliary function that generates a set of covariates for predicting the stochastic prediction error, denoted as $\cA: \cZ^{\ntrain} \times \cZ^{\nval} \to \R^m$, we define tuples $\{ (\ERR_{i}^{\fea}, \ERR_{i}^{\test}) \}_{i \in [K]}$ and $(\ERR_\star^\fea, \ERR_\star^{\test})$ by the equations below:
\begin{equation}\label{eqn:Err_fea_test_definition}
\begin{aligned}
\ERR_{i}^{\fea} =&~ \cA(z_{D_i}, z_{V_i}) \in \R^m, ~~~~~~~~~&
\ERR_{i}^{\test} =&~ \textstyle \frac{1}{|T_i|} \sum_{t \in T_i} \ell(\hat f(x_t; z_{D_{i\star}}), y_t) \in \R, \\
\ERR_\star^{\fea} =&~ \cA(z_{D}, z_{V})\in \R^m,~~~~~~~~~& \ERR_\star^{\test} =&~ \textstyle 
 \frac{1}{|T|} \sum_{t \in T} \ell(\hat f(x_t; z_{D_\star}), y_t)\in \R.
\end{aligned}
\end{equation}
Here, $z_D$ is a condensed notation for $\{ z_t \}_{t \in D}$. It is important to note that $\{ (\ERR_{i}^{\fea}, \ERR_{i}^{\test}) \}_{i \in [K]}$ and $\ERR_\star^\fea$ can be computed from the observed dataset $\{ z_t \}_{t \in [\ntrain]}$, but $\ERR_\star^{\test}$ depends on future unobserved data points. Indeed, $\ERR_\star^{\test}$ is the same as the stochastic prediction error $\ERRTar$, as defined in Eq. (\ref{eqn:ERRTar_simplified}). Assuming the time series is stationary, it is evident that $\{ (\ERR_{i}^{\fea}, \ERR_{i}^{\test}) \}_{i \in [K]}$ and $(\ERR_\star^\fea, \ERR_\star^{\test})$ are identically distributed, though they are not independent due to time correlation. 

The QFCV method, given these definitions, can be summarized in the three steps below. First, calculate $\{ (\ERR_{i}^{\fea}, \ERR_{i}^{\test}) \}$ and $\ERR_\star^\fea$ given by Eq. (\ref{eqn:Err_fea_test_definition}). Then, predict quantiles of $\{\ERR_{i}^{\test}\}$ based on feature vectors $\{\ERR_{i}^{\fea}\}$ by minimizing the empirical quantile loss to obtain quantile functions $\hat f^{\alpha/2}$ and $\hat f^{1 - \alpha/2}$. Finally, the QFCV prediction interval is given by $\hatPI^\alpha_\QFCV = [\hat f^{\alpha/2}(\ERR^{\fea}_\star), \hat f^{1 - \alpha/2}(\ERR^{\fea}_\star)]$. Detailed steps are provided in Algorithm \ref{alg:fcv_qc}. 
%where  $D_i = \{ i, {i+1}, \ldots, {i + D - 1}\}$, $V_i = \{ {i + D}, \ldots, {i + D + V - 1}\}$, $T_i = \{ {i + D + V}, \ldots, {i + D + V + T - 1}\}$, $B_\star = \{\nobs - \ntrain, \nobs\}$, and $T_\star = \{ \nobs + 1, \ldots, \nobs + \ntest \}$. Notice that $\ERR_\star^{\test} = \ERRTar$ is our inference target. 
% Then we provide a prediction interval $\hatCI^\QFCV$ for $\ERR_\star^{\test}$ using Algorithm \ref{alg:fcv_qc}. 
% \begin{assumption}\label{ass:stationary_ergodic}
% Assume that $\{ z_t \}_{t \ge 1}$ is a stationary ergodic stochastic process. 
% \end{assumption}

\begin{figure}[t]
    \centering
    \includegraphics[width = 0.5\linewidth]{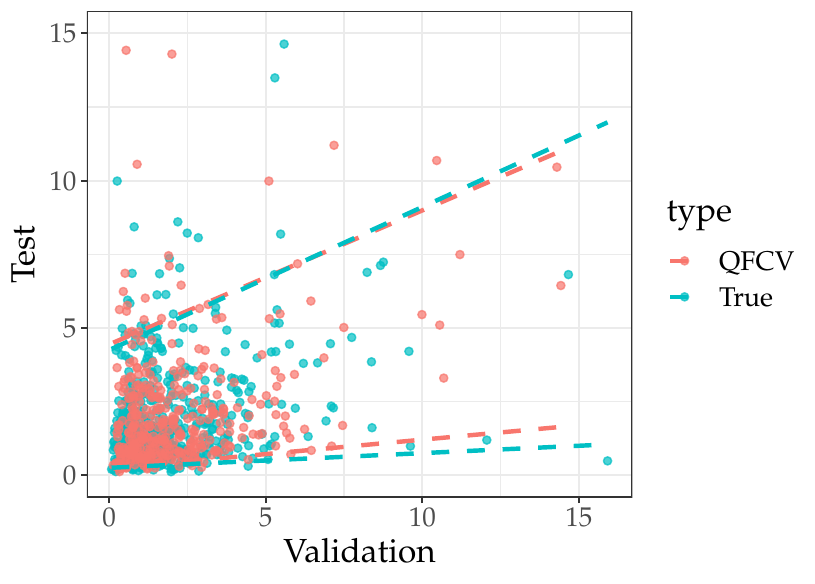}
    \caption{Scatter plot of test and validation errors of the true population and QFCV. We set $\ntrain = 40$ and $\nval = \ntest = 20$. True population points represent validation and test error pairs  $(\ERR_\star^{\fea}, \ERR_\star^{\test})$, derived from $500$ simulation runs. QFCV points are from the $\{(\ERR_i^{\fea}, \ERR_i^{\test})\}_{i \in [K]}$ set, computed from a single time series via (\ref{eqn:Err_fea_test_definition}). The colored dashed lines are the fitted lines for the $5\%$ and $95\%$ quantile regressions, with the red dashed lines representing the QFCV prediction interval. }
    \label{fig:QFCV_biplot}
\end{figure}

\paragraph{Choice of auxiliary function.} There is a certain amount of flexibility in choosing the auxiliary function $\cA$. As we will show in the upcoming section, regardless of the specific choice of the auxiliary function, the QFCV algorithm will yield an asymptotically valid prediction interval, under the assumption of ergodicity. In practice, we aim to choose $\cA(z_{D_i}, z_{V_i})$ such that it is as predictive of $\ERR_i^{\test}$ as possible. Our preferred choice for $\cA$ is given by the validation error, a sample splitting estimator of the test error: 
\begin{equation}\label{eqn:validation_aux_function}
\textstyle \cA(z_{D_i}, z_{V_i}) = \frac{1}{| V_i |} \sum_{t \in V_i} \ell(\hat f(x_t; z_{D_i}), y_t). 
\end{equation}
We illustrate the QFCV method in Figure \ref{fig:QFCV_biplot}, where the QFCV prediction interval is illustrated with red dashed lines.

\subsection{Asymptotic validity of QFCV prediction intervals} \label{sec:validity_QFCV}

We provide theoretical guarantees on the coverage of QFCV prediction intervals, under the assumption of a stationary time series. We present methods for relaxing the stationarity assumption in Section \ref{sec:rolling}.
\begin{assumption}\label{as:stationarity}
The time series $\{ z_t \}_{t \ge 1}$ is a stationary ergodic stochastic process, such that for any function $g: \cZ^k \to \R$ with $\E[| g(z_{1:k})|] < \infty$ and $\varepsilon > 0$, we have 
\[
\lim_{n \to \infty} \P\Big( \Big\vert \frac{1}{n}\sum_{t = 1}^n g(z_{t: t + k - 1}) - \E[g(z_{1:k})] \Big\vert \ge \varepsilon \Big) = 0. 
\]
\end{assumption}
Given this assumption, ${(\ERR_t^{\fea}, \ERR_t^{\test})}_{t \ge 1}$ is likewise stationary ergodic and shares the same distribution as $(\ERR_\star^{\fea}, \ERR_\star^{\test})$. Let $\mathcal{L}$ denote this common distribution. For simplicity, we will assume $\mathcal{L}$ has a density with respect to the Lebesgue measure. While this is a strong assumption, it could be relaxed by more sophisticated analysis. 
\begin{assumption}\label{ass:density}
The distribution $\cL$ of $(\ERR_\star^{\fea}, \ERR_\star^{\test})$ has a density with respect to the Lebesgue measure. 
\end{assumption}
Moreover, we denote $\cF_\beta$ as the class of $\beta$-quantiles of the conditional distribution of $\ERR_\star^{\test}$ given $\ERR_\star^{\fea}$. This class can be expressed as minimizers of the pinball loss, 
\[
\cF_\beta = \Big\{ g^* \in \arg\min_{g: \R^m \to \R} \E_{\cL}\big[\pinball_\beta\big(\ERR^{\test} 
 -g(\ERR^{\fea})\big)\big]  \Big\}. 
\]
We assume the function class $\cF$ used in the QFCV algorithm (Algorithm \ref{alg:fcv_qc}) can realize the $\alpha/2$ and $1 - \alpha/2$ quantile classes, $\cF_{\alpha/2}$ and $\cF_{1-\alpha/2}$. Furthermore, for simplicity, we assume that $\cF$ is finite, a strong assumption that could also be relaxed. 
\begin{assumption}\label{ass:finiteness_realizability}
Assume that $\cF$ is a finite function class, and $\cF_{\alpha/2} \cap \cF \neq \emptyset$, $\cF_{1 - \alpha/2} \cap \cF \neq \emptyset$. 
\end{assumption}
We are ready to state the asymptotically valid coverage for the QFCV method, as detailed in Theorem \ref{thm:conditional_coverage}. The proof of this theorem can be found in Appendix \ref{sec:proof_conditional_coverage}. A more general statement going beyond the finite function class assumption (Assumption \ref{ass:finiteness_realizability}) can be found in Appendix \ref{sec:general_conditional_coverage}.

\begin{theorem}\label{thm:conditional_coverage}
Let Assumption \ref{as:stationarity}, \ref{ass:density} and \ref{ass:finiteness_realizability} hold.  Let $\hatPI^\alpha_\QFCV$ be the output of Algorithm \ref{alg:fcv_qc}. Then we have 
\begin{equation}\label{eqn:asymptotic_validity}
\lim_{n \to \infty} \P\Big( \ERRTar \in \hatPI^\alpha_\QFCV \Big) = 1 - \alpha. 
\end{equation}
\end{theorem}
We remark that the choice of the auxiliary function $\cA: \cZ^{\ntrain \times \nval} \to \R^m$ does not affect the asymptotic validity of QFCV prediction intervals. However, when the dimension $m$ is large, more samples will often be required for the asymptotic regime to take effect. In practice, we would like to choose $m$ to be moderately small and to be some function such that $\ERR_\star^{\fea}$ and $\ERR_\star^{\test}$ have a large mutual information.

\subsection{Comparing QFCV methods via numerical simulation}
\label{sec:simulation_QFCV}

% \begin{align*}
%     & y_t = x_t^\sT \beta + \varepsilon_t\\
%     & x_t \sim \mathcal{N}(0, I_p)\\
%     & \varepsilon_t = \sum_{k=1}^a \phi_k \varepsilon_{t-k} + \eta_t + \sum_{i=1}^b \theta_i \eta_{t-i},
% \end{align*}
% where $\{\eta_t\}$ is a white noise process with mean $0$ and variance $1$, and coefficient $\beta = (1,1,1,1,0,\ldots, 0) \in \mathbb{R}^p$ for $p = 20$ features. Notice that $\{\varepsilon_t\}_{t \geq 1}$ follows a standard autoregressive-moving-average model $\text{ARMA}(a,b)$ with AR parameters $\phi = (\phi_1, \ldots, \phi_a)^\sT$ and MA parameters $\theta = (\theta_1, \ldots, \theta_b)^\sT$. We consider two choices of ARMA models. 
% \begin{itemize}
%     \item $\text{ARMA}(1,0)$ with $\phi \in [0,1]$. 
%     \item $\text{ARMA}(1,20)$ with $\phi \in [0,1]$ and $\theta = (0.1,\ldots 0.9, 1.0, 1.0, 0.9, \ldots, 0.1) \in \mathbb{R}^{20}$. 
% \end{itemize}

\begin{figure}[t]
    \centering
    \includegraphics[width = 0.6\linewidth]{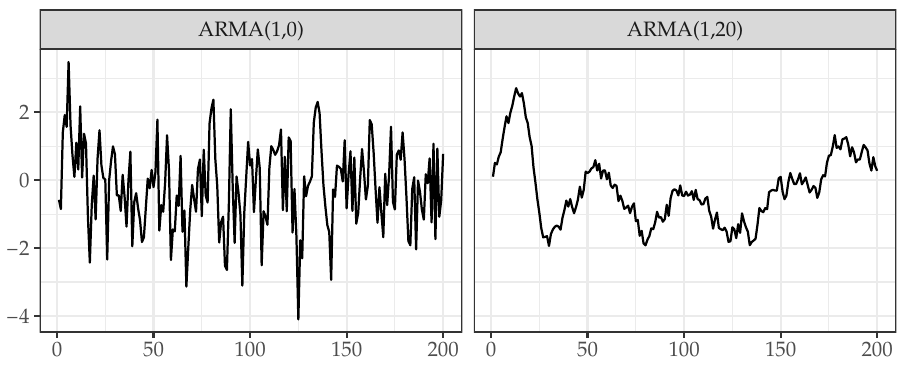}
    \caption{Instances of $\{\varepsilon_t\}_{t \ge 1}$ following two different ARMA(a,b). The left hand side plots ARMA(1,0) and the right hand side plots $0.2 \, \times \,$ARMA(1,20). In both plots, we choose $\phi = 0.5$. In the right plot, $\btheta = (0.1, 0.2,\ldots, 1, 1, \ldots, 0.1)$ is a wedge-shaped list of weights, resulting in a more smooth stationary process.}
    \label{fig:ARMA}
\end{figure}

We evaluate QFCV methods with different choices of auxiliary functions $\cA$ through a simulation study. Recall the setting as in Section \ref{sec:setting}, we generate a time series with $\nobs = 2000$ historical data, train a model on $\ntrain = 40$ training set, and aim to construct prediction intervals for the test error over the next $\ntest \in \{5, 20 \}$ unseen points (see Figure \ref{fig:data} for a diagram of this dataset division). The simulated time series $\{z_t\}_{t\geq 1}$ is generated as follows: for each time step $t$, $z_t = (x_t, y_t) \in \R^p \times \R$ with $p = 20$, where 
\begin{align}\label{eqn:simulation_linear_model}
y_t = x_t^\sT \beta + \varepsilon_t, ~~~~~x_t \sim_{iid} \cN(0, I_p),~~~~~ \{ \varepsilon_t \}_{t \ge 1} \sim {\rm ARMA}(a, b).
\end{align}
We fix the true coefficient vector as $\beta = (1,1,1,1,0,\ldots, 0)^\sT \in \mathbb{R}^p$, where the first four coordinates are $1$, and the rest of the coordinates are $0$. The noise process $\{\varepsilon_t\}_{t \geq 1}$ follows a standard autoregressive-moving-average model $\text{ARMA}(a, b)$ with AR coefficients $\bphi = (\phi_1, \ldots, \phi_a)^\sT$ and MA coefficients $\btheta = (\theta_1, \ldots, \theta_b)^\sT$. That is 
\begin{equation}
\textstyle \varepsilon_t = \sum_{k=1}^a \phi_k \cdot  \varepsilon_{t-k} + \eta_t + \sum_{i=1}^b \theta_i \cdot  \eta_{t-i},~~~~~~~~~ \{ \eta_t\}_{t \ge 1} \sim_{iid} \sim \cN(0, 1). 
\end{equation}
We consider two ARMA parameter settings: 
\begin{itemize}
    \item[(1)] $\text{ARMA}(1,0)$ with $\bphi = \phi \in [0,1]$. 
    \item[(2)] $\text{ARMA}(1,20)$ with $\bphi = \phi \in [0,1]$ and $\btheta = (0.1, 0.2, \ldots 0.9, 1, 1, 0.9, \ldots, 0.1)^\sT \in \mathbb{R}^{20}$. 
\end{itemize}
Both processes are stationary, but setting (2) exhibits smoother sample paths than setting (1), as visualized in Figure \ref{fig:ARMA} which plots realizations of $\{\varepsilon_t\}_{t\ge 1}$ for the first 200 time steps. We posit that setting (2), as a smoother noise process, will yield a higher correlation between validation and test errors $(\ERR_\star^{\fea}, \ERR_\star^{\test})$.

% \begin{figure}[t]
% \centering
% \begin{minipage}[c]{.45\linewidth}
%   \centering
%   \includegraphics[width = \linewidth]{forward_simulation/ARMA(1,0,0).pdf}
%   \subcaption{ARMA(1,0)}
% \end{minipage}%
% \begin{minipage}{.45\linewidth}
%   \centering
%   \includegraphics[width = \linewidth]{forward_simulation/ARMA(1,0,20).pdf}
%   \subcaption{$0.2 \,\times\,$ARMA(1,20)}
% \end{minipage}
% \caption{Instances of $\{\varepsilon_t\}_{t \ge 1}$ following two different ARMA(a,b). In both plots, we choose $\phi = 0.5$. In the right plot, $\btheta = (0.1, 0.2,\ldots, 1, 1, \ldots, 0.1)$ is a wedge-shaped list of weights, resulting in a more smooth stationary process.}
% \label{fig:ARMA}
% \end{figure}

We consider QFCV methods utilizing the following auxiliary functions. For $m \in \{1, \ldots, \nval\}$, we split the validation set $V_i$ into $m$ continuous and non-overlapping subsets $V_{i1}, \ldots, V_{im} \subseteq V_i$. We define the auxiliary function $\cA_m$ of dimension $m$ as the vector of validation errors on each subset:  
\[
\cA_m(z_{D_i}, z_{V_i}) = \left(\frac{1}{|V_{i1}|}\sum_{t \in V_{i1}}\ell(\hat{f}(x_t;z_{D_i}), y_t), \ldots  \frac{1}{|V_{im}|}\sum_{t \in V_{im}}\ell(\hat{f}(x_t;z_{D_i}), y_t)\right)^\sT \in \mathbb{R}^m.
\]
For $m = 0$, we use a trivial auxiliary function $\cA_0(z_{D_i}, z_{V_i}) \equiv 1$. We denote by $\text{QFCV}(m)$ the QFCV method utilizing auxiliary function $\cA_m$. We provide a few special examples as follows. 
\begin{itemize}
    \item $\text{QFCV}(0)$: $\cA_0(z_{D_i}, z_{V_i}) = 1$. Here, QFCV essentially calculates the empirical quantile of $\{ \ERR_i^{\test} \}$, equivalent to the intercept-only quantile regression. 
    \item $\text{QFCV}(1)$: $\cA_1(z_{D_i}, z_{V_i}) = \frac{1}{|V_i|} \sum_{t \in V_i} \ell(\hat{f}(x_t;z_{D_i}), y_t) \in \mathbb{R}$. Here, QFCV uses quantile regression with the single feature given by the average validation error.   
    \item $\text{QFCV}(\nval)$: $\cA_{\nval}(z_{D_i}, z_{V_i}) = ( \ell(\hat{f}(x_t;z_{D_i}), y_t) )_{t \in V_i} \in \mathbb{R}^\nval$. Here, QFCV uses quantile regression with $\nval$ features given by the individual validation errors. 
\end{itemize}
In all cases, we use linear quantile regression of $\{\ERR_i^{\test}\}$ against $\{\ERR_i^{\val}\}$, choosing $\cF$ as the class of linear functions.

% \sm{Setting: $ARMA(p, q)$ for different choice of $(p, q)$ and parameters. }

% \sm{A figure: 3 methods, FCV, NFCV, QFCV. X-axis: $\phi$ as the parameter of ARMA. Fix $\theta$. Y-axis: coverage, length. }

% \sm{Sample size small: NFCV fail? }
\begin{figure}[t]
\centering
\begin{minipage}[c]{.5\linewidth}
  \centering
  \includegraphics[width = \linewidth]{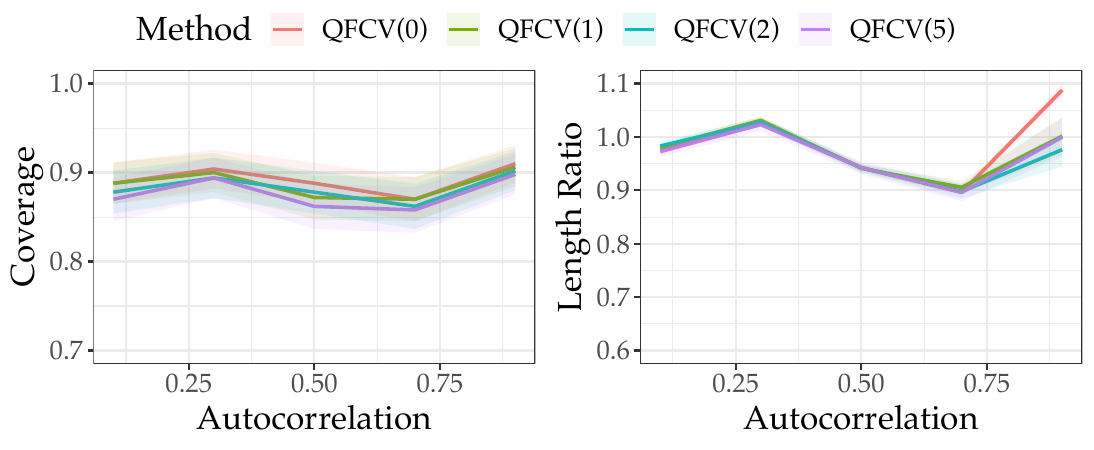}
  \subcaption{ARMA(1,0), $\nval = 5$.}
\end{minipage}%
\begin{minipage}{.5\linewidth}
  \centering
  \includegraphics[width = \linewidth]{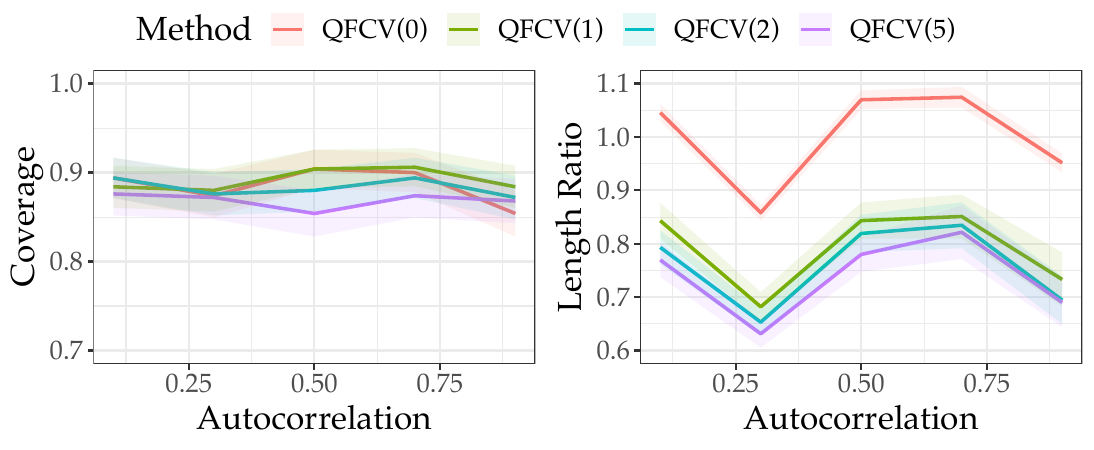}
  \subcaption{ARMA(1,20), $\nval = 5$.}
\end{minipage}
\vskip0.5cm
\begin{minipage}[c]{.5\linewidth}
  \centering
  \includegraphics[width = \linewidth]{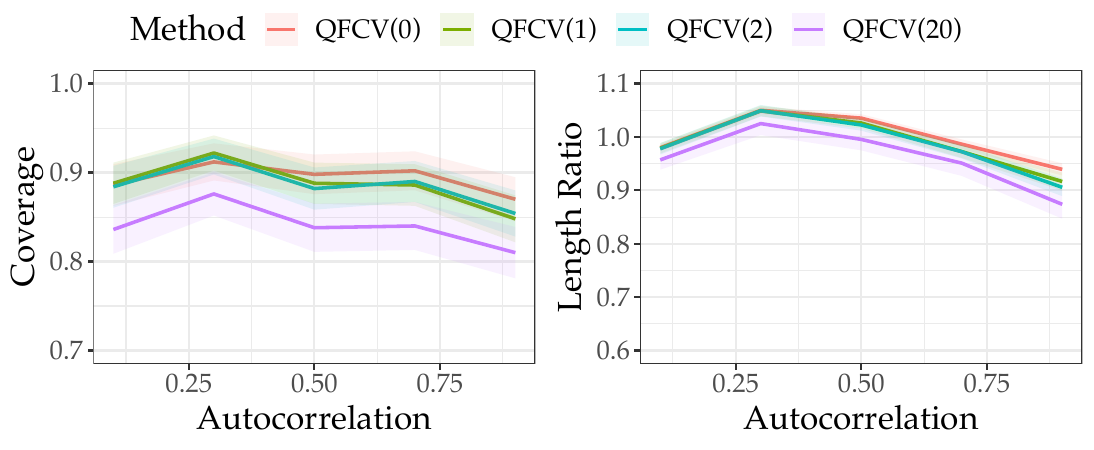}
  \subcaption{ARMA(1,0), $\nval = 20$.}
\end{minipage}%
\begin{minipage}{.5\linewidth}
  \centering
  \includegraphics[width = \linewidth]{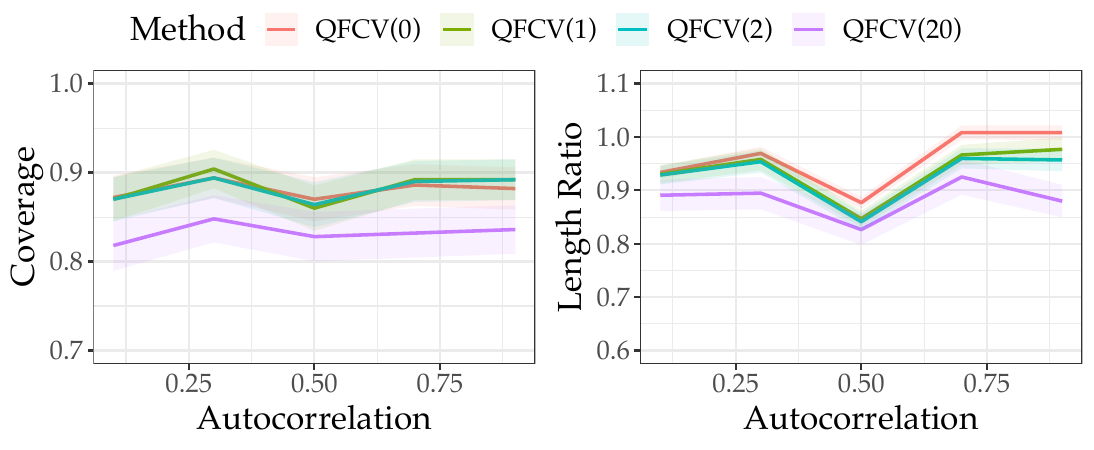}
  \subcaption{ARMA(1,20), $\nval= 20$.}
\end{minipage}
\caption{Performance comparison of QFCV prediction intervals on simulated time series (\ref{eqn:simulation_linear_model}). The time series length is fixed at $\nobs = 2000$, with $\ntrain = 40$, $\nval = \ntest \in \{5, 20\}$, and $p = 20$. The x-axis represents the autocorrelation parameter $\phi \in [0, 1]$. Each panel displays two plots: (i) actual coverage of the $90\%$ nominal QFCV intervals, and (ii) ratio of the QFCV interval length to the true marginal quantiles interval length. The true marginal quantiles interval is defined by the $5\%$ and $95\%$ quantiles of the true test error distribution. Each curve displays the mean and standard error over $500$ simulation instances. }
\label{fig:sim_QFCV}
\end{figure}

Figure \ref{fig:sim_QFCV} reports the coverage and length of QFCV intervals under different simulation settings. The time series is generated according to model (\ref{eqn:simulation_linear_model}), where the noise $\{\varepsilon_t\}$ follows either an $\text{ARMA}(1,0)$ process (panels (a) and (c)) or $\text{ARMA}(1,20)$ process (panels (b) and (d)). For both noise processes, we fix the MA coefficients $\btheta$ and vary the AR coefficient $\phi$ along the x-axis over $[0,1]$ to modulate autocorrelation. The nominal coverage level is $90\%$. To normalize the interval lengths, we divide the QFCV lengths by the true marginal quantiles interval length, defined as the difference between the $95\%$ and $5\%$ quantiles of the true test error distribution. Each curve displays the mean and standard error over $500$ simulation instances.

% The simulation results comparing coverage and length of prediction intervals for different variants of QFCV methods over $500$ simulation runs are presented in Figure \ref{fig:sim_QFCV}. For both $\text{ARMA}(1,0)$ and $\text{ARMA}(1,20)$, we fix moving-average parameters $\btheta$ and vary autocorrelation parameters $\phi$ along the x-axis. Length of prediction interval on the y-axis is normalized by the true marginal quantiles prediction interval length, which is the difference between the $95\%$ and $5\%$ quantiles of true population level target test errors. In other words, the denominator of true marginal quantiles length is the shortest possible fixed interval that can achieve nominal coverage rate. When the length ratio is at $1$, we achieve similar performance as a fixed true marginal quantiles interval. But when the error process is more smooth, we expect larger correlation between validation errors and test errors, and therefore some advantage in terms of smaller prediction interval length after quantile regression. 

\begin{figure}[t]
    \centering
    \includegraphics[width = 0.9\linewidth]{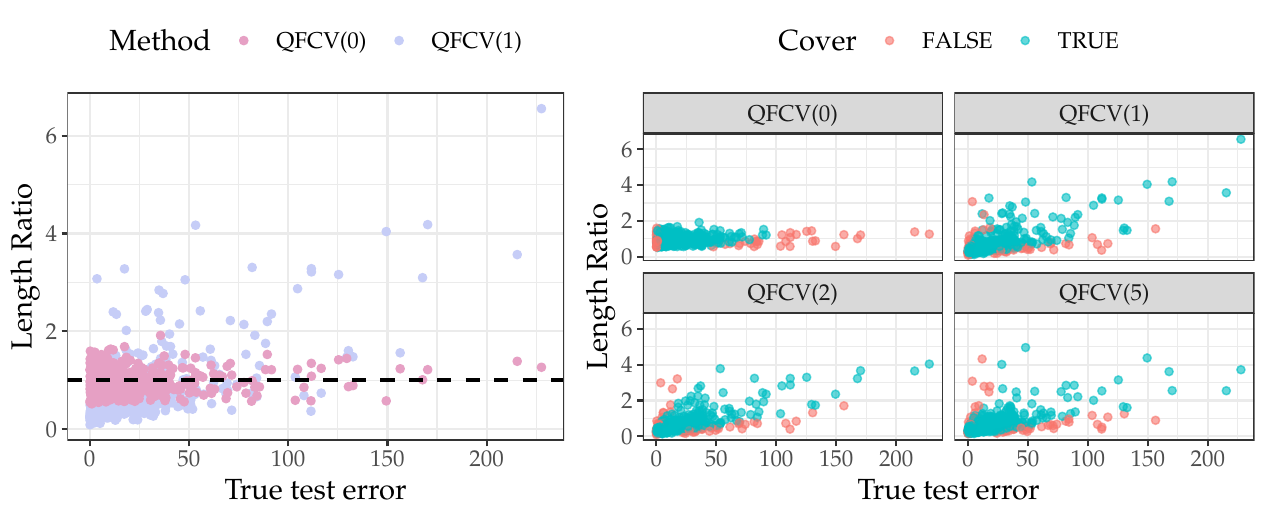}
    \caption{Scatter plots comparing interval length ratios versus true test error over $500$ simulation instances. The noise process is $\text{ARMA}(1,20)$ with $\phi = 0.9$ and $\nval = 5$. The left panel compares $\text{QFCV(0)}$ and $\text{QFCV}(1)$. The right panel illustrates coverage patterns for each QFCV method, with points colored by whether the true test error falls inside or outside of the constructed interval. }
    \label{fig:QFCV_adapt}
\end{figure}

Figure \ref{fig:sim_QFCV} shows that $\text{QFCV}(m)$ achieves close to $90\%$ coverage at the nominal level when $m$ is less than or equal to $2$. However, for $m=20$ in panels (c) and (d), the coverage deviates further from the nominal level, likely because quantile regression requires more samples to be consistent with a $20$-dimensional feature vector. In terms of interval lengths, all QFCV methods produce similar lengths to the true marginal quantiles on data with non-smooth $\text{ARMA}(1,0)$ noise (panels (a) and (c)). However, for smooth $\text{ARMA}(1, 20)$ noise (panels (b) and (d)), $\text{QFCV}(m>0)$ can substantially outperform the true marginal quantiles length, especially when $\nval=5$ (panel (b)). This occurs because with smooth noise, the validation error $\ERR_\star^{\val}$ becomes more predictive of the actual test error $\ERR_\star^{\test}$.

% It can be seen from Figure \ref{fig:sim_QFCV} columns 1 and 3 that $\text{QFCV}(m)$ generally results in correct coverage at nominal level of $90\%$ when $m$ is not too large. Coverage seems problematic when $m = 20$ because of slower mixing rate with larger feature dimensions. This is not a contradiction to asymptotic validity, but an evidence of insufficient sample size. However, we do not recommend using large values of $m$ because of inefficient data usage and slow mixing rate without too much gain in power. In terms of length ratio, we can see from column $2$ that all variants of QFCV methods have approximately the same prediction interval length when the noise follows a non-smooth AR process. On the other hand, for the relatively more smooth process in column $4$, we can see that $\text{QFCV}(0)$ has worse average length ratio than the other QFCV methods with both intercept and slope. Due to potentially higher predictive power of validation error for true test error, QFCV methods using validation errors as features in quantile regression can gain some advantage, as seen from smaller average prediction interval length for QFCV methods with $m > 0$.

To explain why $\text{QFCV}(m>0)$ produces smaller average interval lengths than $\text{QFCV}(0)$, Figure \ref{fig:QFCV_adapt} presents scatter plots of the length ratios versus true test errors over $500$ simulation instances. The noise process is taken to be $\text{ARMA}(1,20)$ with $\phi = 0.9$ and $\nval = 5$. The left panel shows that $\text{QFCV}(0)$ (the empirical quantile method) generates interval lengths close to the true marginal quantiles for all ranges of the true test error. In contrast, $\text{QFCV}(1)$ assigns smaller intervals for small true test errors. The right panel colors points by whether they are mis-covered. We see that while $\text{QFCV}(0)$ tends to mis-cover points with large test errors, the mis-covered points for $\text{QFCV}(> 0)$ are more uniformly distributed across the test error range. This indicates that by adapting to the validation error, $\text{QFCV}(> 0)$ produces prediction intervals that are overall shorter while maintaining coverage. 

% To better explain the phenomenon that $\text{QFCV}(m)$ for $m > 1$ has a better length than $\text{QFCV}(0)$, we present a scatter plot of the interval length ratios against true test errors in Figure \ref{fig:QFCV_adapt}. On the left panel, it can be observed that $\text{QFCV}(0)$ (i.e. empirical quantile method) produces interval lengths approximately the same as the fixed true marginal quantiles interval length for all ranges of true test errors, while $\text{QFCV}(1)$ can cover small true test errors with much smaller prediction interval lengths. Notice that the behavior of $\text{QFCV}(m)$ for $m > 0$ are similar in terms of length saving and we omit cases $m=2$ and $m=5$ on the left panel for simplicity. On the right panel, we look at the scatter plots from a different perspective, by visualizing the location of points that are miscovered (i.e. outside of the prediction interval). It can be observed that while almost all points with large true test errors are miscovered by $m=0$, true test errors of miscovered points for $m>0$ are more uniformly distributed on the test error range. 

\begin{figure}[t]
    \centering
    \includegraphics[width = 0.45\linewidth]{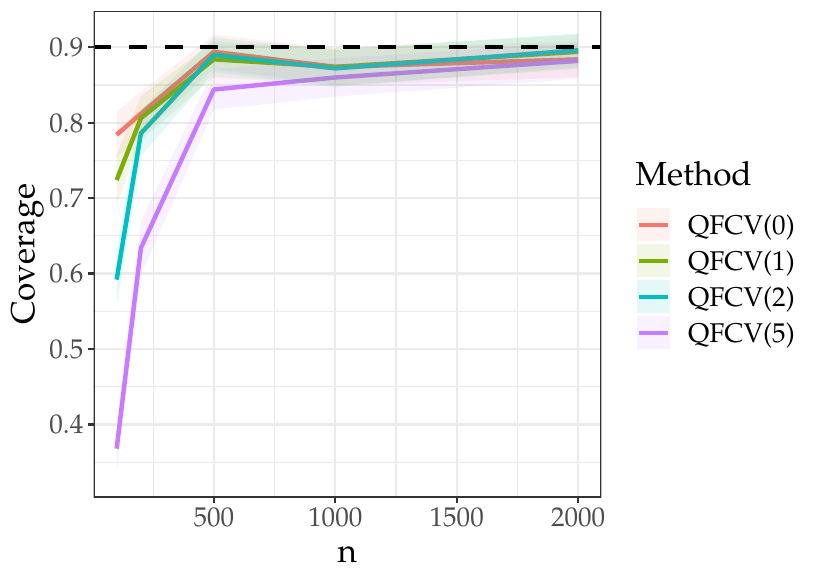}
    \caption{Actual coverage of QFCV methods as a function of the time series length $\nobs$. The noise process is $\text{ARMA}(1,0)$ with autocorrelation parameter $\phi=0.5$ and $\nval=5$.}\label{fig:QFCV_coverage_n}
\end{figure}

% Notice that convergence rate depends on the time series correlation structure, model fitting, training and test sample size, etc. Here we fix all hyperparameters and compare convergence rate for different variants of QFCV methods. 

Figure \ref{fig:QFCV_coverage_n} shows how the actual coverage of QFCV methods varies with the time series length $\nobs$, for an $\text{ARMA}(1,0)$ noise process with $\phi=0.5$ and $\nval=5$. We find that all QFCV variants achieve close to the nominal $90\%$ coverage when $\nobs > 1000$. However, using a larger number $m$ of features for quantile regression leads to a larger estimation error, requiring more samples to attain the nominal coverage level. This highlights the need to avoid excessively large $m$ values in practice.

\subsection{Discussion}\label{sec:discussion_QFCV}

Let us take a moment to reflect on and discuss our main proposal of QFCV prediction intervals for time series forecasting.

\paragraph{Quantile-based methods versus CLT-based methods.} 

QFCV produces intervals using quantiles, which provides asymptotic validity relying on the law of large numbers for the ergodic process of test errors. While quantile-based methods are a natural choice in hindsight, CLT-based methods like forward cross-validation (FCV; see Section \ref{sec:FCV}) may seem like a more instinctive first approach for practitioners. However, as we will see in Section \ref{sec:FCV}, FCV prediction intervals have certain limitations that quantile-based QFCV avoids. By modeling the conditional quantiles of the test error distribution, QFCV does not make Gaussian assumptions and can provide valid coverage where naive CLT-based methods fail. 

% QFCV intervals are quantile-based methods that give asymptotically valid prediction intervals for test errors, whose validity is based on the law of large numbers (LLN) of the ergodic process of test errors. While using quantile-based methods seems to be the natural choice in hindsight, we note that many practitioners may take variance-based methods as their first choice, for example, the forward cross-validation (FCV) method that we will discuss in the next section. We will show that variance-based methods have certain limitations in time series forecasting, which quantile-based methods can avoid. 

\paragraph{$\text{QFCV}(1)$ as the advocated approach.} Our simulations illustrate the tradeoff in selecting the number of features $m$ for the auxiliary function $\cA$. Adding more features can yield shorter intervals but requires more samples to attain valid coverage. We advocate using $\text{QFCV}(1)$, which takes the average validation error (\ref{eqn:validation_aux_function}) as the single feature. As shown in our experiments, $\text{QFCV}(1)$ improves on the interval length of $\text{QFCV}(0)$ while matching the length of $\text{QFCV}(2)$. Critically, $\text{QFCV}(1)$ maintains valid coverage across all settings. In later sections, we will refer to the QFCV method as $\text{QFCV}(1)$ by default.

% Numerical simulations have demonstrated the tradeoff of choosing the number of features $m$ in the auxiliary function $\cA$. While adding features will provide shorter prediction intervals, it will take more samples for the intervals to have valid coverage. Our advocated choice is $\text{QFCV}(1)$, which amounts to taking the auxiliary function to be the single feature of the validation error (\ref{eqn:validation_aux_function}). As have demonstrated in simulations, $\text{QFCV}(1)$ outperforms $\text{QFCV}(0)$ on interval length and is on par with $\text{QFCV}(2)$. More importantly, $\text{QFCV}(1)$ maintains valid coverage. 

\paragraph{QFCV point estimators.} While we have focused on QFCV for constructing prediction intervals, point estimators can also be naturally derived from the QFCV framework. We include more details of QFCV point estimators, as well as other algorithm design choices of QFCV intervals, in Appendix \ref{appendix:QFCV}.

% Before moving to confidence intervals and adaptive methods for non-stationary time series, we take a moment to reflect and discuss our main proposal of QFCV prediction interval methods in time series forecasting. Quantile-based forward cross-validation (QFCV) gives asymptotically valid prediction intervals for true test errors under the assumption of ergodicity. The key of using the quantile-based method is that while autocovariance structure may be difficult to characterize for an arbitrary time series, the law of large numbers (LLN) is often valid, which can be used to construct statistically meaningful prediction intervals. 

% There are different variants of QFCV methods, with flexibility in algorithmic design (i.e., training and retraining set structure), number of features for quantile regression, and quantile regression basis functions. While increasing the number of features can lead to shorter interval length, the corresponding tradeoff is slower convergence rate to nominal coverage. Lastly, we have introduced the QFCV method as a prediction interval without a point estimator. However, point estimators can also be constructed naturally via QFCV methods. We include more details of other algorithmic designs and point estimators of QFCV methods in Appendix \ref{appendix:QFCV}.

\section{CLT-based forward cross-validation}\label{sec:FCV}
% \hx{Comparison of LLN and CLT for ergodic time series (check assumptions). }

Many inference methods for prediction error under the IID assumption produce confidence intervals aiming to cover the expected test error $\ERR$ based on the central limit theorem (CLT). Such methods typically have the form:
\[
\hatCI = \big(\widehat{\ERR} - z_{1-\alpha/2} \hatSE, \widehat{\ERR} + z_{1-\alpha/2} \hatSE \big), 
\]
where $\widehat{\ERR}$ is a point estimate of $\ERR$, $\hatSE$ is the estimated standard error of $\widehat{\ERR}$, and $z_\alpha$ is the $\alpha$-quantile of a standard normal distribution. In time series forecasting, practitioners usually adopt a naive adaptation of the cross-validation called forward cross-validation (FCV) as natural candidates for constructing confidence intervals targeting expected error $\ERR$.

% Many inference methods for prediction error under IID assumption are confidence intervals aiming to cover the expected test error $\ERR$ based on the central limit theorem (CLT). Such methods usually take the form of 
% \[
% \hatCI = \big(\widehat{\ERR} - z_{1-\alpha/2} \hatSE, \widehat{\ERR} + z_{1-\alpha/2} \hatSE \big), 
% \]
% where $\widehat{\ERR}$ is a point estimator for $\ERR$, $\hatSE$ is an estimator of the standard error of $\widehat{\ERR}$, and $z_\alpha$ is the $\alpha$-th quantile of a standard normal distribution. In time series forecasting, practitioners usually adopt a naive adaptation of the cross-validation (CV) method, namely the forward cross-validation (FCV) method.

In this section, we examine FCV for constructing confidence and prediction intervals. We illustrate when it succeeds or fails, comparing it to QFCV. We show that as confidence intervals covering expected error $\ERR$, naive FCV lacks valid coverage unless adjusted by an autocovariance correction. Furthermore, as prediction intervals aim to cover stochastic error $\ERRTar$, scaling-corrected FCV only provides valid coverage under strong Gaussianity assumptions, unlike QFCV. 

% In this section, we will examine the FCV method for providing confidence and prediction intervals, illustrating when it works or not, and compare it with the QFCV method. We will show that, as a confidence interval, a naive FCV method does not provide valid coverage unless adjusted by an autocovariance correction. Furthermore, as prediction intervals, the scaling-corrected FCV method does not give valid coverage unless the strong Gaussianity assumption holds. 

% A caveat of confidence intervals in the time series context is that the assumption of CLT may not necessarily hold due to time correlation. The conditions for CLT to hold are usually more stringent than ergodicity as in LLN. Besides, unbiased estimators for standard error $\hatSE$ are more difficult to obtain due to time correlation and can only be approximated through autocovariance estimation and truncation of dependence structure assumptions.

\subsection{Forward cross-validation procedures}\label{sec:FCV_procedure}

\begin{figure}[t]
    \centering
    \includegraphics[width = 0.7\linewidth]{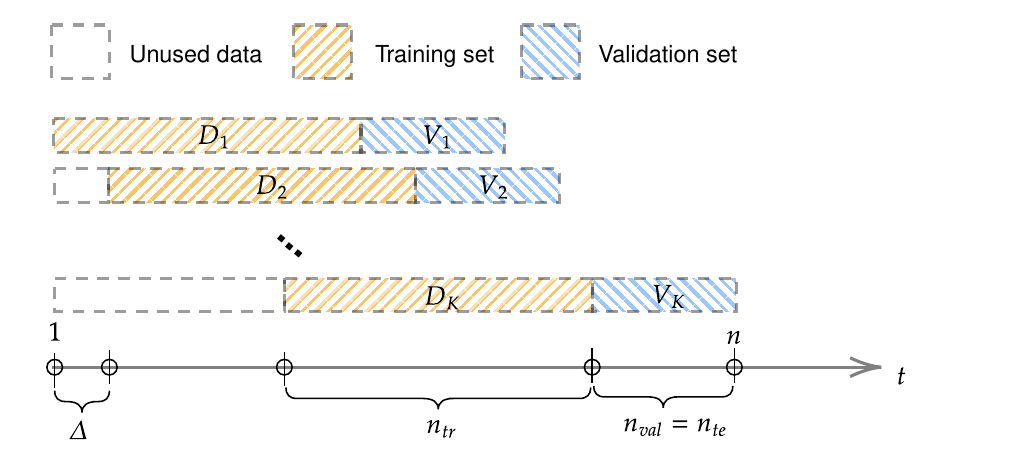}
    \caption{Diagram illustration for forward cross-validation (FCV)}
    \label{fig:FCV_diagram}
\end{figure}

In FCV, we split the historical data sequence $z_{1:\nobs}$ into $K$ possibly overlapping time windows. Each window is shifted $\Delta$ time steps from the previous one. For the $i$-th window, we further divide it into a training set $D_i = \{(i - 1) \Delta + 1, \ldots, (i - 1) \Delta + \ntrain\}$ of size $\ntrain$ and a validation set $V_i = \{(i - 1) \Delta + \ntrain + 1, \ldots, (i - 1) \Delta + \ntrain + \nval \}$ of size $\nval$. See Figure \ref{fig:FCV_diagram} for a diagram illustration. The FCV point estimate of the prediction error is:
\begin{equation}\label{eqn:hatERR-FCV}
\hatERR_K^\FCV =  \frac{1}{K}\sum_{i = 1}^K \frac{1}{\nval}\sum_{t \in V_i} \ell(\hat f(x_t; z_{D_i}), y_t) =\frac{1}{K} \sum_{i = 1}^K E_i , ~~~~ \text{ where } E_i = \frac{1}{\nval}\sum_{t \in V_i} \ell(\hat f(x_t; z_{D_i}), y_t). 
\end{equation}
Under the stationary assumption, $\hatERR_K^\FCV$ is an unbiased estimator for the expected target error $\ERR$, as stated in Lemma \ref{lem:unbiased} below. 
\begin{lemma}[Unbiasedness of FCV estimator]
\label{lem:unbiased}
Assume $\{ z_t \}_{t \ge 1}$ is stationary, and let $\ntest = \nval$. Then the FCV estimator is unbiased for the expected test error (\ref{eqn:ERR}), that is, 
\[
\E[\hatERR_K^\FCV] = \ERR.
\]
\end{lemma}
\begin{proof} By direct calculation,  
\[\textstyle     \E[\hatERR_K^\FCV] =  \E[K^{-1} \sum_{i = 1}^K E_i ] \overset{(1)}{=} \E [ \ERRTar ]  \overset{(2)}{=} \ERR,
\]
where (1) uses stationarity and (2) is by definition of $\ERR$. 
\end{proof}

% Unlike in the IID setting described in \cite{bates2021cross}, where the cross-validation estimand is difficult to characterize and data splitting confidence interval $\hatCI^{\DS}$ gives valid coverage, the situation is reversed in forward cross-validation. 
% \[
% \ERRFCV(\tilde{D}) = \frac{1}{k}\sum_{i = 1}^k \frac{1}{|V_i|} \sum_{t \in V_i} l(\hat{y}_t(x_t; D_i), y_t).
% \]
% If each fold is of the same sample size $|V_i| = V$ for some $V$, then 
% \begin{align*}
    %     \mathbb{E}[\hatERR_K^\FCV] &=  \mathbb{E}\left[\frac{1}{K V}\sum_{i = 1}^K \sum_{t \in V_i} \ell(\hat f(x_t; z_{D_i}), y_t)\right] \overset{(1)}{=} \mathbb{E}\left[\frac{1}{V}\sum_{t \in V_i}\ell(\hat f(x_t; z_{D_i}), y_t)\right]  \overset{(2)}{=} \ERR,
    % \end{align*}

\paragraph{Naive FCV interval.} Following the strategy of constructing confidence intervals in the IID setting, the naive FCV interval based on CLT is: 
\begin{equation}
\label{eqn:naive_fcv_interval}
\hatCI_K^{\FCV} = \Big( \hatERR_K^\FCV - z_{1-\alpha/2} \cdot \hatSE_{K}^\FCV, \hatERR_K^\FCV + z_{1-\alpha/2} \cdot \hatSE_{K}^\FCV \Big),
\end{equation}
where
\begin{equation}
\label{eqn:naive_fcv}
\textstyle    \hatSE_{K}^\FCV = \frac{1}{\sqrt{K}} \sqrt{\frac{1}{K} \sum_{i = 1}^K (E_i - \bar E)^2}.
\end{equation}
We present the algorithm for computing the naive FCV interval in Algorithm \ref{alg:fcv}. 

The naive FCV interval has some limitations in serving as both a confidence interval for the expected test error $\ERR$ and a prediction interval for the stochastic test error $\ERRTar$. When treated as a confidence interval for expected test error $\ERR$, naive FCV often underestimates the true variance of the FCV point estimator since it doesn't account for the time correlation. When treated as a prediction interval for stochastic test error $\ERRTar$, naive FCV does not capture the variance of $\ERRTar$, but instead the variance of the FCV point estimator. Therefore, the naive FCV interval will undercover both $\ERR$ as a confidence interval and $\ERRTar$ as a prediction interval. 

To address these issues, we describe two modifications to the naive FCV intervals. In the first modification, an autocovariance correction is adopted to derive a consistent variance estimate for the FCV point estimator. This results in an autocovariance-corrected FCV that becomes a valid confidence interval for expected test error $\ERR$. In the second modification, the standard error estimator is rescaled, making it an estimate of the standard error of the stochastic test error. This results in a scaling-corrected FCV interval whose length is on the scale of the standard deviation of the stochastic prediction error.

\paragraph{Autocovariance-corrected FCV confidence interval.} Assuming that the time series is stationary, the variance of the FCV point estimator gives
\begin{align*}
\Var(\hatERR_K^\FCV)  & = \frac{1}{K^2} \Var  \Big[\sum_{i=1}^K E_i \Big] = \frac{1}{K^2}\sum_{i=1}^K \Var [ E_i ] + \frac{1}{K^2} \sum_{i \neq j} \Cov [E_i, E_j ] = \frac{1}{K} \Big[ \gamma(0) + 2 \sum_{s = 1}^{K-1} \Big( 1 - \frac{s}{K} \Big) \gamma(s) \Big].
\end{align*}
Here, $\gamma(s) = \Cov(E_1, E_{1 + s})$ is the autocovariance function of the stationary process $\{ E_i \}$, which can be estimated from the data using the sample autocovariance $\hat \gamma(s)$, 
\[
\hat \gamma(s) = \frac{1}{K - s} \sum_{i = 1}^{K - s} (E_i - \bar E) (E_{i + s} - \bar E), ~~~~~ \bar E = \frac{1}{K} \sum_{i = 1}^K E_i = \hatERR_K^\FCV. 
\]
The adjusted standard error estimator is thus a truncated sum of the sample autocovariance function, denoted as $\hatSE^{\FCV(c)}_{K}$, where the $c$ indicates covariance adjustment. 
\begin{align}\label{eqn:SE-FCV}
\textstyle \hatSE^{\FCV(c)}_{K} = \frac{1}{\sqrt{K}} \sqrt{ \hat \gamma(0) + 2 \sum_{s = 1}^{\Ktrun} \Big( 1 - \frac{s}{K} \Big) \hat \gamma(s)}.  %=\frac{1}{\sqrt{K}} \sqrt{\frac{1}{K} \sum_{i = 1}^K (E_i - \bar E)^2 + 2\sum_{s = 1}^\Ktrun \frac{1}{K} \sum_{i = 1}^{K - s} (E_i - \bar E) (E_{i + s} - \bar E) }, 
\end{align}
Here, $\Ktrun < K$ truncates the summation to the correlation length of the stationary process $\{ E_i \}_{i \in [K]}$. This is needed since $\hat \gamma(s)$ converges to $\gamma(s)$ very slowly for large $s$. We expect that when $s$ exceeds some threshold $\Ktrun$, the autocovariance becomes very small and can be neglected. %That is, we assume that $\Ktrun$-dependence assumption that for all $T_0 \in \mathbb{N}$, the sequence $\{E_i: i \in [T_0]\}$ and $\{E_i: i >  \min(T_0 + \Ktrun, K)\}$ are independent. A special case of the $\Ktrun$-dependence condition is $\Ktrun=0$, which corresponds to the i.i.d setting. \sm{I will rephrase this. }

Given the adjusted standard error estimator, the autocovariance-adjusted FCV confidence interval is 
\[
\hatCI_K^{\FCV(c)} = \Big( \hatERR_K^\FCV - z_{1-\alpha/2} \cdot \hatSE_{K}^{\FCV(c)}, \hatERR_K^\FCV + z_{1-\alpha/2} \cdot \hatSE_{K}^{\FCV(c)} \Big).
\]
Under certain assumptions on the stationary process $\{E_i\}_{i \in [K]}$, it can be shown via the central limit theorem that $\hatERR_K^{\FCV(c)}$ is asymptotically normal with asymptotic standard deviation consistently estimated by $\hatSE^{\FCV(c)}_{K}$. Therefore, $\hatCI_K^{\FCV(c)}$ is an asymptotically valid confidence interval for $\ERR$. This leads to the following informal proposition, with a rigorous statement presented in Appendix \ref{sec:proof_consistency_variance_FCV}. 
\begin{proposition}[Asymptotic normality of the autocovariance-corrected FCV estimator; Informal]\label{prop:consistency_variance_FCV} Under certain assumptions of the stationary process $\{ E_i \}_{i \in [K]}$, as $K \to \infty$, we have convergence in distribution
\[
\big(\hatERR_K^\FCV - \ERR \big) / \hatSE^{\FCV(c)}_{K} \stackrel{d}{\longrightarrow} \cN(0, 1). 
\]
Consequently, $\hatCI_K^{\FCV(c)}$ has asymptotically  $1 - \alpha$ coverage over $\ERR$. 
\end{proposition}

\begin{algorithm}[t]
  \caption{Forward cross-validation: naive, autocovariance-corrected, and scaling-corrected}
  \label{alg:fcv}
  \small
  \begin{algorithmic}[1]
    \REQUIRE Dataset $\{ z_t\}_{t \in [\nobs]} \subseteq \cZ$, loss $\ell$, size of training set $\ntrain$, size of validation set $\nval$, number of fold $K$, prediction algorithm $\hat{f}: \cX \times \cZ^{\ntrain} \to \cY$. Parameter $\Ktrun$ for autocovariance-corrected FCV.
    %\STATE Set time lag $\Delta = \lfloor (n-\ntrain-\nval )/ (K - 1) \rfloor$.
    %\STATE Let $D_i = \{ (i-1) \Delta + 1, \cdots,  (i - 1) \Delta + \ntrain\}$, $V_i = \{(i - 1) \Delta + \ntrain + 1, \ldots, (i - 1) \Delta + \ntrain + \nval \}$. 
    % \FOR{$i = \{1,\ldots, K\}$}
    \STATE Compute $E_i = (1/\nval)\sum_{t\in V_i}\ell(\hat{f}(x_t;z_{D_i}), y_t)$ for $i \in [K]$.
    % \ENDFOR
    \STATE Compute $\hatERR_K = K^{-1} \sum_{i = 1}^K E_i$ as in Eq. (\ref{eqn:hatERR-FCV}).
    \STATE Compute $\hatSE_K$ using Eq. (\ref{eqn:naive_fcv}) for naive FCV, using Eq. (\ref{eqn:SE-FCV}) for autocovariance-corrected FCV, or using Eq. (\ref{eq:fcv_pi}) for scaling-corrected FCV. 
    \STATE \textbf{Return} $\hatERR_K \pm z_{1 - \alpha/2} \hatSE_K$.
  \end{algorithmic}
\end{algorithm}

\paragraph{Scaling-corrected FCV prediction interval. } Both $\hatSE_{K}^\FCV$ in (\ref{eqn:naive_fcv}) and $\hatSE^{\FCV(c)}_{K}$ in (\ref{eqn:SE-FCV}) estimate the variance of $\hatERR_K^\FCV$ instead of the variance of $\ERRTar$, and thus do not provide prediction intervals for $\ERRTar$. To estimate the variance of $\ERRTar$, a commonly used approach is a scaling correction. In particular, we define the scaling-corrected standard error estimator as $\hatSE_K^{\FCV(p)}$, where $p$ indicates this is for a prediction interval: 
\begin{equation}
 \textstyle   \hatSE_K^{\FCV(p)} = \sqrt{K} \cdot \hatSE_K^{\FCV}= \sqrt{\frac{1}{K} \sum_{i = 1}^K (E_i - \bar E)^2}.
    \label{eq:fcv_pi}
\end{equation}
The scaling-corrected FCV prediction interval is then:
\begin{equation}
\hatPI_K^{\FCV(p)} = \Big( \hatERR_K^\FCV - z_{1-\alpha/2} \cdot \hatSE_{K}^{\FCV(p)}, \hatERR_K^\FCV + z_{1-\alpha/2} \cdot \hatSE_{K}^{\FCV(p)} \Big). 
\end{equation}
Under the strong assumption that $\{ E_i \}_{i \in [K]}$ follows a Gaussian process, we can show $\hatPI_K^{\FCV(p)}$ gives approximately the $\alpha/2$ and $1-\alpha/2$ quantiles of the stochastic test error distribution. This gives the following proposition, with the proof in Appendix \ref{sec:proof_fcv_p}. 
\begin{proposition}
\label{prop:fcv_p}
Assume $\{ E_i \}_{i \in [K]}$ is a stationary ergodic Gaussian process. Let $\hat V_K = K^{-1}\sum_{i = 1}^K (E_i - \bar E)^2$. Then as $K \to \infty$, we have convergence in probability 
\begin{equation}\label{eqn:V_K_convergence}
 \hat V_K \stackrel{p}{\longrightarrow} \Var(E_1).  
\end{equation}
In this case, we have convergence in probability
\begin{equation}\label{eqn:quantile_converge_FCV_p}
\hatERR_K^\FCV - z_{1-\alpha/2} \cdot \hatSE_{K}^{\FCV(p)} \stackrel{p}{\longrightarrow} q_{\alpha/2}(\ERRTar),  ~~~~ \hatERR_K^\FCV + z_{1-\alpha/2} \cdot \hatSE_{K}^{\FCV(p)} \stackrel{p}{\longrightarrow} q_{1-\alpha/2}(\ERRTar).  
\end{equation}
That is, $\hatPI_K^{\FCV(p)}$ is an asymptotically valid $1-\alpha$ prediction interval for $\ERRTar$,
\begin{equation}\label{eqn:Errtar_FCV_p_coverage}
\lim_{K \to \infty} \P \Big( \ERRTar \in \hatPI_K^{\FCV(p)}  \Big) = 1 - \alpha. 
\end{equation}
\end{proposition}
It is worth noting that $\text{QFCV}(0)$, the empirical quantile method, will also give the $\alpha/2$ and $1-\alpha/2$ quantiles of the stochastic test error distribution. Consequently, under the Gaussian process assumption, the scaling-corrected FCV prediction interval is asymptotically equivalent to $\text{QFCV}(0)$. As shown in Section \ref{sec:QFCV}, $\text{QFCV}(0)$ has a larger interval length than our advocated $\text{QFCV}(1)$ method. Thus, we would also expect scaling-corrected FCV to have a larger interval length than $\text{QFCV}(1)$.

\begin{figure}[t]
\centering
  \includegraphics[width = 0.46 \linewidth]{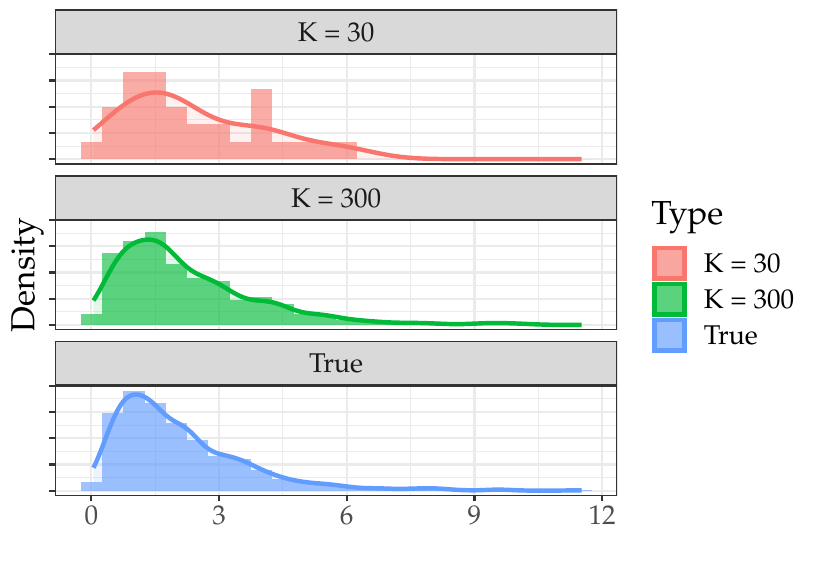}
  % \caption{Distribution of test error}
\caption{Empirical error distributions of $E_i$ for $K=30$ number of validation errors, $K = 300$ validation errors, and true population level stochastic test error distribution. Solid lines are density estimations of the histograms. }
\label{fig:histogram}
\end{figure}

We note that the Gaussian process assumption for $\{E_i\}_{i \in [K]}$ is a strong assumption that may not hold in practice. In Figure \ref{fig:histogram}, we provide an example where the Gaussianity assumption is violated. The bottom panel shows the population-level stochastic test error distribution obtained with 1000 samples, where the setting matches Figure \ref{fig:motivation} except with a smaller test set size of $5$. With small test sets, errors are no longer Gaussian distributed - the distribution is truncated at zero and skewed. Therefore, the scaling-corrected FCV interval may not have valid coverage due to violating the Gaussian assumption. 

In contrast, QFCV methods do not require the Gaussian assumption. As seen in the top and middle panels of Figure \ref{fig:histogram}, the empirical validation error distribution from a single FCV run converges to the population test error distribution as the number of folds $K$ increases. The $\text{QFCV}(0)$ method bypasses the Gaussianity assumption by approximating the oracle stochastic test error distribution with the empirical distribution of validation errors and reporting empirical quantiles for prediction intervals. Other QFCV variants, such as $\text{QFCV}(1)$, exploit the time correlation between past validation error and future test error to yield shorter intervals than $\text{QFCV}(0)$. 

Given that Gaussianity often does not hold in practice, scaling-corrected FCV has no guarantees of asymptotically valid coverage because it is using Gaussian quantiles for uncertainty interval construction. Therefore, we do not advocate it due to the strong assumptions required. We recommend using QFCV, which does not rely on this assumption.

\subsection{Comparing FCV and QFCV via numerical simulations}\label{sec:QFCV_FCV_comparison}

\begin{figure}[t]
\centering
\begin{minipage}[c]{.5\linewidth}
  \centering
  \includegraphics[width = \linewidth]{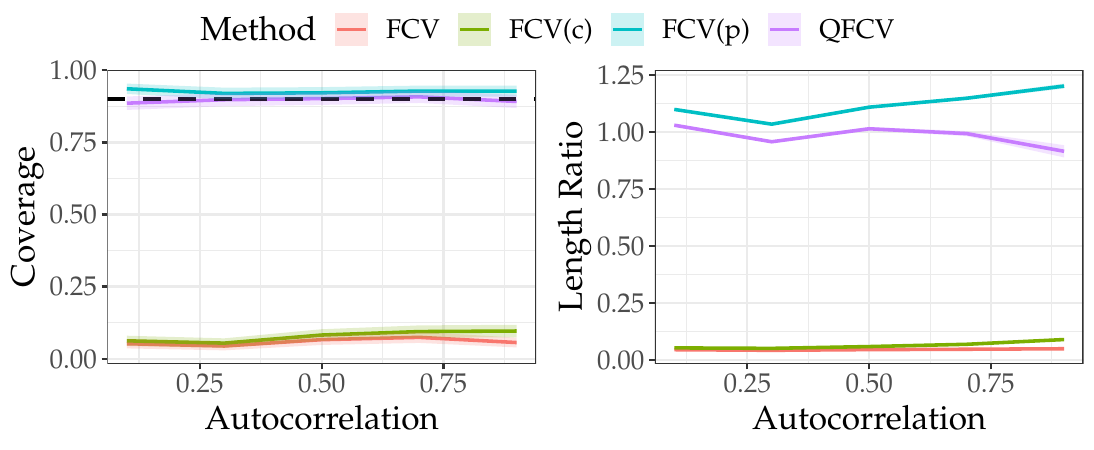}
  \subcaption{ARMA(1,0), $\nval = 5$.}
\end{minipage}%
\begin{minipage}{.5\linewidth}
  \centering
  \includegraphics[width = \linewidth]{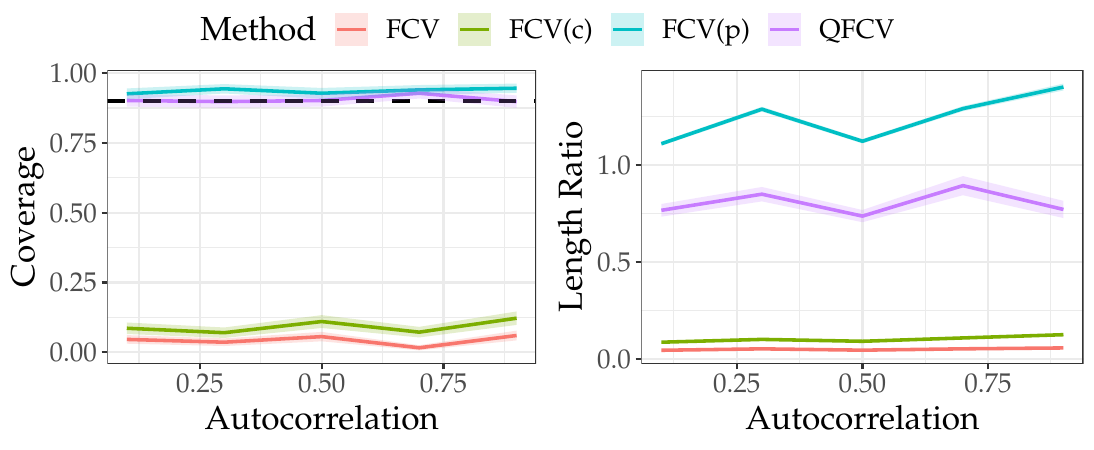}
  \subcaption{ARMA(1,20), $\nval = 5$.}
\end{minipage}
\vskip0.5cm
\begin{minipage}[c]{.5\linewidth}
  \centering
  \includegraphics[width = \linewidth]{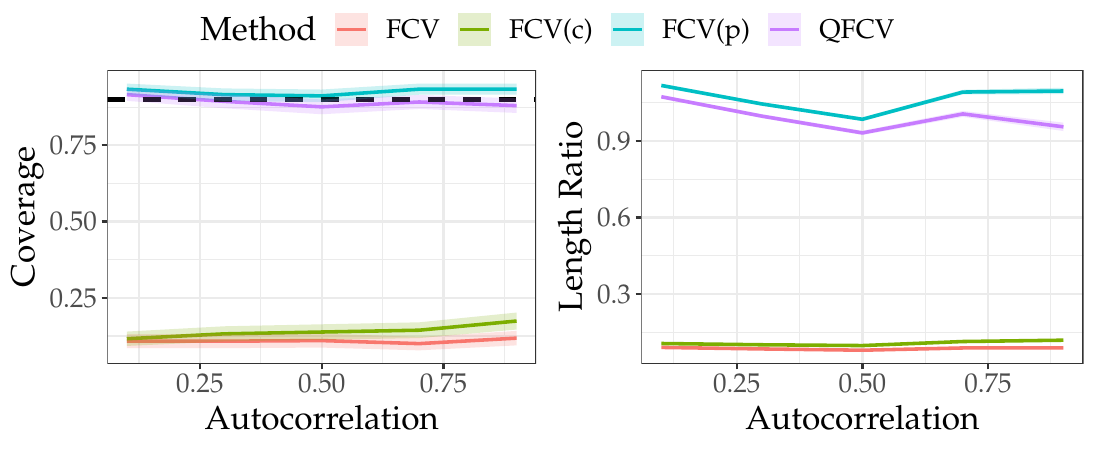}
  \subcaption{ARMA(1,0), $\nval = 20$.}
\end{minipage}%
\begin{minipage}{.5\linewidth}
  \centering
  \includegraphics[width = \linewidth]{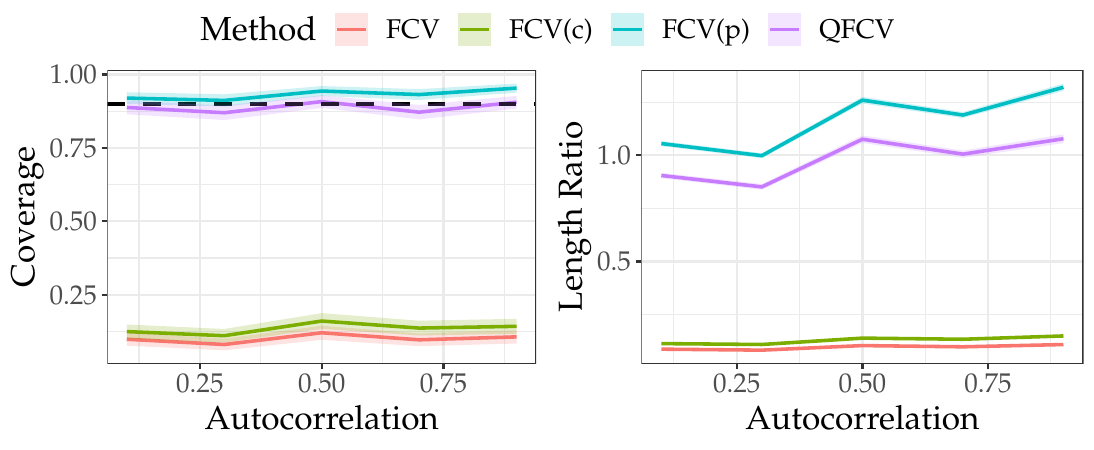}
  \subcaption{ARMA(1,20), $\nval= 20$.}
\end{minipage}
\caption{
% Performance comparison of QFCV prediction intervals on simulated time series (\ref{eqn:simulation_linear_model}). The time series length is fixed at $\nobs = 2000$, with $\ntrain = 40$, $\nval = \ntest \in \{5, 20\}$, and $p = 20$. The x-axis represents the autocorrelation parameter $\phi \in [0, 1]$. Each panel displays two plots: (i) actual coverage of the $90\%$ nominal QFCV intervals, and (ii) ratio of the QFCV interval length to the oracle interval length. The oracle interval is defined by the $5\%$ and $95\%$ quantiles of the true test error distribution. Each curve displays the mean and standard error over $500$ simulation instances. 
Performance comparison of QFCV and CLT-based FCV prediction intervals on simulated time series (\ref{eqn:simulation_linear_model}). We examine naive FCV (FCV), autocovariance-corrected FCV ($\text{FCV}(c)$), scaling-corrected FCV ($\text{FCV}(p)$), and $\text{QFCV}(1)$. The time series generative process follows the same setup as in Figure \ref{fig:sim_QFCV}. The x-axis represents the autocorrelation parameter $\phi \in [0, 1]$. We plot the coverage and length ratio averaged over $500$ simulation runs. The nominal coverage level is $90\%$, and the length ratio is the prediction interval length divided by the true marginal quantiles interval length based on the true test error distribution quantiles. 
}
\label{fig:FCV}
\end{figure}

We perform numerical simulations to compare CLT-based FCV intervals (Algorithm \ref{alg:fcv}) with QFCV methods (Algorithm \ref{alg:fcv_qc}). Specifically, we examine naive FCV intervals (FCV), autocorrelation-corrected FCV intervals (FCV(c)), and scaling-corrected FCV intervals (FCV(p)) for the CLT-based methods. For the QFCV method, we use $\text{QFCV}(1)$, where the auxiliary function is the validation error (\ref{eqn:validation_aux_function}). The simulations follow the same setup as in Section \ref{sec:simulation_QFCV}, with time series generated from a linear model (\ref{eqn:simulation_linear_model}) with $\text{ARMA}(a, b)$  noise. We investigate the coverage and length ratio of these intervals. The results, presented in Figure \ref{fig:FCV}, illustrate the performance of the different methods. 

The results in Figure \ref{fig:FCV} demonstrate that the naive FCV and FCV(c) methods undercover the stochastic test error $\ERRTar$ across the simulation settings. In contrast, the FCV(p) and QFCV methods achieve valid coverage. However, the QFCV intervals are shorter than those from FCV(p), particularly for time series with smoother noise processes. This aligns with our expectations, as we anticipated FCV(p) would have similar performance to QFCV(0). Since QFCV(1) utilizes features informative to the stochastic test error during quantile regression, it outperforms QFCV(0), and hence should outperform FCV(p).

% Figure \ref{fig:FCV} shows that naive FCV and FCV(c) undercover, whereas FCV(p) and QFCV provide valid coverage for $\ERRTar$ in all simulation settings. However, QFCV has shorter prediction intervals than FCV(p), especially when the time series is smoother. This coincides with our expectation since we expect FCV(p) will perform similarly to QFCV(0), whereas QFCV(1) outperforms QFCV(0) since QFCV(1), when performing quantile regression, uses some features that are informative to the stochastic test error. 

Table \ref{table:FCV} provides further comparison of the four methods across the simulation settings. The first row for each setting shows the mis-coverage percentages of the stochastic test error $\ERRTar$ for both the upper and lower ends. This illustrates that QFCV has more balanced mis-coverage than FCV(p), which is expected since QFCV relies on quantiles, without the need of Gaussian assumptions like FCV(p). The second row for each setting gives the mis-coverage percentages of the expected test error $\ERR$. It demonstrates that FCV(c) provides better coverage for $\ERR$ and serves as a better confidence interval than naive FCV through its autocovariance correction. The fourth row for each setting illustrates the mean squared error of the FCV and QFCV point estimators for estimating the stochastic test error $\ERRTar$. It shows that QFCV point estimator has a smaller MSE than the FCV point estimator.

\begin{table}[t]
\centering
\begin{tabular}{ 
p{0.5cm} p{3.2cm}||  p{0.6cm} p{0.6cm} | p{0.6cm} p{0.6cm}|p{0.6cm} p{0.6cm}|p{0.6cm} p{0.6cm}}
\toprule
\multicolumn{2}{c||}{Setting} & \multicolumn{8}{c}{Miscoverage (\%) \& Performance} \\
&& \multicolumn{2}{c}{FCV} & \multicolumn{2}{c}{FCV(c)} &\multicolumn{2}{c}{FCV(p)}& \multicolumn{2}{c}{QFCV} \\
 \multicolumn{1}{c}{Data} & \multicolumn{1}{c||}{Target} & \multicolumn{1}{c}{Hi} & \multicolumn{1}{c}{Lo} & \multicolumn{1}{c}{Hi} & \multicolumn{1}{c}{Lo} & \multicolumn{1}{c}{Hi} & \multicolumn{1}{c}{Lo} & \multicolumn{1}{c}{Hi} & \multicolumn{1}{c}{Lo} \\ \midrule
 \midrule
 {\multirow{3}{4em}{(a)}} & $\text{Miscoverage of }\ERRTar$  & 37.2 & 56.2 & 36.8 & 55.0 & 7.8 & 0.0 & \textbf{5.8} & \textbf{4.0}\\
 & $\text{Miscoverage of } \ERR$  & 18.8 & 4.8 & 13.4 & 2.4 & 0.0 & 0.0 & 0.0 & 0.0 \\
  & $\text{Interval Length}$ & \multicolumn{2}{c|}{0.211} & \multicolumn{2}{c|}{0.273} & \multicolumn{2}{c|}{5.13} & \multicolumn{2}{c}{4.69} \\
 & $\text{MSE of }\ERRTar$ & \multicolumn{2}{c|}{2.39} & \multicolumn{2}{c|}{-} & \multicolumn{2}{c|}{-} & \multicolumn{2}{c}{2.40} \\
 \midrule 
  {\multirow{3}{4em}{(b)}} & $\text{Miscoverage of }\ERRTar$ & 30.4 & 64.0 & 27.8 & 61.2 & 7.2 & 0.0 & \textbf{5.0} & \textbf{4.8}\\
 & $\text{Miscoverage of } \ERR$ & 34.6 & 10.4 & 15.6 & 0.6 & 0.0 & 0.0 & 0.4 & 0.0 \\
  & $\text{Interval Length}$ & \multicolumn{2}{c|}{0.287} & \multicolumn{2}{c|}{0.572} & \multicolumn{2}{c|}{6.99} & \multicolumn{2}{c}{4.59} \\
 & $\text{MSE of }\ERRTar$ & \multicolumn{2}{c|}{5.24} & \multicolumn{2}{c|}{-} & \multicolumn{2}{c|}{-} & \multicolumn{2}{c}{3.30} \\
  \midrule 
  {\multirow{3}{4em}{(c)}} & $\text{Miscoverage of }\ERRTar$ & 39.2 & 49.8 & 38.4 & 47.8 & 7.8 & 1.0 & \textbf{5.6} & \textbf{6.8}\\
 & $\text{Miscoverage of } \ERR$ & 18.2 & 4.4 & 11.8 & 1.8 & 0.0 & 0.0 & 0.0 & 0.0 \\
  & $\text{Interval Length}$ & \multicolumn{2}{c|}{0.239} & \multicolumn{2}{c|}{0.293} & \multicolumn{2}{c|}{2.91} & \multicolumn{2}{c}{2.75} \\
 & $\text{MSE of }\ERRTar$ & \multicolumn{2}{c|}{0.881} & \multicolumn{2}{c|}{-} & \multicolumn{2}{c|}{-} & \multicolumn{2}{c}{0.841} \\
   \midrule 
  {\multirow{3}{4em}{(d)}} & $\text{Miscoverage of }\ERRTar$ & 30.2 & 57.8 & 28.4 & 55.6 & 5.6 & 0.0 & \textbf{4.8} & \textbf{4.4}\\
 & $\text{Miscoverage of } \ERR$ & 13.8 & 13.8 & 7.8 & 6.6 & 0.0 & 0.0 & 0.0 & 0.2 \\
  & $\text{Interval Length}$ & \multicolumn{2}{c|}{0.492} & \multicolumn{2}{c|}{0.660} & \multicolumn{2}{c|}{5.99} & \multicolumn{2}{c}{5.12} \\
 & $\text{MSE of }\ERRTar$ & \multicolumn{2}{c|}{3.22} & \multicolumn{2}{c|}{-} & \multicolumn{2}{c|}{-} & \multicolumn{2}{c}{2.99} \\
\bottomrule
\end{tabular}
\caption{
Comparison of point estimates and prediction intervals for QFCV and CLT-based procedures. The simulations follow the same settings as Figure \ref{fig:FCV}, with autocorrelation $\phi = 0.5$. The four cases (a)-(d) correspond to the same settings as in Figure \ref{fig:FCV}. The values are averaged over $500$ simulations. The nominal coverage level is $90\%$, with $5\%$ nominal mis-coverage on each end. The point estimator for QFCV is computed by linear regression of  $\{ (\ERR_{i}^{\fea}, \ERR_{i}^{\test}) \}$ as detailed in Appendix \ref{appendix:QFCV}. 
% prediction interval and confidence interval miscoverages for both above and below are reported for a nominal level of 90\% coverage. 
%True test error (True), naive forward cross-validation (FCV), FCV with covariance correction (FCV(c)), FCV with predictive scaling correction (FCV(p)), and quantile-based forward cross-validation (QFCV). 
% Table of comparison for point estimates and prediction intervals under different correlation structure. Data settings correspond to  simulations settings in Figure \ref{fig:FCV} with autocovariance $\phi = 0.5$. True test error (True), naive forward cross-validation (FCV), FCV with covariance correction (FCV(c)), FCV with predictive scaling correction (FCV(p)), and quantile-based forward cross-validation (QFCV) are computed over 500 simulations. prediction interval and confidence interval miscoverages for both above and below are reported for a nomial level of 90\% coverage. Point estimator for QFCV is computed as linear regression fit of $\{ (\ERR_{i}^{\fea}, \ERR_{i}^{\test}) \}$ with details in Appendix \ref{appendix:QFCV}.
}
\label{table:FCV}
\end{table}

\section{Rolling intervals under nonstationarity}\label{sec:rolling}

In previous sections, we provided inference methods for test errors in stationary time series. However, constructing valid prediction intervals becomes more difficult when the time series is non-stationary. With arbitrary non-stationarity, the observed data may contain little information about the distribution of future data points, and obtaining prediction intervals is therefore challenging. A more feasible goal is to provide \emph{rolling intervals} for the test error that have asymptotically valid \emph{time-average coverage}. %Fortunately, a feasible goal is to provide \emph{rolling prediction intervals} of the test error with asymptotically valid time-average coverage. 

\subsection{ACI with delayed feedback for rolling intervals}

\paragraph{Rolling intervals.} Suppose that we observe the time series $\{ z_t = (x_t, y_t) \}_{t \ge 1} \subseteq \cZ = \cX \times \cY$ arriving sequentially, where the $\{ z_t\}_{t \ge 1}$ are treated as deterministic, unlike the stochastic settings in previous sections. We are provided with a forecasting function $\hat f: \cX \times \cZ^{\ntrain} \to \cY$ and a loss function $\ell: \cY \times \cY \to \R$. Define $\ERRTar^{t + 1}$ as the $(t + 1)$-th test error, 
\begin{equation}
\ERRTar^{t + 1} = \frac{1}{\ntest} \sum_{s = t +1}^{t + \ntest} \ell(\hat f ( x_s; z_{t - \ntrain + 1 : t}), y_t). 
\end{equation}
This is the rolling version of $\ERRTar$ in Eq. (\ref{eqn:ERRTar_simplified}), where we add the superscript $t + 1$ to emphasize its dependence on the time index. Our goal is to construct rolling intervals $\{ \hatRI^{\alpha, t}\}_{t \ge 1} \subseteq \R$ where each $\hatRI^{\alpha, t}$ depends on $z_{1:t-1}$, the time series up to time $t - 1$. We hope these intervals will have \emph{time-average coverage} $1 - \alpha$ over the set of times indices $S_{T, \Delta} = \{ \Delta, 2 \Delta, 3 \Delta, \ldots, \lfloor T / \Delta \rfloor \cdot \Delta \}$ spaced by $\Delta \ge 1$, 

% \sm{$\Delta$ makes the algorithm more complicated. Consider to set $\Delta = 1$. }
\begin{equation}
\lim_{T \to \infty} \frac{1}{\vert S_{T, \Delta} \vert} \sum_{t \in S_{T, \Delta}} 1\{ \ERRTar^t \in \hatRI^{\alpha, t} \} = 1 - \alpha.
\end{equation}
We emphasize that time-average coverage differs from statistical coverage in (\ref{eqn:PI_CI}). We will revisit this distinction later. 

\paragraph{Adaptive conformal inference with delayed feedback (ACI-DF).}

We introduce the ACI-DF algorithm for constructing rolling prediction intervals, described in Algorithm \ref{alg:ACI_DF}. The ACI-DF algorithm adapts the ACI algorithm \cite{gibbs2021adaptive} to the delayed feedback setting. The algorithm takes as input a sequence of prediction interval constructing functions $\hatPI^t: \cZ^{t-1} \times \R \subseteq \R$, and sets the rolling interval at step $t$ to be $\hatPI^t(z_{1:t-1}; \theta_t)$. The parameter $\theta_t$ is updated every $\Delta$ steps using online gradient descent on the pinball loss (line \ref{algline:ogd}), but with delayed feedback since $\ERRTar^t$ depends on $z_{t:\ntest+t-1}$. So we update $\theta_t$ based on the most recent computable 0-1 loss of coverage.

\begin{algorithm}[t]
  \caption{ACI-DF for rolling intervals}
  \label{alg:ACI_DF}
  \small
  \begin{algorithmic}[1]
    \REQUIRE Dataset $\{ z_t = (x_t, y_t) \}_{t \ge 1}$. Coverage level $\alpha$. prediction interval constructing functions $\hatPI^t: (\cZ^{t-1}, \R) \to 2^\cY$. Number of test samples $\ntest$. Lag parameter $\Delta$. Stepsize $\gamma > 0$. 
    \STATE Initialize $\theta_0 = 0$. Take $k = \arg\min_{s \in \N_+}\{s \Delta \ge \ntest \}$. 
    \FORALL{$t = 1, 2, \ldots, T$}
    \STATE Construct prediction set: $\hatRI^{\alpha, t} = \hatPI^t(z_{1:t-1};\theta_t)$. 
    \STATE Obtain $z_t = (x_t, y_t)$. 
    \STATE Compute $c_{t-\ntest+1} = 1\{ \ERRTar^{t-\ntest+1}  \in \hatRI^{t-\ntest+1} \}$. // $\ERRTar^{t-\ntest+1}$ depends on $z_{t-\ntest+1 : t}$ and is computable.  
    \IF{mod$(t, \Delta) =0$ and $t > k \Delta$}
    \STATE $\theta_{t+1} = \theta_{t-\Delta+1} + \gamma (1  - \alpha - c_{t- k \Delta+1})$.  \label{algline:ogd}
    \ELSE
    \STATE $\theta_{t+1} = \theta_t$. 
    \ENDIF
    \ENDFOR
  \end{algorithmic}
\end{algorithm}

We show that the ACI-DF algorithm has the following time-average coverage guarantee, whose proof follows similarly to \cite{gibbs2021adaptive} and is given in Appendix \ref{sec:proof_ACI_delayed}.
\begin{theorem}[ACI with delayed feedback]\label{thm:ACI_delayed}
Suppose $\hatPI^t: \cZ^{t-1} \times \R \to \R$ is a sequence of set constructing functions. Suppose there exists $m$ and $M$ such that for all $z_{1:t-1} \in \cZ^{t-1}$, we have $\hatPI^t(z_{1:t-1}, \theta) = \R$ for all $\theta > M$, and $\hatPI^t(z_{1:t-1}, \theta) = \emptyset$ for all $\theta < m$. Then the rolling intervals $\{ \hatRI^{\alpha, t}\}_{t \ge 1}$ from Algorithm \ref{alg:ACI_DF} satisfy
\begin{equation}
\lim_{T \to \infty} \frac{1}{| S_{T, \Delta} |} \sum_{t \in S_{T, \Delta}} 1 \{ \ERRTar^t \in  \hatRI^{\alpha, t}\} = 1 - \alpha. 
\end{equation}
\end{theorem}
We note that, similar to adaptive conformal inference, we have flexibility in choosing the prediction interval constructing functions. In particular, to achieve the time-average coverage guarantee, we are not required to use a statistically valid prediction interval construction function. Indeed, any prediction interval $\hatPI^t(z_{1: t-1}) = [\lo(z_{1:t-1}), \hi(z_{1:t - 1})]$ can serve as the base procedure, by simply taking $\hatPI^t(z_{1: t-1}; \theta) = [\lo(z_{1:t-1}) - \theta, \hi(z_{1:t - 1}) + \theta]$ to be the constructing function. However, if we hope for the average interval length to be small, we need a good base procedure. One suitable choice is the QFCV intervals from Section \ref{sec:QFCV_algorithm}. That is, at iteration $t$, we take $\hatPI^t(z_{1:t-1}; \theta_{t}) = \hatPI^{\alpha + \theta_t}_{\QFCV}$, where $\hatPI^{\alpha + \theta_t}_{\QFCV}$ is the QFCV interval (Algorithm \ref{alg:fcv_qc}) with $\nobs = t$ and nominal mis-coverage $\alpha + \theta_t$. We call the resulting algorithm Adaptive QFCV, and the rolling intervals are denoted as $\{ \hatRI_{\QFCV}^{\alpha, t} \}_{t \ge 1}$. 

% We remark that, similar to adaptive conformal inference, the choice of prediction interval constructing functions is flexible in ACI-DF. In particular, in order to have time-average coverage guarantee, we are not required to use a prediction interval constructing function that has statistical validity. Indeed, any prediction interval $\hatPI^t(z_{1: t-1}) = [\lo(z_{1:t-1}), \hi(z_{1:t - 1})]$ works by taking $\hatPI^t(z_{1: t-1}; \theta) = [\lo(z_{1:t-1}) - \theta, \hi(z_{1:t - 1}) + \theta]$. However, if we hope that the average length of the rolling prediction intervals is small, we need a good prediction interval constructing function $\hatPI^t$, for example, the QFCV procedure. 

\paragraph{Time-average coverage versus instance-average coverage.}

We note that the time-average coverage differs from the frequentist notion of instance-average coverage. To understand the difference, we define $c_{iT} = 1\{ \ERRTar^{t, i} \in \hatRI^{\alpha, t, i} \}$, where $i \in [\nsim]$ is the index of the simulation instance, and $t \in [T]$ is the time index. Different simulation instances are assumed to be IID. Then the coverage indicators form the following matrix: 
\[
\begin{bmatrix}
    c_{11} & \ldots & c_{1T}\\
    &\ldots & \\
   c_{\nsim 1} & \ldots & c_{\nsim T}
\end{bmatrix},
\]
where each row is the coverage of a fixed simulation instance across different time indices, and each column is the coverage of a fixed time index across different simulation instances. The two different coverage notions are then:
\begin{itemize}
    \item Time-average coverage of instance $i$: the row average $\lim_{T \to \infty}\frac{1}{T} \sum_{t=1}^T c_{it}$. 
    \item Instance-average coverage of time $t$: the column average $\lim_{\nsim \to \infty} \frac{1}{\nsim} \sum_{i=1}^{\nsim} c_{it} = \P(\ERRTar^t \in \hatRI^{\alpha, t})$. 
\end{itemize}
In practice, both notions of coverage are useful depending on the application scenario. 

We further remark that valid time-average coverage does not imply valid instance-average coverage, and vice versa. Note that ACI-DF method yields valid time-average coverage as shown in Theorem \ref{thm:ACI_delayed}, while QFCV prediction intervals have asymptotically valid instance-average coverage under ergodicity as shown in Theorem \ref{thm:conditional_coverage}. So we expect AQFCV to have valid coverage of both notions, though there is no guarantee of valid instance-average coverage.

% Notice that the time-average-coverage validity provides that the time average of the coverage $1\{ \ERRTar^t \in \hatRI^{\alpha, t} \}$ is nearly $(1 - \alpha) 100 \%$ for a fixed time series. On the other hand, the frequentist validity implies that, suppose we have $100$ independent time series, the instance average of coverage $1\{ \ERRTar^t \in \hatRI^{\alpha, t} \}$ is nearly $(1 - \alpha) 100 \%$ for any fixed time index. 
% To be more specific, consider the following matrix of coverage indicators, 
% \[
% \begin{pmatrix}
%     c_{11} & \ldots & c_{1T}\\
%     &\ldots & \\
%    c_{\nsim 1} & \ldots & c_{\nsim T}
% \end{pmatrix},
% \]
% where $c_{it}$ is the indicator for whether prediction interval at time $t$ covers the true target error $1\{\ERRTar^t \in \hatRI^t\}$ for simulation run indexed by $i$. 

%Notice that the reverse does not hold theoretically. While in practice we may observe empirical evidence otherwise, there is no theoretical guarantee and counterexamples can be produced. 

\subsection{Numerical simulations of AQFCV}\label{sec:simulation_AQFCV}
\begin{figure}[t]
\centering
\begin{minipage}[c]{.5\linewidth}
  \centering
  \includegraphics[width = 0.7 \linewidth]{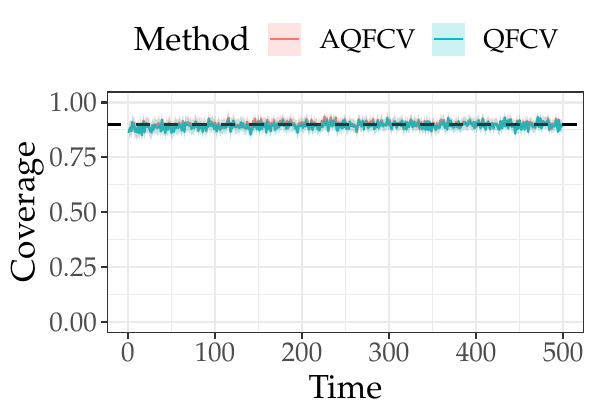}
  \subcaption{Instance-average coverage under stationarity}
\end{minipage}%
\begin{minipage}{.5\linewidth}
  \centering
  \includegraphics[width = 0.7 \linewidth]{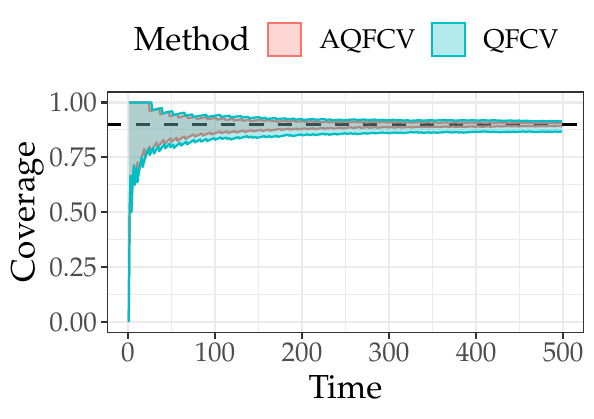}
  \subcaption{Time-average coverage under stationarity}
\end{minipage}
\begin{minipage}[c]{.5\linewidth}
  \centering
  \includegraphics[width = 0.7 \linewidth]{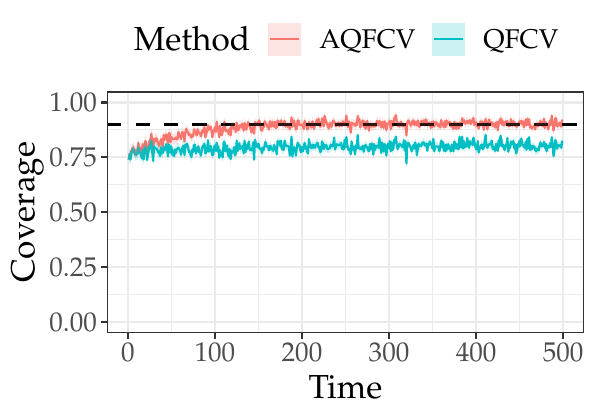}
  \subcaption{Instance-average coverage under nonstationarity}
\end{minipage}%
\begin{minipage}{.5\linewidth}
  \centering
  \includegraphics[width = 0.7 \linewidth]{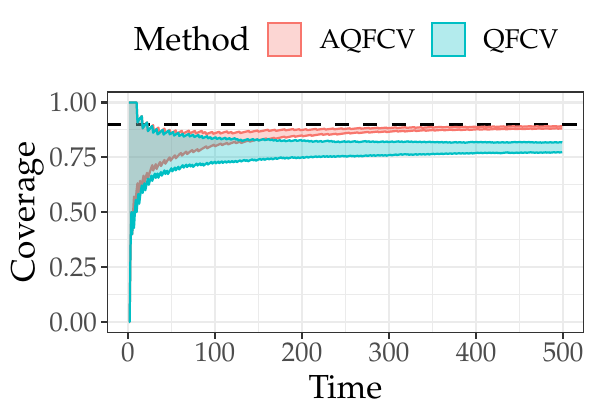}
  \subcaption{Time-average coverage under nonstationarity}
\end{minipage}
\caption{Instance-average and time-average coverages for a fixed time and rolling window under stationary and nonstationary time series.}
\label{fig:AQFCV}
\end{figure}

\paragraph{Comparing QFCV and AQFCV} 
We perform simulations to illustrate the coverage properties of QFCV and AQFCV (Adaptive conformal inference wrapped with QFCV) in both stationary and nonstationary time series. We generate $\nsim = 500$ IID instances of time series with $T = 3000$ steps each. The training set size for the (fixed window size) forecasting function is $\ntrain = 40$, and we evaluate the test error for the next $\ntest = 5$ steps. Furthermore, we generate the rolling interval every $\Delta = 5$ steps, starting from step $500$. So in total, there are $(3000 - 500) / 5 = 500$ rolling intervals. The time series are generated from the linear model (\ref{eqn:simulation_linear_model}) with $p = 20$ and $\beta = (1,1,1,1,0,\ldots, 0)^\sT \in \mathbb{R}^p$. In the stationary setting, the noise process $\varepsilon_t$ is an $\text{AR}(1)$ process with parameter $\phi = 0.5$. In the non-stationary setting, $\varepsilon_t$ is the sum of an $\text{ARIMA}(1,1,0)$ process with AR parameter $\phi = 0.99$ and an independent Gaussian sequence $\eta_t \sim_{iid} \mathcal{N}(0, t^4)$. This choice results in a noise process far from stationarity. In Figure \ref{fig:AQFCV}(a) and \ref{fig:AQFCV}(c), we report the instance-average coverage $\frac{1}{\nsim}\sum_{i=1}^\nsim c_{i t}$ for $t \in S_{\Delta, T}$, under the stationary and nonstationary settings. In Figure \ref{fig:AQFCV}(b) and \ref{fig:AQFCV}(d), we calculate the time-average coverage $\frac{1}{|S_{\Delta, t}|}\sum_{t \in S_{\Delta, t}} c_{it}$ for each instance $i \in [\nsim]$ and report the $5\%$ and $95\%$ quantiles over the $500$ instances.

Figure \ref{fig:AQFCV} shows that AQFCV results in asymptotically valid time-average coverage for both stationary and non-stationary settings. In stationary settings where QFCV is already instance-average valid, ACI-DF does not affect the instance-average validity. In nonstationary settings where QFCV slightly undercovers, ACI-DF can correct the interval to give correct time-average coverage. We note that AQFCV can also obtain valid instance-average coverage in nonstationary settings, an interesting phenomenon beyond our theories that deserves further investigation. 

\paragraph{Comparing AFCV and AQFCV} We compare the performance of ACI-DF wrapped with two base methods: CLT-based FCV intervals (AFCV) and QFCV prediction intervals (AQFCV). Theorem \ref{thm:ACI_delayed} provides asymptotic time-average coverage guarantee when ACI-DF is combined with any base method satisfying mild assumptions. However, AQFCV may have advantages over AFCV despite both having valid asymptotic coverage.

\begin{figure}[t]
    \centering
    \includegraphics[width = 0.6\linewidth]{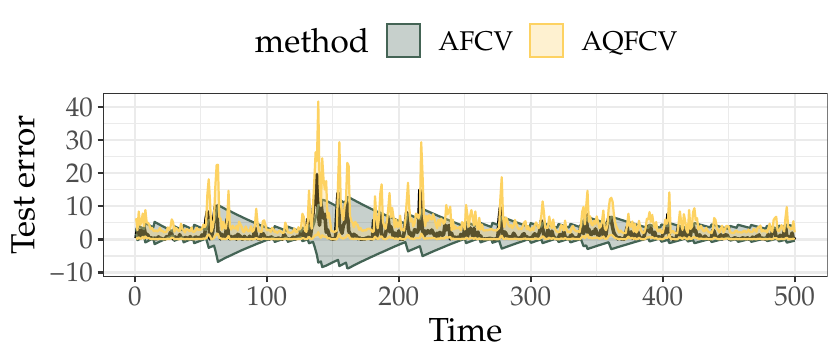}
    \caption{Comparison of adaptive QFCV and naive FCV intervals by wrapping around adaptive conformal inference over one time series instance. }
    \label{fig:AFCV}
\end{figure}

Through simulation, we aimed to illustrate why practitioners may prefer AQFCV over wrapping ACI-DF with a less carefully constructed prediction interval such as FCV based interval. 
In Figure \ref{fig:AFCV}, we compared the rolling intervals from AFCV and AQFCV on a generated time series with $T=3000$ steps. The forecasting function used a fixed training window of $\ntrain = 40$ steps to generate forecasts for the next $\ntest = 5$ steps, with rolling intervals for prediction error produced every $\Delta = 5$ steps starting at step $500$, for $500$ intervals total. The time series comes from a $20$-dimensional linear model (\ref{eqn:simulation_linear_model}) with an $\text{ARMA}(1, 20)$ noise process at $\phi = 0.5$. This is the same generative process used in prior simulation as in case (b) of Figure \ref{fig:sim_QFCV}, \ref{fig:FCV}, and Table \ref{table:FCV}. 

Figure \ref{fig:AFCV} shows that AQFCV provides more adaptive intervals than AFCV, with wider intervals in volatile regions and narrower intervals when volatility is low. This adaptiveness of AQFCV stems from the base QFCV method exploiting time correlation between validation and test errors. We note this phenomenon may not always occur, especially if error correlations are weak. Still, AQFCV is recommended since a good base method at low computational cost brings benefits such as potential instance-average coverage. Quantile-based methods also naturally handle arbitrary nonstationarity through flexible quantile-based prediction interval constructing function.  

% \begin{table}[t]
%     \centering
%     \begin{tabular}{c|ccccc}
%     \toprule
%     \text { Date and time } & \text { Price } & \text { Price D-1 } & \text { Price D-7 } & \text { For. cons. } & \text { DOW } \\ \midrule
%     \text{11/01/16 0PM} & 21.95 & 15.58 & 13.78 & 58800 & \text { Monday } \\ 
% \text {11/01/16 1PM} & 20.04 & 19.05 & 13.44 & 57600 & \text { Monday } \\ 
% \vdots & \vdots & \vdots & \vdots & \vdots & \vdots \\
% \text{12/01/16 0PM} & 21.51 & 21.95 & 25.03 & 61600 & \text { Tuesday } \\ 
% \text{12/01/16 1PM} & 19.81 & 20.04 & 24.42 & 59800 & \text { Tuesday } \\ 
% \vdots & \vdots & \vdots & \vdots & \vdots & \vdots \\ 
% \text{18/01/16 0PM} & 38.14 & 37.86 & 21.95 & 70400 & \text { Monday } \\ 
% \text{18/01/16 1PM} & 35.66 & 34.60 & 20.04 & 69500 & \text { Monday } \\ 
% \vdots & \vdots & \vdots & \vdots & \vdots & \vdots \\
%     \bottomrule
%     \end{tabular}
%     \caption{Extract of the dataset for French electricity spot price forecasting.}
%     \label{tab:french_electricity}
% \end{table}

% \subsection{Discussion}
% In this section, we combined ACI with QFCV to result in an adaptive method called AQFCV, with provable asymptotic time-average coverage under arbitrary nonstationarity. Two interesting open questions remain. 
% \begin{itemize}
%     \item Whether AQFCV has instance-average coverage? More generally speaking, will any base method combined with ACI has instance-average coverage? We need more understanding in this direction
%     \item Whether QFCV has time-average coverage under stationarity? 
% \end{itemize}

% \begin{table}[!h]
% \centering
% \begin{tabular}{p{4cm} | p{1cm}  p{1cm}}
% \toprule
% & QFCV & AQFCV \\
% \midrule
% Instance-average coverage & \multicolumn{1}{c}{\checkmark} &  \multicolumn{1}{c}{?} \\
% Time-average coverage & \multicolumn{1}{c}{?} & \multicolumn{1}{c}{\checkmark} \\
% \bottomrule
% \end{tabular}
% \end{table}

% \begin{figure}
%     \centering
%     \includegraphics[width = 0.5\linewidth]{length.pdf}
%     \caption{Comparison of prediction interval lengths for stationary and nonstationary settings. }
%     \label{fig:aci_length}
% \end{figure}

\section{Real data examples}

% \sm{Prediction without context: ARIMA noise without feature; 1. ARIMA fit, or Linear regression fit. }

%\sm{ARIMA feature + iid noise}

% \sm{Prediction with context: iid feature + ARIMA noise; 1. Linear regression (X + Y)}

% %\sm{If time permit: ARIMA feature + ARIMA ground truth + ARIMA noise}

% %\sm{ARIMA to SARIMA? }

% \sm{Prediction with enlarged window: same as the above setting, but using the dataset with enlarged window (weighted linear regression). } \sm{Explore. }

% \sm{Adaptive; non-stationary data. }

% \subsection{Real dataset}

\subsection{French electricity dataset}

We first evaluate QFCV and AQFCV using the French electricity price forecast dataset from \cite{zaffran2022adaptive}. The dataset contains French electricity spot prices from 2016 to 2019, with $(3 \times 365 + 366) \times 24 = 35064$ hourly observations of forecast consumptions and prices. The goal of the forecaster is to predict at day $D$ the prices of the next 24 hours of the day $D + 1$. The forecaster uses the following features: day-ahead forecast consumption, day-of-the-week, $24$ prices of the day $D - 1$, and $24$ prices of the day $D - 7$. 

% Table \ref{tab:french_electricity} provides an extract of the dataset. 

\begin{figure}[t]
\centering
\begin{minipage}[c]{.5\linewidth}
  \centering
  \includegraphics[width = \linewidth]{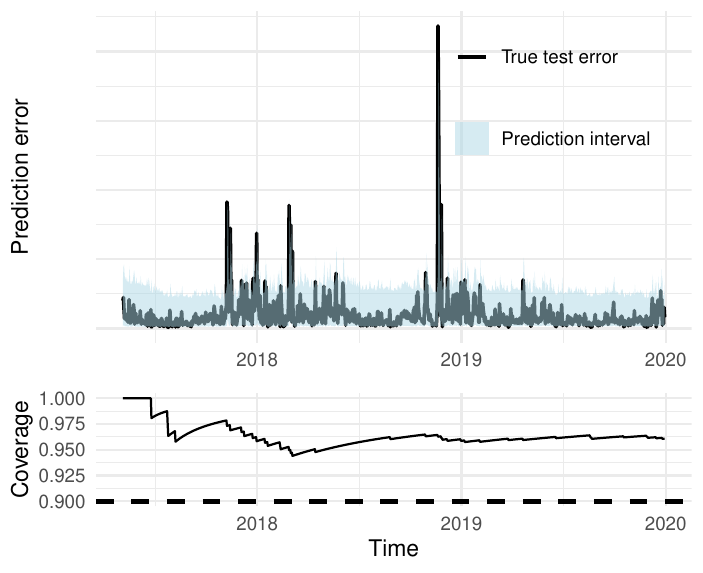}
  \subcaption{QFCV}
\end{minipage}%
\begin{minipage}{.5\linewidth}
  \centering
  \includegraphics[width = \linewidth]{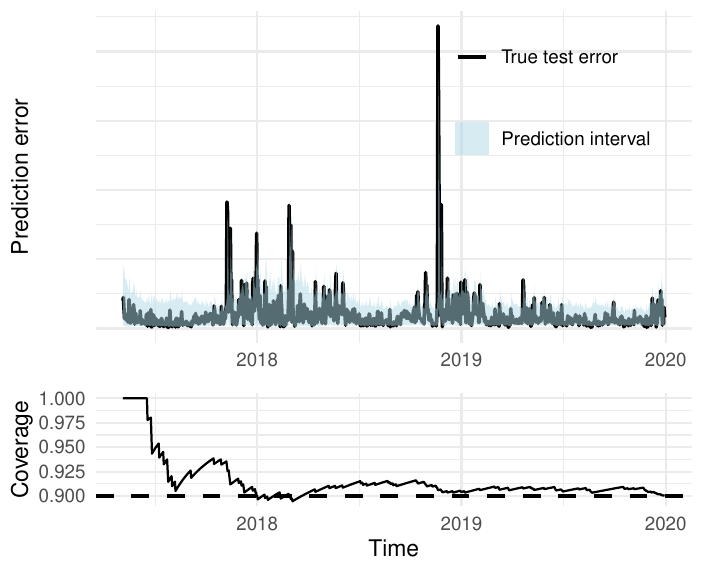}
  \subcaption{AQFCV}
\end{minipage}
\caption{French electricity prediction error time-average coverage with QFCV and AQFCV methods. Linear ridge regression is trained on data in the last 7 months to predict the next 24 hours of electricity prices so that $\ntrain = 24 \times 7 \times 30 = 5040$. We choose $\nval = \Delta = \ntest = 24$, $\gamma = 0.01$. Top row plots true prediction errors and calculated prediction intervals across time. The bottom row plots the time-average coverage of uncertainty intervals of prediction error.}
\label{fig:electricity}
\end{figure}

We use linear ridge regression with appropriate fixed ridge parameters as the prediction algorithm $\hat f$, trained on the data of the last $7$ months, so that $\ntrain = 24 \times 7 \times 30 = 5040$. The test error is evaluated on the next $24$ hours of electricity price, so $\ntest = 24$. We use {\QFCV} and {\AQFCV} to provide the prediction interval for the test error. In methods, we set $\nval = \Delta = \ntest = 24$, $\gamma = 0.01$. The result is provided in Figure \ref{fig:electricity}. 

By visual inspection, the error process of French electricity dataset is not stationary due to large spikes of varying strengths. The AQFCV method, which wraps ACI-DF with QFCV, yields the asymptotically nominal $90\%$ coverage rate as promised in Theorem \ref{thm:ACI_delayed}. Although QFCV does not have a time-average coverage guarantee under nonstationarity, it has enough coverage in this particular example, despite some over-coverage. This contrasts the simulation example in Section \ref{sec:simulation_AQFCV}, where QFCV undercovers in nonstationary time series forecasting (see Figure \ref{fig:AQFCV}). The simulation example is a special case where forecasting difficulty increases over time, so intervals using historical data underestimate future errors, causing under-coverage. In the French electricity dataset, nonstationarity arises from error spikes available to the QFCV quantile regression step, so QFCV assumes spikes occur regularly under stationarity. Thus, QFCV outputs wide intervals, resulting in over-coverage.

\subsection{Stock Market Volatility dataset}
\begin{figure}[t]
\centering
\begin{minipage}[c]{.5\linewidth}
  \centering
  \includegraphics[width = 0.9 \linewidth]{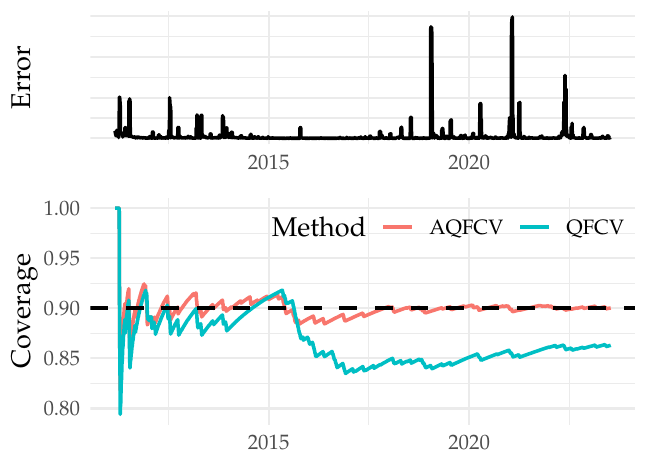}
  \subcaption{Nvidia}
\end{minipage}%
\begin{minipage}{.5\linewidth}
  \centering
  \includegraphics[width = 0.9 \linewidth]{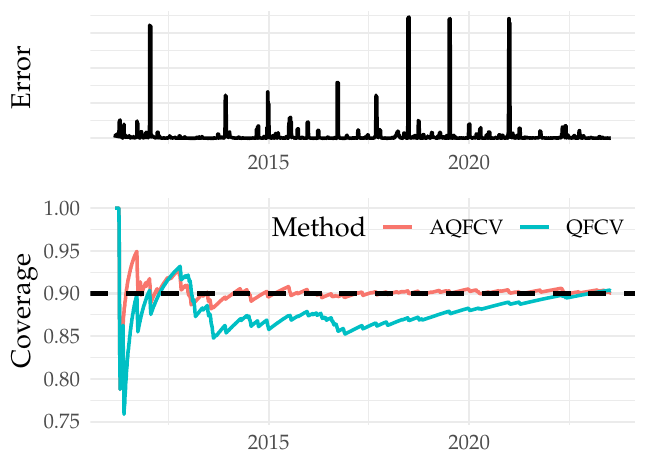}
  \subcaption{AMD}
\end{minipage}
\begin{minipage}[c]{.5\linewidth}
  \centering
  \includegraphics[width = 0.9 \linewidth]{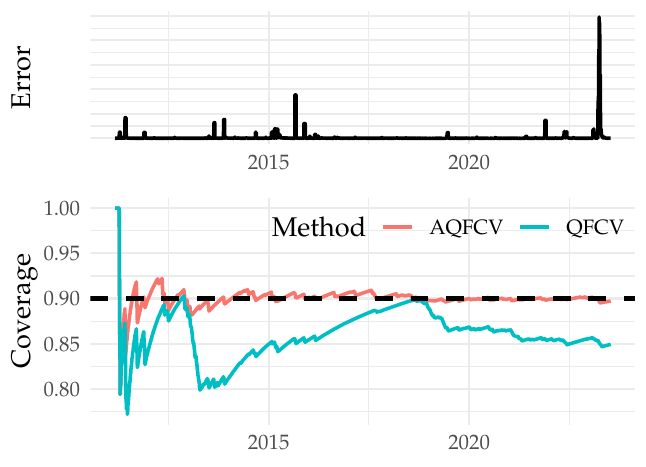}
  \subcaption{BlackBerry}
\end{minipage}%
\begin{minipage}{.5\linewidth}
  \centering
  \includegraphics[width = 0.9 \linewidth]{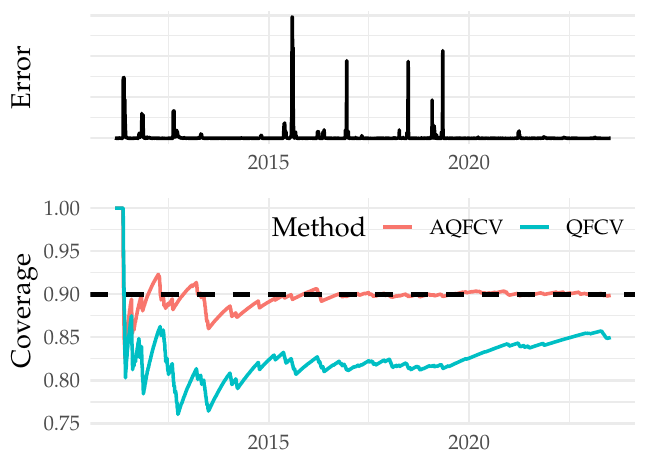}
  \subcaption{Fannie Mae}
\end{minipage}
\caption{Stock market volatility prediction error time-average coverage with QFCV and AQFCV methods. $\text{Garch}(1,1)$ model is trained on data in the last 1000 business days to predict the next $7$ days of realized volatility, such that $\ntrain = 1000$. We choose $\nval = \ntest = 7$, $\Delta = 1$, and $\gamma = 0.01$. Top row plots true prediction errors across time. Bottom row plots the time-average coverage of uncertainty intervals.}
\label{fig:stock}
\end{figure}

Next, we evaluate QFCV and AQFCV using the publicly available stock market volatility dataset from \textit{The Wall Street Journal}, the same dataset adopted in \cite{gibbs2021adaptive}. As in \cite{gibbs2021adaptive}, we select the same four stocks and use their daily minimum prices $\{P_t\}_{1 \leq t\leq T}$ from January 2005 to July 2023. For each stock we calculate daily returns $R_t = (P_t - P_{t-1})/P_{t-1}$ for $t \geq 2$ and thus the realized volatility $V_t = R_t^2$. We consider the task of multiperiod forecasting of market volatility $y_{t+1:t+\ntest} := (V_{t+1}, \ldots, V_{t+\ntest})$ using previously observed returns $x_t := \{R_s\}_{1\leq s \leq t}$.

Similar to \cite{gibbs2021adaptive}, we adopt the $\text{Garch}(1,1)$ model \cite{bollerslev1986generalized} as the prediction algorithm $\hat{f}$, which assumes $R_t = \sigma_t \varepsilon_t$ where $\varepsilon_t$ are IID sampled from $\mathcal{N}(0,1)$ and $\sigma_t^2 =  \omega + \tau V_{t-1} + \beta \sigma_{t-1}^2$ with $(\omega, \tau, \beta)$ as trainable parameters. For a given time stage $t$, we train on the data from the last $\ntrain$ business days $\{R_s\}_{t - \ntrain < s \leq t}$ to obtain coefficient estimates $(\hat{\omega}_t, \hat{\tau}_t, \hat{\beta}_t)$ and past standard deviation estimates $\{\hat{\sigma}^t_s\}_{t-\ntrain < s\leq t}$. Then we give point estimates of multiperiod volatility in the next $\ntest$ days as follows,
\begin{equation*}
\hat{f}(r;\{R_s\}_{t-\ntrain < s \leq t}) = (\hat{\sigma}_{t+r}^t)^2 :=
    \begin{cases}
    \hat{\omega}_t + \hat{\tau}_t V_{t-1} + \hat{\beta}_t (\hat{\sigma}_{t-1}^t)^2, & r = 1, \\
    \hat{\omega}_t + (\hat{\tau}_t + \hat{\beta}_t) (\hat{\sigma}_{t+r-1}^t)^2, & 1 < r \leq \ntest.
\end{cases}
\end{equation*}
We then want to give a prediction interval for the test error incurred in predicting the volatility over the next $\ntest$ days. That is, we want to find intervals $\{ \text{RI}_t^\alpha \}_{t \in [T]}$ such that 
\[
\frac{1}{T}\sum_{t = 1}^T 1 \Big\{ \frac{1}{\ntest}\sum_{r=1}^\ntest \big((\hat{\sigma}_{t+r}^t)^2 - V_{t+r} \big)^2 \in \text{RI}_t^\alpha\Big\} \approx 1-\alpha. 
\]
% where $\ell$ is taken to be squared error loss.

% We apply both QFCV and AQFCV (ACI-DF wrapped with QFCV) methods to produce rolling prediction intervals for test error as described in Algorithm \ref{alg:fcv_qc} and Algorithm \ref{alg:ACI_DF}, respectively. 

For both QFCV and AQFCV, we set $\ntrain = 1000$, $\nval = \ntest = 7$, $\Delta = 1$, and $\gamma = 0.01$. Time-average coverages for all four stocks are presented in Figure \ref{fig:stock}. For all four stock prices, the AQFCV method provides the asymptotically nominal $90\%$ coverage rate as established in Theorem \ref{thm:ACI_delayed}. On the other hand, we do not expect QFCV to have a time-average coverage guarantee under nonstationarity. In settings (a), (c), and (d) of Figure \ref{fig:stock} (corresponding to Nvidia, BlackBerry, and FannieMae), QFCV undercovers with an approximately $85\%$ coverage rate. This aligns with simulation results in Section \ref{sec:rolling}. In setting (c), we can see that QFCV restores the nominal coverage rate in years with fewer spikes in true error evolution. And in setting (b) (corresponding to AMD), QFCV achieves a nominal coverage rate in more recent years due to fewer spikes in frequency and magnitude of error evolution trends compared to historical trends. 

% Notice that different stocks exhibit different levels of nonstationarity, as can be seen from the evolution of prediction errors over time with spikes of varying strengths and frequencies. 
\section{Conclusion and discussion}

This paper provides a systematic investigation of inference for prediction errors in time series forecasting. We propose the quantile-based forward cross-validation (QFCV) method to construct prediction intervals for the stochastic test error $\ERRTar$, which have asymptotic valid coverage under stationarity assumptions. Through simulations, we find that QFCV provides well-calibrated prediction intervals that can be substantially shorter than naive empirical quantiles when the validation error is predictive of the test error. For non-stationary settings, we propose an adaptive QFCV procedure (AQFCV) that combines QFCV intervals with adaptive conformal prediction using delayed feedback. We prove AQFCV provides prediction intervals with asymptotic time-average coverage guarantees. Experiments on simulated and real-world time series demonstrate that AQFCV adapts efficiently during periods of low volatility. Overall, we advocate the use of QFCV and AQFCV procedures due to their statistical validity, computational simplicity, and flexibility to different time series characteristics. 

Our work opens up several promising research avenues. One direction is proving that AQFCV maintains its time-average coverage guarantee when the underlying data-generating process is stationary. Another direction is developing statistically principled inference methods for prediction errors that are compatible with the expanding window forecasting setup. More broadly, this work underscores the value of rigorous uncertainty quantification when evaluating predictive models on temporal data. We hope that our proposed methodology helps catalyze further research into validated approaches for quantifying uncertainty in time series forecasting tasks.

\section*{Acknowledgement}

The authors would like to thank Ying Jin, Trevor Hastie, Art Owen, Anthony Norcia, and Shuangping Li for their helpful discussions. Song Mei was supported in part by NSF DMS-2210827 and CCF-2315725. Robert Tibshirani was supported by the National Institutes of Health (5R01 EB001988-16) and the National Science Foundation (19 DMS1208164).

\bibliographystyle{alpha}
\bibliography{main_arXiv}

\clearpage

\appendix

\section{Proof of Theorem \ref{thm:conditional_coverage}}\label{sec:proof_conditional_coverage}

\begin{proof}[Proof of Theorem \ref{thm:conditional_coverage}]
Throughout the proof, we take $\Delta = 1$. For general $\Delta > 1$, the proof follows similarly. The proof uses the uniform convergence argument and relies on the following lemma.
% \begin{lemma}[Birkhoff's ergodic theorem]\label{lem:ergodic_LLN}
% Let $\{ z_t \}_{t \ge 1}$ be an ergodic stochastic process. Then for any function $g: \cZ^k \to \R$ with $\E[\vert g(z_{1:k}) \vert] < \infty$, any $\varepsilon > 0$, and any integer $k \ge 1$, we have 
% \[
% \lim_{n \to \infty} \P\Big( \Big\vert \frac{1}{n}\sum_{t = 1}^n g(z_{t: t + k - 1}) - \E[g(z_{1:k})] \Big\vert \ge \varepsilon \Big) = 0. 
% \]
% \end{lemma}

\begin{lemma}\label{lem:risk_growth}
Let $(E_1, E_2)$ be random variables that have a joint density with respect to the Lebesgue measure. Define the risk function
\[
R_\beta(f) = \E_{(E_1, E_2)}\Big[ \pinball_\beta\Big( E_2 - f(E_1) \Big) \Big]. 
\]
Then for any function $f_\star \in \arg\min_{g: \R^m \to \R} R_\beta(g)$, we have, 
\[
\P_{(E_1, E_2)}(E_2 \le f_\star(E_1)) = 1 - \beta. 
\]
\end{lemma}
We take $K = \nobs - \ntrain - \ntest - \nval + 1$ and define the empirical risk associated to the $\beta$-pinball loss as
\[
\widehat R_{\beta}(f) = \frac{1}{K} \sum_{i = 1}^{K} \pinball_\beta\Big(\textup{Err}_{i}^{\test} - f(\textup{Err}_{i}^{\fea}) \Big).
\]
For $\beta = \alpha/2$ and $\beta = 1 - \alpha/2$, and for any quantile function $q_\beta \in \cF \cap \cF_\beta$, we define 
\begin{equation}\label{eqn:def_varepsilon_beta}
\varepsilon_\beta = \frac{1}{3} \Big[\inf_{f \in \cF \setminus \cF_\beta} R_\beta(f) - R_\beta(q_\beta)\Big].
\end{equation}
By the definition of $\cF_\beta$, we have $\varepsilon_\beta$ is well-defined. By Assumption \ref{ass:finiteness_realizability} that $\cF$ is a finite set, we have $\varepsilon_\beta > 0$. By Assumption \ref{as:stationarity}, for any fixed function $f \in \cF$, we have point convergence of $\widehat R_\beta(f)$ to $R_\beta(f)$, 
\begin{equation}
\lim_{n \to \infty} \P\Big( \Big\vert \widehat R_\beta(f) - R_\beta(f) \Big\vert \le \varepsilon_\beta \Big) = 1. 
\end{equation}
By Assumption \ref{ass:finiteness_realizability} that $\cF$ is a finite set, applying union bound gives uniform convergence 
\begin{equation}\label{eqn:uniform_convergence}
\lim_{n \to \infty} \P\Big( \sup_{f \in \cF} \Big\vert \widehat R_\beta(f) - R_\beta(f) \Big\vert \le \varepsilon_\beta \Big) = 1. 
\end{equation}
As a consequence, by Assumption \ref{ass:finiteness_realizability} that $\exists q_{\alpha/2}, q_{1 - \alpha/2} \in \cF$ and by the definition of $\varepsilon_\beta$ as in Eq. (\ref{eqn:def_varepsilon_beta}), when the good event in Eq. (\ref{eqn:uniform_convergence}) happens, we have 
\[
\begin{aligned}
\inf_{f \in \cF \setminus \cF_\beta} \widehat R_\beta(f) - \widehat R_\beta(q_\beta) 
\ge&~ \Big[ \inf_{f \in \cF \setminus \cF_\beta}  R_\beta(f) - R_\beta(q_\beta) \Big] - 2  \sup_{f \in \cF} \Big\vert \widehat R_\beta(f) - R_\beta(f) \Big\vert 
\ge  3 \varepsilon_\beta - 2 \varepsilon_\beta = \varepsilon_\beta. 
\end{aligned}
\]
In words, any minimizer of the empirical risk should be contained in the set $\cF_\beta$. By the fact that $\hat f^\beta_n$ is a minimizer of the empirical risk $\widehat R_\beta(f)$, we get that $\hat f^\beta_n$ is contained in the set $\cF_\beta$ with high probability. That is, for $\beta = \alpha/2$ or $\beta = 1 - \alpha/2$, we have
\[
\lim_{n \to \infty} \P\Big( \hat f_n^\beta \in \cF_\beta \Big) = 1. 
\]
Then by Lemma \ref{lem:risk_growth},  we have 
\[
\lim_{n \to \infty} \P\Big( \ERRTar \in  [\hat f_n^{\alpha/2}(\ERR_\star^{\fea}), \hat f_n^{1 - \alpha/2}(\ERR_\star^{\fea})] \Big) = 1 - \alpha. 
\]
This finishes the proof of the theorem. 
\end{proof}

% That is, we have
% \[
% \lim_{n \to \infty} \P\Big( \inf_{f \in \cF \setminus \cF_\beta} \widehat R_\beta(f) - \widehat R_\beta(q_\beta) \ge \varepsilon_\beta  \Big) = 1. 
% \]
% \begin{proposition}
% For any $\alpha \in (0,1)$, we have 
% \[
% \lim_{n \to \infty} \sup_{e \in \R} | \hat f^\alpha_n (e) - f^\alpha(e) | = 0, 
% \]
% where $f^\alpha(e)$ is the $\alpha$-quantile of $\textup{Err}^{test}$ given $\textup{Err}^{val} = e$, when $(\textup{Err}^{val}, \textup{Err}^{test}) \sim \cL$. 
% \end{proposition}

% While Lemma \ref{lem:ergodic_LLN} is a textbook lemma, w

We next give a proof of Lemma \ref{lem:risk_growth}. 

\begin{proof}[Proof of Lemma \ref{lem:risk_growth}]
\def\supp{{\rm supp}}

Let $p_{(E_1, E_2)}(e_1, e_2)$ denote the density of $(E_1, E_2)$, $p_{E_1}(e_1)$ denote the marginal density of $E_1$, and $p_{E_2 | E_1}(e_2 | e_1)$ denote the conditional density of $E_2$ given $E_1 = e_1$. 

\noindent
{\bf Step 1. Show that $\P( E_2 \le f_\star(e_1) | E_1 = e_1) = 1 - \beta$ for almost every $e_1 \in \supp\{ p_{E_1} \}$. } For any $f_\star \in \arg\min_{g: \R^m \to \R} R_\beta(g)$, we claim that $\P( E_2 \le f_\star(e_1) | E_1 = e_1) = 1 - \beta$ for almost every $e_1 \in \supp\{ p_{E_1} \}$. Otherwise, suppose that for certain $D \subseteq \R$ with $\P(E_1 \in D) > 0$, we have $\P( E_2 \le f_\star(e_1) | E_1 = e_1) \neq 1 - \beta$, by the fact that the minimizer of the expected $\beta$-pinball loss gives the $\beta$-quantile, we have 
\[
\E_{E_2 | E_1}\Big[ \pinball_\beta\Big( E_2 - f_\star(e_1) \Big)  \Big| E_1 = e_1 \Big] > \inf_{h} \E_{E_2 | E_1}\Big[\pinball_\beta\Big( E_2 - h \Big)  \Big| E_1 = e_1 \Big].
\]
Then we can take $g: D \to \R$ with $g(e_1) \in \arg\min_{h \in \R} \E_{E_2 | E_1}[ \pinball_\beta( E_2 - h ) | E_1 = e_1]$ for $e_1 \in D$, and define $\tilde f(e_1) = f_\star(e_1) 1\{ e_1 \not\in D \} + g(e_1) 1\{ e_1 \in D\}$. Then we have 
\[
\begin{aligned}
&~\E_{(E_1, E_2)}\Big[ \pinball_\beta\Big( E_2 - \tilde 
 f(E_1) \Big) \Big] - \E_{(E_1, E_2)}\Big[ \pinball_\beta\Big( E_2 - f_\star(E_1) \Big) \Big] \\
 =&~ \int_{D} \Big\{ \E_{E_2 | E_1}\Big[ \pinball_\beta\Big( E_2 - \tilde 
 f(e_1) \Big)  \Big| E_1 = e_1 \Big] - \E_{E_2 | E_1}\Big[\pinball_\beta\Big( E_2 - f_\star(e_1) \Big)  \Big| E_1 = e_1 \Big] \Big\} p_{E_1}(e_1) d e_1 < 0, \\
 \end{aligned}
\]
which contradicts the fact that $f_\star \in \arg\min_{g: \R^m \to \R} R_\beta(g)$. 

\noindent
{\bf Step 2. Concludes the proof. } We have
\[
\P(E_2 \le f_\star(E_1))= \int \P( E_2 \le f_\star(e_1) | E_1 = e_1) p_{E_1}(e_1) d e_1 = 1 - \beta, 
\]
where the last equality is by Step 1. This proves the lemma. 
\end{proof}

\subsection{General result going beyond finite function class}\label{sec:general_conditional_coverage}

In this section, we state and prove a more general version of Theorem \ref{thm:conditional_coverage}, which goes beyond the assumption that $\cF$ is a finite function class. We start by stating the additional required assumptions. 
% \begin{assumption}
% The stochastic process $\{ z_t \}_{t \ge 1}$ is stationary and ergodic.   
% \label{as:stationarity_general}
% \end{assumption}

\begin{assumption}\label{ass:independence-errtar}
$(\ERR_\star^{\fea}, \ERR_\star^{\test})$ is independent of $\{ (\ERR_t^{\fea}, \ERR_t^{\test}) \}_{t \ge 1}$. 
\end{assumption}

For a distribution $Q$ and for $\alpha \in (0, 1)$, the $\alpha$-quantile of $Q$ is the set 
\[
q_\alpha(Q) = \{ t \in \R: Q((-\infty, t]) \ge \alpha, Q([t, + \infty) ) \ge 1 - \alpha \}. 
\]
We write $q_{\alpha, \min}(Q) \equiv \min q_\alpha(Q)$ and $q_{\alpha, \max}(Q) \equiv \max q_\alpha(Q)$. Consider the stationary distribution $\cL \in \cP(\R \times \R)$ of $\{ (\ERR_t^{\fea}, \ERR_t^{\test}) \}_{t \ge 1}$ and let $(X, Y) \sim \cL$. We denote $\cL_X$ to be the marginal distribution of $X$, and $\cL_{Y| x}$ to be the conditional distribution of $Y$ given $X = x$. 
\begin{assumption}\label{ass:density_bound_general}
For $\cL_X$-almost-every $x \in \R$, $\cL_{Y| x}$ has a density that is uniformly upper-bounded by $C$. 
\end{assumption}

\begin{assumption}[Definition 2.6 of \cite{steinwart2011estimating}]\label{ass:quantile_condition_general}
We have $\supp \cL_{Y| x} \in [0, B]$ for $\cL_X$-almost-every $x \in \R$. Furthermore, for $\cL_X$-almost-every $x \in \R$, there exists $c(x) \in (0, 2]$ and $b(x) > 0$ such that for all $s \in [0, c(x)]$, we have 
\begin{align*}
&~\cL_{Y| x}((q_{\beta, \min}(\cL_{Y| x}) - s, q_{\beta, \min}(\cL_{Y| x}))) \ge b(x) s, \\
&~\cL_{Y| x}((q_{\beta, \max}(\cL_{Y| x}) , q_{\beta, \max}(\cL_{Y| x})+ s)  ) \ge b(x) s,~~~~~\forall \beta \in \{ \alpha/2, 1 - \alpha/2\}.
\end{align*}
Moreover, denoting $\gamma(x) = 1 / [b(x) \cdot c(x)]$, we have $\gamma \in L^1(\cL_X)$. 
\end{assumption}

Recall that we have 
\[
\cF_\beta = \Big\{ g^* \in \arg\min_{g: \R^m \to \R} \E_{\cL}\big[\pinball_\beta\big(\ERR^{\test} 
 -g(\ERR^{\fea})\big)\big]  \Big\}. 
\]
\begin{assumption}\label{ass:finiteness_realizability_general}
Assume that $(\cF, \| \cdot \|_\infty )$ is a compact space, and $\cF_{\alpha/2} \cap \cF \neq \emptyset$, $\cF_{1 - \alpha/2} \cap \cF \neq \emptyset$. 
\end{assumption}

Given these assumptions, we are ready to state the general asymptotic validity result. 

\begin{theorem}[General version of Theorem \ref{thm:conditional_coverage}]\label{thm:conditional_coverage_general}
Let Assumption \ref{as:stationarity}, \ref{ass:independence-errtar}, \ref{ass:density_bound_general}, \ref{ass:quantile_condition_general} and \ref{ass:finiteness_realizability_general} hold.  Let $\hatPI^\alpha_\QFCV$ be the output of Algorithm \ref{alg:fcv_qc}. Then we have 
\begin{equation}\label{eqn:asymptotic_validity_general}
\lim_{n \to \infty} \P\Big( \ERRTar \in [\hat f^{\alpha/2}(\ERR^{\fea}_\star), \hat f^{1 - \alpha/2}(\ERR^{\fea}_\star)] \Big) = 1 - \alpha. 
\end{equation}
\end{theorem}

\begin{proof}[Proof of Theorem \ref{thm:conditional_coverage_general}]
For $K = \nobs - \ntrain - \ntest - \nval + 1$, define 
\[
\widehat R_{\beta}(f) = \frac{1}{K} \sum_{i = 1}^{K} \pinball_\beta\Big(\textup{Err}_{i}^{\test} - f(\textup{Err}_{i}^{\fea}) \Big).
\]
For any $\varepsilon > 0$, by the fact that $(\cF, \| \cdot \|_\infty)$ is compact, there exists an $\varepsilon$-cover $N(\varepsilon; \cF, \| \cdot \|_\infty)$ which is a finite set that satisfies the following: for any $f \in \cF$, there exists $f' \in N(\varepsilon; \cF, \| \cdot \|_\infty)$, such that $\| f - f' \|_\infty < \varepsilon$. By Assumption \ref{as:stationarity}, for any fixed function $f \in N(\varepsilon; \cF, \| \cdot \|_\infty)$, we have 
\begin{equation}
\lim_{n \to \infty} \P\Big( \Big\vert \widehat R_\beta(f) - R_\beta(f) \Big\vert \le \varepsilon \Big) = 1. 
\end{equation}
Applying union bound, we get uniform convergence over the $\varepsilon$-cover
\begin{equation}
\lim_{n \to \infty} \P\Big( \sup_{f \in N(\varepsilon; \cF, \| \cdot \|_\infty)} \Big\vert \widehat R_\beta(f) - R_\beta(f) \Big\vert \le \varepsilon \Big) = 1. 
\end{equation}
Notice that we have $| \widehat R_\beta(f) - \widehat R_\beta(f')| \le \| f - f' \|_\infty$ and $| R_\beta(f) - R_\beta(f')| \le \| f - f' \|_\infty$ for any $f, f'$. Then we have uniform convergence over the whole space $\cF$
\begin{equation}\label{eqn:uniform_convergence_general}
\lim_{n \to \infty} \P\Big( \sup_{f \in \cF} \Big\vert \widehat R_\beta(f) - R_\beta(f) \Big\vert \le 3 \varepsilon \Big) = 1. 
\end{equation}
Denote $\hat f_\beta \in \arg\min_{f \in \cF} \widehat R_\beta(f)$ and $f^\star_\beta \in \arg\min_{f \in \cF} R_\beta(f)$ (choose one if there are multiple of them). Then by standard decomposition, we have
\[
\begin{aligned}
R_\beta(\hat f_\beta) =&~ R_\beta(\hat f_\beta) - \widehat R_\beta(\hat f_\beta) + \widehat R_\beta(\hat f_\beta) - \widehat R_\beta( f_\beta) + \widehat R_\beta(f_\beta) - R_\beta(f_\beta) + R_\beta(f_\beta)\\
\le &~ 2 \sup_{f \in \cF} |R_\beta(f) - \widehat R_\beta(f)| + R_\beta(f_\beta).
\end{aligned}
\]
Combining with Eq. (\ref{eqn:uniform_convergence_general}), we have 
\[
\lim_{n \to \infty} \P\Big(  R_\beta(\hat f_\beta) \le \min_{f \in \cF} R_\beta(f) +  6 \varepsilon \Big) = 1. 
\]
Furthermore, by Theorem 2.7 of \cite{steinwart2011estimating} and by Assumption \ref{ass:finiteness_realizability_general}, we have 
\[
\min_{q \in \cF_\beta} \| \hat f_\beta - q \|_{L^1(\cL_X)} \le 2 \sqrt{\| \gamma \|_{L^1(\cL_X)} } \sqrt{ R_\beta(\hat f_\beta) - \inf_{f \in \cF} R_\beta(f) }. 
\]
This implies that for any $\varepsilon > 0$ (may be different from the $\varepsilon$ above), we have
\[
\lim_{n \to \infty} \P\Big(  \min_{q \in \cF_\beta} \| \hat f_\beta - q \|_{L^1(\cL_X)} \le  \varepsilon \Big) = 1. 
\]
Finally, by Assumption \ref{ass:density_bound_general} that the density of $\cL_{|x}$ is uniformly bounded by $C$, we have the inequality that $|\P_{X, Y}(Y \le f(X)) - \P_{X, Y}(Y \le g(X)) | \le C \| f - g \|_{L^1(\cL_X)}$ for any fixed $f$ and $g$. By Assumption \ref{ass:independence-errtar} that $(\ERR_\star^{\fea}, \ERR_\star^{\test} = \ERRTar)$ is independent of $\{ (\ERR_t^{\fea}, \ERR_t^{\test}) \}_{t \ge 1}$ but has the same distribution $\cL$, taking $X = \ERR_\star^{\val}$ and $Y = \ERR_\star^{\test} =\ERRTar$, we have
\[
\begin{aligned}
&~\Big| \P_{X, Y}(Y \in [\hat f_{\alpha/2}(X), \hat f_{1 - \alpha/2}(X)]) - \P_{X, Y}(Y \in [ f_{\alpha/2}(X), f_{1 - \alpha/2}(X)]) \Big| \\
\le&~ C \Big[\| \hat f_{\alpha/2} - f_{\alpha/2} \|_{L^1(\cL_X)} + \| \hat f_{1-\alpha/2} - f_{1-\alpha/2} \|_{L^1(\cL_X)}\Big].
\end{aligned}
\]
Notice that $\P_{X, Y}(Y \in [ f_{\alpha/2}(X), f_{1 - \alpha/2}(X)]) = 1 - \alpha$. This gives that for any $\varepsilon > 0$, we have
\[
\P\Big( \Big| \P_{X, Y}(Y \in [\hat f_{\alpha/2}(X), \hat f_{1 - \alpha/2}(X)]) - (1 - \alpha) \Big| \le \varepsilon \Big) = 1. 
\]
This concludes the proof of Theorem \ref{thm:conditional_coverage}. 
\end{proof}

% \sm{Complete the proof of Lemma \ref{lem:ergodic_LLN} and \ref{lem:risk_growth}. }

% \begin{remark}
% We may also prove that 
% \[
% \lim_{n \to \infty} \P\Big( \overline{\textup{Err}}_{XY} \in C_n^\alpha \Big\vert \textup{Err}^{val}  \Big) = 1 - \alpha. 
% \]
% However, the proof of this may not be as easy. The reason is that, conditional on $\textup{Err}^{val}$, the process $\{ z_t \}_{t \ge 1}$ is no longer distributed as a stationary process. 
% \end{remark}

% \begin{remark}
% We can conditional on statistics of $z_{n - D - V + 1:n}$ other than $\textup{Err}^{val}$. A good statistics (which has a large power) should be a good point estimator of $\overline{\textup{Err}}_{XY}$. We believe that $\textup{Err}^{val}$ is a good point estimator of $\overline{\textup{Err}}_{XY}$. 
% \end{remark}

\section{Proofs for Section \ref{sec:FCV}}\label{sec:proof_FCV}

\subsection{Formal statement and proof of Proposition \ref{prop:consistency_variance_FCV}}\label{sec:proof_consistency_variance_FCV}

\begin{proposition}[Asymptotic normality of FCV(c) (restatement)]\label{prop:consistency_variance_FCV_restate} Assume that $\{ E_i \}_{i \ge 0}$ as defined in \eqref{eqn:hatERR-FCV} is stationary ergodic, and satisfies 
\begin{itemize}
\item[(1)] $\Ktrun$-independence: $\{ E_i \}_{i \in [k]}$ and $\{ E_i\}_{i \ge k + \Ktrun + 1}$ are independent for any $k \in \N$. 
\item[(2)] The variance is finite: $\Var(E_1) < \infty$. 
\item[(3)] Let $\cF_t = \sigma(E_s, s \le t)$ be a filtration. We have $\lim_{s \to \infty} \Var[ \E[E_{t+s} | \cF_t]  ] = 0$ for any $t \in \N$. 
\item[(4)] We have $\sum_{s = 0}^\infty \Var ( \E[ E_{t + s} | \cF_{t}] - \E[E_{t + s} | \cF_{t - 1}])^{1/2} < \infty$ for any $t \in \N$.   
\end{itemize}
Then we have convergence in distribution as $K \to \infty$
\begin{equation}\label{eqn:FCVc_CLT}
\big(\hatERR_K^\FCV - \ERR \big) / \hatSE^{\FCV(c)}_{K} \stackrel{d}{\longrightarrow} \cN(0, 1),
\end{equation}
where $\hatERR_K^\FCV$ is as defined in \eqref{eqn:hatERR-FCV}, and $\hatSE^{\FCV(c)}_{K}$ is as defined in \eqref{eqn:SE-FCV}. 
\end{proposition}

\begin{proof}[Proof of Proposition \ref{sec:proof_consistency_variance_FCV}]

The proof of Proposition \ref{sec:proof_consistency_variance_FCV} relies on the following two lemmas. 

\begin{lemma}[Convergence of sample autocovariance]\label{lem:sample_auto_convergence}
Suppose that $\{ E_t\}_{t \in [n]}$ is stationary ergodic with average $\bar E = n^{-1} \sum_{t = 1}^n E_t$. Denote its autocovariance function and sample autocovariance function as
\[
\textstyle \gamma(k) = \E[(E_1 - \E[E_1])( E_{k+1} - \E[E_{k+1}])], ~~~~ \hat \gamma_n(k) = \frac{1}{n - k} \sum_{t = 1}^{n - k} (E_t - \bar E) (E_{t + k} - \bar E). 
\]
Then for any fixed $k \in \N$, as $n \to \infty$, we have convergence in probability
\[
\hat \gamma_n(k) \stackrel{p}{\rightarrow} \gamma(k). 
\]
\end{lemma}

\begin{lemma}[Gordin's central limit theorem (\cite{hayashi2011econometrics} Page 402-405) ]\label{lem:Gordin_CLT}
Suppose that $\{ M_t \}_{t \in \Z}$ is mean-zero stationary ergodic, with autocovariance function $\gamma(k) = \E[M_1 M_{k+1}]$. Assume that 
\begin{itemize}
\item[(1)] The variance is finite: $\Var[M_t] < \infty$. 
\item[(2)] Let $\cF_t = \sigma(M_s, s \le t)$ be a filtration. We have $\lim_{s \to \infty} \Var[ \E[M_{t+s} | \cF_t]  ] = 0$ for any $t \in \N$. 
\item[(3)] We have $\sum_{s = 0}^\infty \Var ( \E[ M_{t + s} | \cF_{t}] - \E[M_{t + s} | \cF_{t - 1}])^{1/2} < \infty$ for any $t \in \N$.   
\end{itemize}
Then we have $\sum_{k = 0}^\infty | \gamma(k) | < \infty$, and as $n \to \infty$, we have convergence in distribution
\[
\frac{1}{\sqrt{n}} \sum_{t = 1}^n M_t \stackrel{d}{\rightarrow} \cN\Big(0, \gamma(0) + 2 \sum_{k = 1}^\infty \gamma(k)\Big). 
\]
\end{lemma}

By the assumption of Proposition \ref{sec:proof_consistency_variance_FCV} and by Gordin's central limit theorem (Lemma \ref{lem:Gordin_CLT}), as $K \to \infty$, we have convergence in probability
\begin{equation}\label{eqn:hatERR_CLT_in_proof}
\sqrt{K} \Big(\hatERR_K^\FCV - \ERR \Big) = K^{-1/2} \Big(\sum_{t = 1}^K E_t - \E[E_t] \Big) \stackrel{d} {\longrightarrow} \cN(0, \SE). 
\end{equation}
Here, taking $\gamma(k)$ to be the autocovariance function of $\{ E_i \}_{i \in [K]}$, we have
\[
\SE = \gamma(0) + 2 \sum_{k = 1}^\infty \gamma(k) = \gamma(0) + 2 \sum_{k = 1}^{\Ktrun} \gamma(k),
\]
where the last equality is by the assumption that $\{ E_i \}_{i \in [K]}$ is $\Ktrun$-independent. 

Furthermore, by the definition of $\hatSE^{\FCV(c)}_{K}$ as in \eqref{eqn:SE-FCV} and by Lemma \ref{lem:sample_auto_convergence}, as $K \to \infty$, we have
\begin{equation}\label{eqn:KhatSE_convergence}
K \cdot \big( \hatSE^{\FCV(c)}_{K} \big)^2 = \hat \gamma(0) + 2 \sum_{s = 1}^{\Ktrun} \Big( 1 - \frac{s}{K} \Big) \hat \gamma(s) \stackrel{p}{\rightarrow} \gamma(0) + 2 \sum_{s = 1}^{\Ktrun} \gamma(s) = \SE^2. 
\end{equation}
As a consequence, by Slutsky's theorem, \eqref{eqn:hatERR_CLT_in_proof} and \eqref{eqn:KhatSE_convergence} imply that \eqref{eqn:FCVc_CLT} holds. 
\end{proof}

\subsection{Proof of Proposition \ref{prop:fcv_p}}\label{sec:proof_fcv_p}

Since $\{ E_i\}_{i \in [K]}$ is a stationary ergodic process as in Assumption \ref{as:stationarity}, we have that for any $\varepsilon > 0$, 
\[
\lim_{K \to \infty} \P\Big( \Big\vert \frac{1}{K}\sum_{i = 1}^K E_i - \E[E_1] \Big\vert \ge \varepsilon \Big) = 0, ~~~~ \lim_{K \to \infty} \P\Big( \Big\vert \frac{1}{K}\sum_{i = 1}^K \big(E_i - \E[E_1] \big)^2 - \Var(E_1) \Big\vert \ge \varepsilon \Big) = 0. 
\]
This proves Eq. \eqref{eqn:V_K_convergence}. By the fact that $E_i$'s follow Gaussian distributions, we immediately have Eq. \eqref{eqn:quantile_converge_FCV_p} and \eqref{eqn:Errtar_FCV_p_coverage}.

\section{Proof of Theorem \ref{thm:ACI_delayed}}\label{sec:proof_ACI_delayed}

\begin{proof}[Proof of Theorem \ref{thm:ACI_delayed}]
Throughout the proof, we assume that $\Delta =1$. For general $\Delta > 1$, the proof follows similarly. Our proof is similar to Proposition 4.1 in \cite{gibbs2021adaptive} and Theorem 1 in \cite{feldman2022conformalized}. 

\begin{lemma}\label{lem:bounded_theta}
Under the condition of Theorem \ref{thm:ACI_delayed}. Then for all $t \in \N$, we have $\theta_t \in [m -  \ntest \gamma, M +  \ntest \gamma]$. 
\end{lemma}

\begin{proof}[Proof of Lemma \ref{lem:bounded_theta}]

Assume for the sake of contradiction that there exists $t \in \N$ such that $\theta_t > M + \ntest \gamma$ (the complementary case is similar). Further assume that for all $ t' < t$, we have $\theta_{t'} \le M +  \ntest \gamma$. Since $c_t, \alpha \in [0, 1]$, we get that: 
\[
\textstyle \theta_{t-\ntest} = \theta_t - \gamma\sum_{s = t - 2\ntest + 1}^{t-\ntest}(1 - \alpha - c_s) \ge \theta_t - \ntest \gamma > M + \ntest \gamma - \ntest \gamma = M.
\]
Therefore, $\theta_{t-\ntest} > M$. Since $\hatPI^t( * ; \theta) = \R$ for $\theta > M$, we get that $c_{t-\ntest} = 1\{ \ERRTar^{t - \ntest} \in \hatPI^{t - \ntest}( * ; \theta_{t-\ntest}) \} = 1\{ \ERRTar^{t - \ntest} \in\R \} = 1 > 1- \alpha$. As a result: $\theta_t = \theta_{t-1} + \gamma(1 - \alpha - c_{t-\ntest}) < \theta_{t-1} \le M + \ntest \gamma$. Which is a contradiction to our assumption.
\end{proof}

We next prove Theorem \ref{thm:ACI_delayed}. By applying Lemma \ref{lem:bounded_theta} we get that $\theta_t \in [m - \ntest \gamma , M + \ntest \gamma]$ for all $t \in \N$. Denote $m' = m -\ntest \gamma $, and $M' = M + \ntest \gamma$. We follow the proof of \cite{gibbs2021adaptive} and expand the recursion:
\[
[m', M'] \ni \theta_{T + 1} = \theta_\ntest + \sum_{s = \ntest}^T \gamma (1 - \alpha - c_s).
\]
By rearranging this we get that:
\[
\frac{m' - \theta_\ntest}{T \gamma} \le \frac{1}{T} \sum_{s = \ntest}^T (1 - \alpha - c_s) = \frac{\theta_{T+1} - \theta_\ntest}{T\gamma } \le \frac{M - \theta_\ntest}{T \gamma}.
\]
Therefore we get 
\[
\Big\vert \frac{1}{T} \sum_{s = \ntest}^T (1 - \alpha - c_s) \Big\vert \le \frac{(M - m) + 2 \ntest \gamma}{T \gamma},
\]
and 
\[
\Big\vert \frac{1}{T} \sum_{s = 1}^T (1 - \alpha - c_s) \Big\vert \le \frac{(M - m) + 3 \ntest \gamma}{T \gamma}. 
\]
Lastly, the definition of the loss, $c_t = 1 \{ \ERRTar^t \in \hatRI^{\alpha, t} \}$, gives us the risk statement. 
\end{proof}

\section{The QFCV point estimator and variations of QFCV intervals}\label{appendix:QFCV}

% In this section, we give different variations or application of the quantile-based forward cross-validation (QFCV) method. First we present natural point estimators that can be derived from QFCV method to estimate true test error. Next we discuss potential modifications in the structure of QFCV method, particularly in the computation of training, validation, and test errors for each rolling time window. Lastly, we demonstrate the application of QFCV methods in the commonly seen time series setting with enlarged time windows, where training data size increases. 

In this section, we first provide QFCV point estimators for the stochastic test error. We then discuss QFCV methods in the ``expanding window'' setting, which utilizes an incrementally expanding training dataset to fit the forecaster over time.

\subsection{QFCV point estimators}

We adopt the notations as in Section \ref{sec:QFCV_algorithm}. Recall that the tuples $\{ (\ERR_{i}^{\fea}, \ERR_{i}^{\test}) \}_{i \in [K]}$ and $(\ERR_\star^{\fea}, \ERR_\star^{\test})$ are defined in Eq. (\ref{eqn:Err_fea_test_definition}). Importantly, $\{ (\ERR_{i}^{\fea}, \ERR_{i}^{\test}) \}_{i \in [K]}$ and  $\ERR_\star^{\fea}$ are computable from the observed dataset, and our goal is to estimate $\ERR_\star^{\test}$. To provide a point estimator, we compute
\[
\textstyle \hat f = \arg\min_{f \in \cF} \frac{1}{K} \sum_{i = 1}^{K} \ell\big(\ERR_{i}^{\test}, f(\ERR_{i}^{\fea}) \big), 
\]
where $\ell(e, \hat e)$ is a loss function, such as the absolute or square loss. The QFCV point estimator is then given by
\[
\hatERR_K^\QFCV = \hat{f}(\ERR_\star^{\fea}).
\]
Note that $\ERR_\star^{\test}$ is a random variable, so this estimator is not a point estimator in the statistical decision-theoretic sense. 

In simulation examples of Section \ref{sec:QFCV_FCV_comparison}, we adopt the square loss to produce point estimators. Table \ref{table:FCV} shows that QFCV point estimators provide lower mean squared error (MSE) compared to the classical forward cross-validation estimator $\hatERR_K^\FCV$, especially when the time series is smoother so that past validation errors give stronger signals about future test error.

% One way to give point estimator is to slightly modify the quantile parameter in QFCV algorithm. In particular, we compute
% \[
%     \hat f^{\beta} = \arg\min_{f \in \cF} \frac{1}{K} \sum_{i = 1}^{K} \pinball_\beta\Big(\ERR_{i}^{\test} - f(\ERR_{i}^{\fea}) \Big), 
%     \]
%     where $\pinball_\beta(t) = - t \cdot (1\{ t \le 0 \} - \beta)$ is the pinball loss with quantile parameter $\beta$, for the choice of $\beta = 0.5$. A point estimator can then be given by 
% \[
% \hatERR_K^\QFCV = \hat{f}^{0.5}(\ERR_\star^{\fea}).
% \]

% Another way to give point estimator is to compute a linear regression fit $\hat{f}^{lr}$ with tuples $\{ (\ERR_{i}^{\fea}, \ERR_{i}^{\test}) \}$, which gives 
% \[
% \hatERR_K^\QFCV = \hat{f}^{lr}(\ERR_\star^{\fea}).
% \]
% In simulation examples, we adopt the second method to give point estimators and it can be seen from Table \ref{table:FCV} that QFCV derived point estimators can result in smaller mean squared error (MSE) as compared to classical forward cross-validation estimator $\hatERR_K^\FCV$ especially when the time series is more smooth so that past validation errors provide stronger signals for future test error. 

\subsection{Variations of QFCV method}
\subsubsection{Longer memory span}

% An illustrative diagram for the QFCV method is given in Figure \ref{fig:diagram_QFCV}. 

The construction of validation and test sets is flexible and can be tailored to practitioners' needs (recall the diagram for QFCV in Figure \ref{fig:diagram_QFCV}). Our theoretical guarantee for QFCV will hold as long as the covariates and targets are the same across folds. For instance, we can modify the number of covariates in quantile regression by including more past time windows. That is, we can replace step 2 in Algorithm \ref{alg:fcv_qc} with:
\[
\hat f^{\beta} = \arg\min_{f \in \cF} \frac{1}{K} \sum_{i = m}^{K} \pinball_\beta\Big(\ERR_{i}^{\test} - f(\ERR_{i}^{\fea}, \ldots, \ERR_{i-m+1}^{\fea}) \Big), 
\]
and replace the algorithm output with:
\[\hatPI^\alpha_\QFCV = [\hat f^{\alpha/2}(\ERR^{\fea}_\star,\ldots, \ERR_{K-m+2}^{\fea}), \hat f^{1 - \alpha/2}(\ERR^{\fea}_\star,\ldots, \ERR_{K-m+2}^{\fea})].
\]
The original QFCV method is the special case with $m=1$. Increasing the number of past windows $m$ may improve efficiency when there is a long-range correlation between past validation errors and future test errors. However, the tradeoff is slower mixing and convergence rates for quantile regression.

\begin{figure}[t]
    \centering
    \includegraphics[width = 0.6\linewidth]{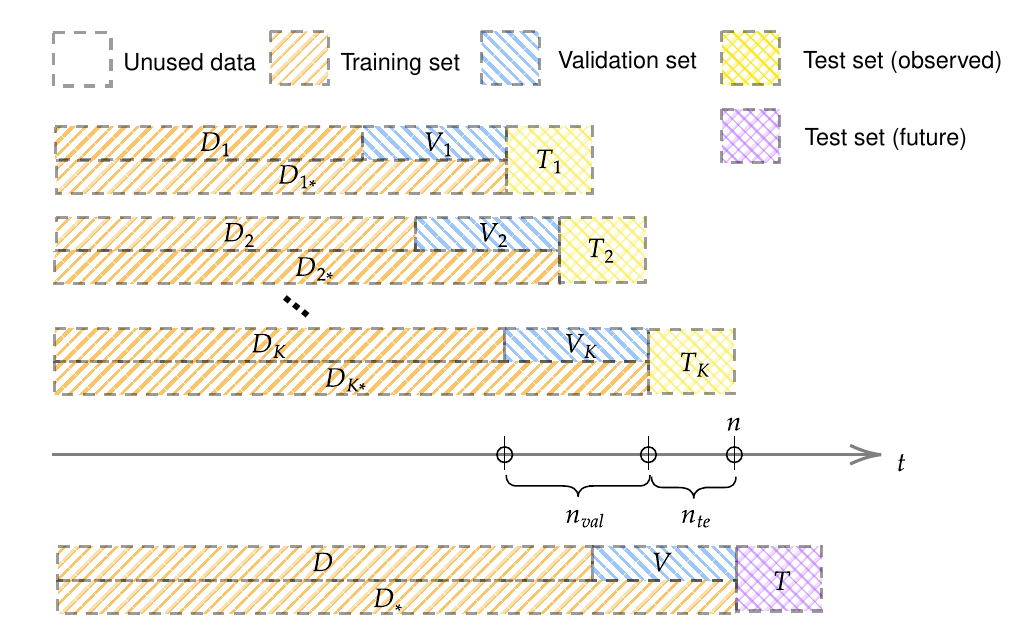}
    \caption{Diagram illustration QFCV with expanding time windows.}
    \label{fig:diagram_QFCV_ew}
\end{figure}

\subsubsection{Expanding window setup}

 \begin{figure}[t]
\centering
\begin{minipage}[c]{.5\linewidth}
  \centering
  \includegraphics[width = \linewidth]{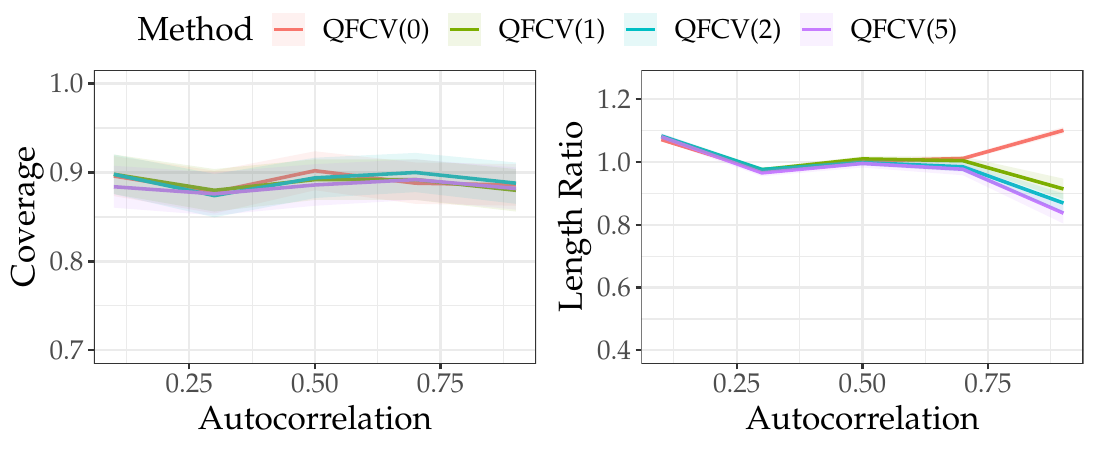}
  \subcaption{ARMA(1,0), $\nval = 5$.}
\end{minipage}%
\begin{minipage}{.5\linewidth}
  \centering
  \includegraphics[width = \linewidth]{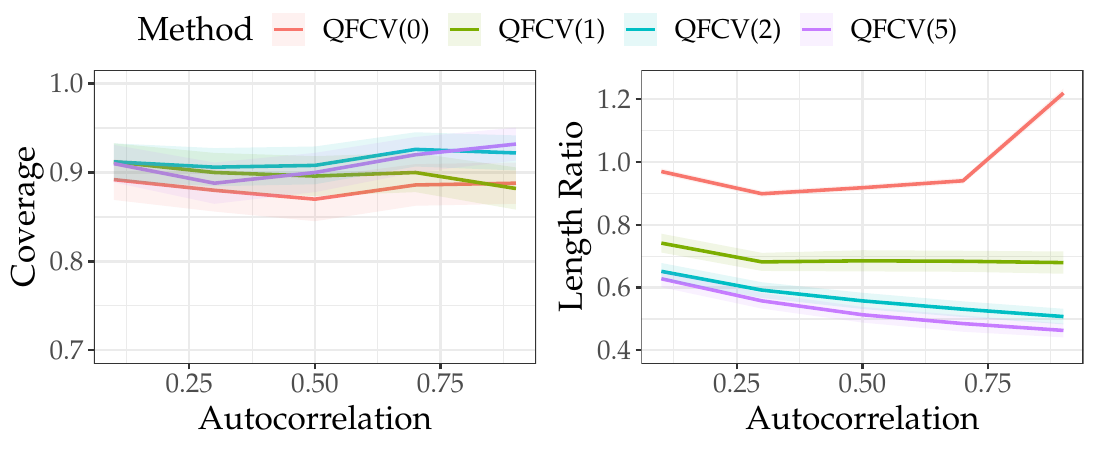}
  \subcaption{ARMA(1,20), $\nval = 5$.}
\end{minipage}
\vskip0.5cm
\begin{minipage}[c]{.5\linewidth}
  \centering
  \includegraphics[width = \linewidth]{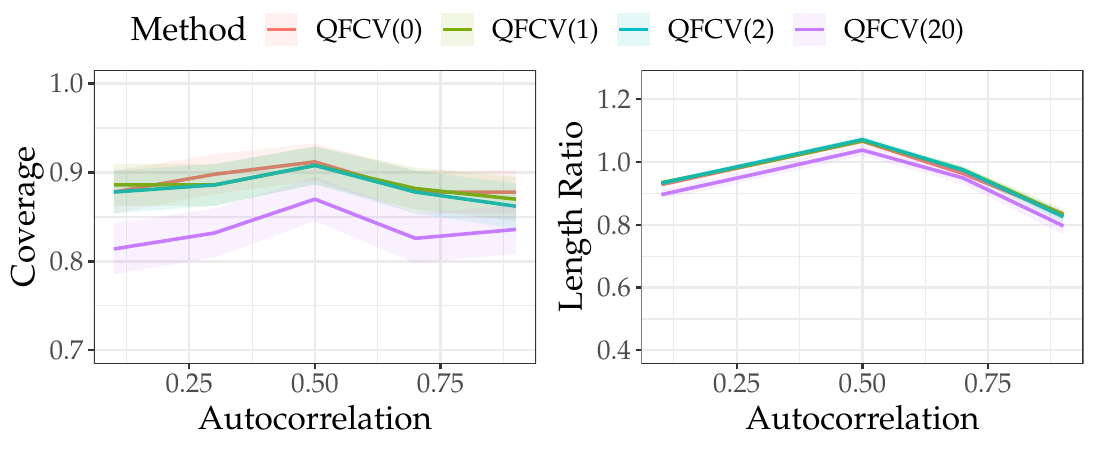}
  \subcaption{ARMA(1,0), $\nval = 20$.}
\end{minipage}%
\begin{minipage}{.5\linewidth}
  \centering
  \includegraphics[width = \linewidth]{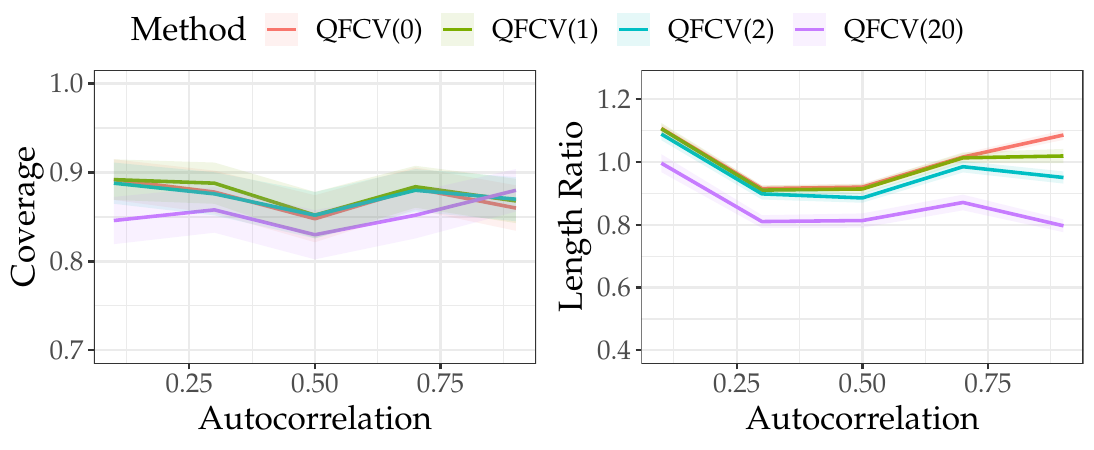}
  \subcaption{ARMA(1,20), $\nval= 20$.}
\end{minipage}

\caption{Comparison of different variants of QFCV methods for expanding time windows. The simulation setting is the same as Figure \ref{fig:sim_QFCV}, where the time series is generated from (\ref{eqn:simulation_linear_model}). Specifically, data is generated by $y_t = x_t^\sT \beta + \varepsilon_t$, where $\{\varepsilon_t\}_{t=1}^{\nobs}$ follows an ARMA(a,b) process. The time series length is fixed at $\nobs = 2000$, with $\ntrain = 40$, $\nval = \ntest \in \{5, 20\}$, and $p = 20$. The x-axis represents the autocorrelation parameter $\phi \in [0, 1]$. Each panel displays two plots: (i) actual coverage of the $90\%$ nominal QFCV intervals, and (ii) ratio of the QFCV interval length to the true marginal quantiles interval length. The true marginal quantiles interval is defined by the $5\%$ and $95\%$ quantiles of the true test error distribution. Each curve displays the mean and standard error over $500$ simulation instances. }
\label{fig:sim_QFCV_ew}
\end{figure}

The QFCV method in Figure \ref{fig:diagram_QFCV} uses a ``rolling window'' approach where a fixed-size training dataset is used to fit the forecaster. Besides this ``rolling window'' approach, another common setup is the ``expanding window'' setting, which uses an incrementally expanding training dataset to fit the forecaster over time. The QFCV method can be adapted to the expanding window setting, where an illustrative diagram is shown in Figure \ref{fig:diagram_QFCV_ew}. Specifically, we consider the following sets:
\begin{equation}
\begin{aligned}
D_i =&~ \{1, \ldots, {(i-1)\Delta + \ntrain}\}, ~~~~~~~~~~~~~~~~~~~~~~~~~~~~~~~~~~~~~~~~~~~~~~~~~~ D = \{1, \ldots,  \nobs - \nval \}, &~\\
V_i =&~ \{ (i-1)\Delta + \ntrain + 1, \ldots, (i-1)\Delta + \ntrain + \nval\},~~~~~~~~~~~~~~~~~~~~ V = \{\nobs - \nval + 1, \ldots,  \nobs \},&~ \\
D_{i\star} =&~ \{1, \ldots, {(i-1)\Delta + \ntrain + \nval}\}, ~~~~~~~~~~~~~~~~~~~~~~~~~~~~~~~~~~~~~~~~~ D_\star = \{1, \ldots,  \nobs\},&~ \\
T_i =&~ \{ (i-1)\Delta + \ntrain + \nval, \ldots, (i-1)\Delta + \ntrain + \nval + \ntest -1 \}, ~~~~~ T = \{ \nobs + 1, \ldots, \nobs + \ntest \}. &~
\end{aligned}
\end{equation}
With these sets $\{ D_i, V_i, D_{i \star}, T_i\}_{i \in [K]}$ and $\{ D, V, D_\star, T\}$, we can compute tuples $\{ (\ERR_{i}^{\fea}, \ERR_{i}^{\test}) \}_{i \in [K]}$ and $\ERR_\star^\fea$ as defined in Equation \eqref{eqn:Err_fea_test_definition}, then perform quantile regression to construct a prediction interval for $\ERR_\star^\test$ as detailed in Algorithm \ref{alg:fcv_qc}. Due to varying training size, we do not have stationarity for $\{ (\ERR_t^{\fea}, \ERR_t^{\test})\}_{t \ge 1}$, even if the original data sequence $\{ z_t \}_{t \ge 1}$ is stationary and ergodic. However, we may have approximate stationarity if the parameter estimates converge over time, providing some justification for using the QFCV method in the expanding window setting.

% Due to varying training size, we do not have stationarity for $\{ (\ERR_t^{\fea}, \ERR_t^{\test})\}_{t \ge 1}$, despite that the original data sequence $\{ z_t \}_{t \ge 1}$ is stationary and ergodic. However, we might have approximate stationarity if parameter estimates converge. In this case, using QFCV method with expanding time window might still give approximately correct coverage. In simulation, we repeat the same simulation experiments as in Figure \ref{fig:sim_QFCV} except that we adopt the expanding time window setting. 

In Figure \ref{fig:sim_QFCV_ew}, we repeat the same simulation experiments as in Figure \ref{fig:sim_QFCV} but using the expanding window setting. Figure \ref{fig:sim_QFCV_ew} shows that QFCV with different choices of $m$ generally achieves about $90\%$ nominal coverage when $m$ is not too large, although we have no theoretical guarantee. Notably, in panel (b) with $\text{ARMA}(1,20)$ model and $\nval = 5$, the QFCV method can produce prediction intervals at the nominal coverage level with only half the interval length compared to the true marginal quantiles interval. Besides, note that AQFCV method still has time-average coverage guarantee in the expanding window setup.

\end{document}